# End-to-end numerical modeling of the *Roman Space Telescope* coronagraph

John E. Krist[a], John B. Steeves[a,b], Brandon D. Dube[a], A. J. Eldorado Riggs[a], Brian D. Kern[a], David S. Marx[a], Eric J. Cady[a], Hanying Zhou[a], Ilya Y. Poberezhskiy[a], Caleb W. Baker[a], James P. McGuire[a], Bijan Nemati[c], Gary M. Kuan[a], Bertrand Mennesson[a], John T. Trauger[a], Navtej S. Saini[a], Sergi Hildebrandt Rafels[a]

[a]Jet Propulsion Laboratory, California Institute of Technology, 4800 Oak Grove Drive, Pasadena, CA
[b]Amazon, Project Kuiper, 8500 Balboa Blvd., Northridge, CA
[c]Tellus1 Scientific, LLC, 8000 Madison Blvd., Ste. D102-265, Madison, AL

**Abstract**. The *Roman Space Telescope* will have the first advanced coronagraph in space, with deformable mirrors for wavefront control, low-order wavefront sensing and maintenance, and a photon-counting detector. It is expected to be able to detect and characterize mature, giant exoplanets in reflected visible light. Over the past decade the performance of the coronagraph in its flight environment has been simulated with increasingly detailed diffraction and structural/thermal finite element modeling. With the instrument now being integrated in preparation for launch within the next few years, the present state of the end-to-end modeling is described, including the measured flight components such as deformable mirrors. The coronagraphic modes are thoroughly described, including characteristics most readily derived from modeling. The methods for diffraction propagation, wavefront control, and structural and thermal finite-element modeling are detailed. The techniques and procedures developed for the instrument will serve as a foundation for future coronagraphic missions such as the *Habitable Worlds Observatory*.

Contents

## 1 Introduction

The resolved imaging and spectroscopy of planets orbiting stars is hindered by the glare of instrumental diffraction and scatter. A coronagraph can be used to suppress the starlight that is diffracted by a telescope's obscurations (e.g., primary mirror edge, secondary mirror and its support struts). Most coronagraphs consist of masks placed at intermediate pupils and/or focal planes. They do little, however, to suppress the light scattered by imperfect optics that can be magnitudes brighter than the planet. This may be reduced by deformable mirrors (DMs) that compensate for the phase and amplitude wavefront errors induced by the imperfections (e.g., polishing errors). The combination of the coronagraphic masks and DMs, in conjunction with wavefront control (WFC) algorithms, produces a *dark hole* around the star largely free of diffracted and scattered light. Post-processing techniques can further reduce the residual starlight by subtracting images of reference stars observed in a similar manner, contingent on the optical stability of the system.

Ground-based telescopes with adaptive optics and the *Hubble Space Telescope* (*HST*) have imaged, and sometimes spectrally characterized, a few dozen young, self-luminous planets[1,2] in the near-infrared using a combination of coronagraphs and post-processing algorithms. *HST* and the *James Webb Space Telescope* (*JWST*) have simple coronagraphs capable of imaging circumstellar dust disks and young planets. However, both lack high-actuator-density DMs, advanced coronagraphic masks, and photon-counting detectors that are necessary to capture images and spectra of mature, reflected light planets. While testbeds and numerical modeling have advanced many of these technologies, none have been demonstrated in space. Such demonstrations[3] will be needed to justify the technological readiness of future missions designed to detect biosignatures on extrasolar planets, such as those evaluated in the recent Habex[4] and LUVOIR[5] studies and endorsed in the National Academies' 2020 Decadal Survey on Astronomy and Astrophysics[6].

In 2012, an opportunity arose to place such an instrument in space as part of the *Wide Field InfraRed Survey* (*WFIRST*) mission. *WFIRST* was the top recommendation of the 2010 Decadal Survey: a space telescope for wide-field infrared imaging to investigate cosmic structure, dark matter distribution, and microlensing exoplanet transits. *WFIRST* was renamed the *Nancy Grace Roman Space Telescope* in 2020. Originally envisioned as a 1.3 m off-axis telescope, it was redesigned to use a donated set of 2.4 m on-axis optics[7]. The larger size enabled the addition of a second instrument besides the Wide Field Imager (WFI), and NASA designated that it would be an advanced coronagraph (herein called the coronagraph instrument, or CGI[8]).

The *Roman* telescope is an anastigmatic design with the primary and secondary mirrors in an on-axis configuration. The secondary is supported by six struts in a tripod arrangement. It is fast (*f*/1.2 primary), with a wide field of view. WFI and CGI view different regions of the field, and each instrument has separate optical trains starting with the tertiary mirrors. The additional light due to the larger aperture versus the original 1.3 m design is a notable advantage for the WFI. However, the secondary mirror and strut obscurations provide copious edges for light to diffract, creating a significant difficulty for coronagraphs that naturally prefer simple, unobscured apertures for optimal starlight suppression with maximum exoplanet throughput.

At the time the decision was made to put a coronagraph on the mission, the maturity was low for any design that could effectively accommodate such an obscured system, and most of the potential techniques were largely conceptual. In 2013, NASA conducted a down-select process over the span of a few months to choose from six submitted designs those that were most promising and might realistically be ready to fly by the mid-2020's. The objective evaluations were primarily based on numerical optical modeling[9,10]. The coronagraph designers provided mask patterns and/or optical surface shapes, and those were run in a computer model of the notional system with realistic errors (optical surface and pointing errors), with WFC provided by DMs. These simulations provided critical performance parameters, including the residual dark hole intensity achieved after WFC, the effective throughput, and tolerances to pointing errors and low-order wavefront error drift. From these were derived estimates of the number of known radial-velocity-detected planets that could be imaged within a given timespan[11]. Together with more subjective judgements regarding technological readiness and complexity, these results formed the basis for selecting two coronagraph designs, the Hybrid Lyot Coronagraph (HLC)[12] for imaging and the Shaped Pupil Coronagraph (SPC)[13] for spectroscopy. A backup design was also identified, the Phase-Induced Amplitude Apodization Complex Mask Coronagraph (PIAACMC)[14], which had much higher throughput than the others; it was later dropped from contention due to issues with fabrication complexity and low-order aberration sensitivities.

Since those early days of concept formulation, the observatory and CGI designs have progressed as more detailed analyses have been conducted, budget and schedule limitations imposed, and technology advanced. The telescope obscuration pattern, the bane of the coronagraph, has evolved as the widths of the baffles on the secondary support struts have increased in response to scattered light analyses for the WFI. Meanwhile, a greater understanding of high-order WFC and improved coronagraphic mask designs have enabled higher performance with greater throughput and reduced aberration sensitivities. Predictions of the impact of time-dependent thermal and dynamic disturbances on wavefront stability have become increasingly accurate using finite element modeling and have directly altered the plans for on-orbit CGI observation planning.

In 2020, a major change implemented by NASA to meet budget and schedule constraints was the reclassification of CGI as a technology demonstrator. This removed multiple top-level requirements that were based on scientific performance and instead defined a single technology demonstration threshold requirement (TTR#5): *Roman shall be able to measure using CGI, with SNR ≥ 5, the brightness of an astrophysical point source located between 6 – 9 λ/D from an adjacent star of V magnitude ≤ 5, with a flux ratio ≥ 10⁻⁷; the bandpass shall have a central wavelength ≤ 600 nm and a bandwidth ≥ 10%.* This does not require measuring such a point source, but rather demonstrating in orbit that the measured throughput and instrumental background noise levels are sufficient to allow it. CGI is not required for *Roman* to meet its mission success criteria – the observatory performance requirements are set by WFI, and they define the telescope interface specifications to which CGI is designed (e.g., maximum errors in pupil position, pointing, aberrations, etc.). However, various operational accommodations have been agreed between Roman and CGI to make the environment for coronagraphy – particularly observatory stability – more benign and predictable without driving the *Roman* design.

The TTR itself does not specify a performance of significant scientific usefulness when compared to the capabilities demonstrated by current ground-based coronagraphs (except perhaps for bright circumstellar disks at visible wavelengths). It also applies only to the HLC due to the wavelength specification. However, the CGI project was asked to maintain the original system design that meets science-informed requirements that can deliver considerably better performance than the TTR, as long as doing so does not exceed the allocated budget and schedule. Therefore, there are more rigorous, but non-critical, performance goals defined with a myriad of low-level system capability specifications. Currently, CGI's expected performance meets or exceeds these more stringent goals. If the combined observatory and CGI system can show sufficiently high performance on-orbit, NASA may choose to devote additional *Roman* mission time to scientific observations, beyond that allocated to the technology demonstrations (2200 hours spread across multiple campaigns during the first 18 months of the mission).

CGI involves many components that have not been previously proven in flight, so estimates of its on-orbit performance are dependent on testbeds and modeling. Testbed experiments[15-,16,17,18,19,20,21] have been conducted over the past decade using masks like the flight designs to validate the basic concepts and operation of the selected CGI coronagraphs. The evaluation of the various possible conditions that may be encountered on-orbit have been constrained by hardware and schedule. In some cases, especially the DMs, the flight components have certain characteristics that differ from those used in the testbed experiments; some others include the filter mechanisms, cameras, and correction mechanisms. Lacking the actual telescope, the testbeds do not reproduce properties such as polarization aberrations. These limitations place much of the burden of tolerancing, wavefront control optimization, and performance prediction on numerical modeling. The diffraction model consists of representations of each optical surface including mirrors, lenses, and masks, with known or likely properties. The wavefront is mathematically propagated from surface to surface, reproducing the diffraction effects that would be encountered in the real system. Wavefront control using DMs is implemented with the same algorithms that will be used on-orbit. Time-dependent system variations are predicted by finite element thermal/structural and dynamic models, and their effects on the performance of post-processing algorithms can be evaluated[22].

Though modeling has been used for numerous ground and proposed space-based coronagraphs, its application to CGI over the past decade is the most detailed and complete so far, something that has only been possible with the resources available for a flight project and the quantity and completeness of data that becomes available after extensive testing of flight hardware. The extreme sensitivity of the coronagraph to wavefront errors has demanded thermal and structural modeling of high accuracy, requiring computational efforts that may have even exceeded those used for *JWST*. The resulting techniques and software developed for CGI are a significant product of the technology demonstration aspect of the mission, and future coronagraphic missions should take advantage of these investments[3].

We present here the pre-flight state of the CGI modeling at the beginning of instrument integration and testing. This incorporates the final telescope design, flight layout, and masks, as well as measurements of actual flight optics, including the DMs. In addition, more accurate representations of the thermal and structural behaviors of the system have been evaluated with more refined observation scenarios. Also detailed are aspects of the coronagraph designs and components that are critical to include in the models or have characteristics that are best described via modeling. We note that this is not a review of all available modeling techniques – there may be other procedures, algorithms, or software that can produce similar results, but we confine ourselves here to those used by the CGI project to generate time series of coronagraphic field variations.

**Notes on conventions used:**

The brightness of the instrumental background (diffracted and scattered light) surrounding a star is often described in terms of *contrast*, usually as the mean within a given region. Technically, *contrast* at a given location is the ratio of the per-pixel field intensity divided by the peak pixel intensity of the star if it were centered at that position. This equals the planet/star brightness ratio of an extrasolar planet whose peak pixel intensity equals the instrumental background's per-pixel intensity. This definition includes the field-dependent reduction in field source intensity due to the coronagraph's focal plane mask (FPM). A common approximation for contrast, *normalized intensity* (NI), omits this dependence. NI is computed using the peak pixel value of the star when the FPM is removed (but with pupil masks and Lyot stops included). This is suitable when discussing the instrumental background intensity as seen on the detector, as the effects of the FPM on that background are already included; if the background brightness is instead being explicitly compared to that of a field source (e.g., exoplanet), then *contrast* is a more appropriate measure. As this document is focused on the instrumental background, "contrast" is colloquially used to refer to the background measured as NI.

Angular separations are specified here in terms of $\lambda/D$ radians (the units of radians are usually implicit), where $\lambda$ is the wavelength and $D$ is the telescope diameter. For CGI, one $\lambda/D$ is equal to 0.050 and 0.072 arcseconds at $\lambda = 575$ nm and 825 nm, respectively. When discussing a finite spectral bandpass, we will refer to the central wavelength as $\lambda_c$ and 50% transmission bandwidth as a percentage of $\lambda_c$.

Low order aberrations are described here in terms of Zernike polynomial coefficients with Noll ordering (e.g., Z4 is defocus, Z6 is 0° astigmatism, etc.). These are orthonormalized for an unobscured circular aperture, despite *Roman* having a central obscuration. In the case of CGI analyses, these Zernikes are used as convenient individual mode shapes and are not used together statistically (e.g., the values are not root-sum-squared to report total low order wavefront error).

A list of acronyms is provided as Appendix A.

## 2 CGI overview

### 2.1 Layout

The final (Phase C) optical layout of the CGI (Figure 1) encompasses a number of modifications made since 2016[23]. The prior layout had separate direct imaging and integral field spectrograph (IFS) channels, each with its own detector and selected using a flip mirror located in the back end of the instrument. To meet mass, power, and cost constraints, the IFS channel was removed and replaced with slits and a dispersing prism that can be inserted into the imaging beamline to create a spectrum of a single source on the imaging detector.

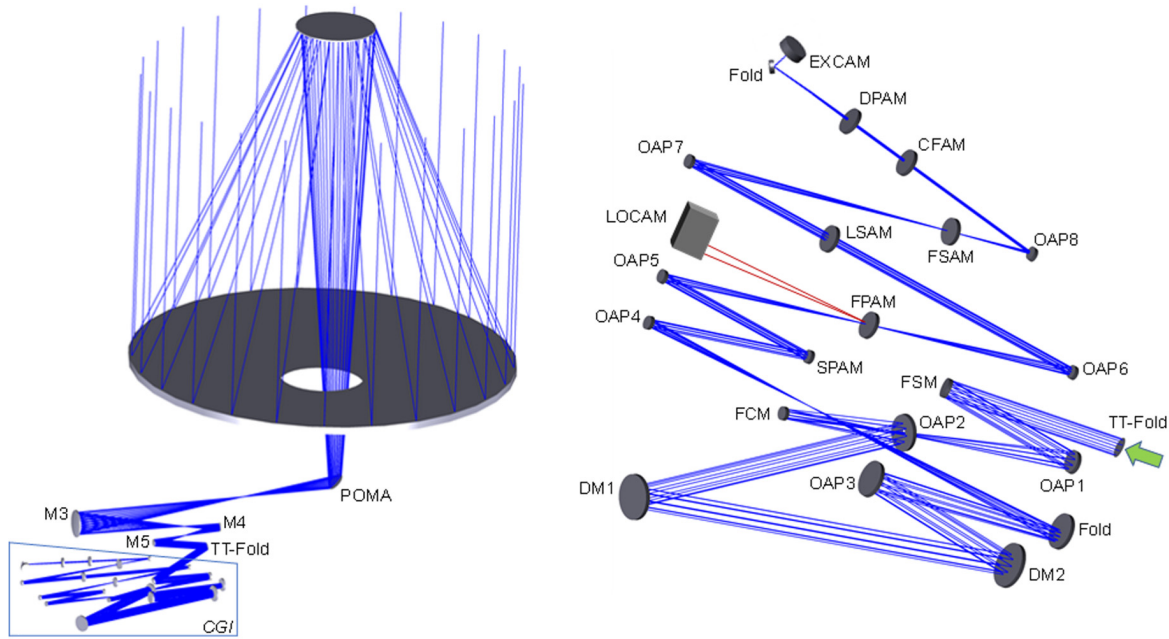

**Figure 1 Layout of the Roman Space Telescope. (Left) The full OTA, TCA, and CGI optical system. (Right) The CGI optical system, beginning with the beam propagating from the TT-Fold (highlighted by the arrow on the right) to the FSM.**

The coronagraph is a very complex optical system. Including the telescope, there are 22 reflective and at least 5 transmissive optics/masks/filters in the main beamline, plus six more in the optical train exclusive to the low-order wavefront sensor (LOWFS)[24], delivering light to its dedicated low-order sensor camera (LOCAM). These are necessary to produce the multiple pupil and focal planes at which masks[25], actuated mirrors, or detectors are placed.

The entrance pupil is defined by the stop at the primary mirror with a clear aperture diameter ($D_{CA}$) of 2.363 m. Over the years the obscuration pattern formed by the secondary baffle and struts has evolved as their sizes changed. There was a simplification of the baffle shape and increases in the strut widths, and the coronagraphs had to be optimized every time the pattern was modified. In the final design the secondary mirror obscuration diameter is 0.3 $D_{CA}$ and the six secondary support struts are each 0.032 $D_{CA}$ wide.

The primary and secondary mirrors compose the Optical Telescope Assembly (OTA)[26]. The beam from the secondary is directed through the central hole in the primary and into the Pickoff Mirror Assembly (POMA). In an early (Phase B) layout the POMA used three flats, each inclined 16° from the chief ray, to bend the beam behind the primary. It was thought that using multiple mirrors with closer-to-normal incidence angles rather than one mirror with a large (~45°) inclination would avoid creating significant polarization cross-terms that would limit the dark hole contrast. Subsequent analysis showed that the aberrations from a 45° fold were not worse than those from the multi-mirror configuration, so the POMA was simplified to a single flat.

The POMA fold is the first component in the Tertiary Collimator Assembly (TCA) that collimates the beam into CGI and corrects the off-axis aberrations (the CGI field is located 0.4° from the telescope's optical axis). After the POMA are three off-axis aspheric mirrors (M3, M4, M5) and a fold mirror (TT-Fold) that is tip/tilt actuated to allow aligning the pupil inside CGI. The TCA is attached to the telescope aft metering structure and is considered exterior to the CGI. The WFI has its own tertiary optics and is located at a different location in the telescope focal plane (the WFI acts as the fine guidance sensor).

The TCA forms a pupil image at the Fast Steering Mirror (FSM) that is used to correct rapid pointing drifts based on measurements from LOCAM. A relay then creates another pupil image on the first deformable mirror (DM1). Within that relay is a flat, the Focus Control Mechanism (FCM), which is mounted on a two-stage actuator that allows both fine adjustment of focus based on measurements from LOCAM and coarse adjustments from phase retrieval measurements. The second DM (DM2) is located 1.0 m downstream from DM1, the separation between them allowing for control of both phase and amplitude wavefront errors over a 360° field. Both DMs are AOA Xinetics Photonex[27] 48×48 actuator modules with 0.9906 mm actuator pitch and a continuous facesheet. The geometric image diameter at the DMs is 46.3 mm, ignoring inclination.

Following DM2, another relay creates a pupil image at the Shaped Pupil Alignment Mechanism (SPAM) that allows the selection of a reflective shaped pupil mask for the SPC or a flat for the HLC (note that each alignment mechanism in CGI is an X-Y translation stage). The beam is then focused onto the Focal Plane Alignment Mechanism (FPAM) on which a variety of focal plane masks (FPMs) are available. An OAP then forms a pupil image on the Lyot Stop Alignment Mechanism (LSAM) containing the pupil stops, after which the beam is focused at the Field Stop Alignment Mechanism (FSAM). The HLC requires a field stop just large enough to pass the dark hole region; the light just outside of the hole would be scattered inside it by subsequent optics and would also saturate the detector if not blocked. The spectroscopic slits are also located in the FSAM.

The last OAP collimates the beam again, sending it to the Color Filter Alignment Mechanism (CFAM) and then to the Dispersion/Polarization Alignment Mechanism (DPAM) that contains imaging and defocusing lenses, Amici zero-deviation dispersing prisms for spectroscopy, and polarizers (Wollaston prisms that simultaneously produce two images on the detector at orthogonal polarizations; the default mode is no polarizers)[28]. One more fold mirror then directs the beam onto the Exoplanetary Camera (EXCAM), an electron-multiplied charge-coupled device[29] (EMCCD) detector that allows both conventional analog CCD imaging, using EM gain to reduce read noise, and photon counting in low-flux conditions with effectively zero read noise. EXCAM is used for both two-dimensional imaging and the capture of the dispersed spectrum when the slit and prism are inserted.

A separate channel is devoted to LOCAM which uses the Zernike phase contrast technique[30] to measure time-dependent low-order wavefront aberration changes (tip/tilt from pointing errors;

defocus, coma, astigmatism, trefoil, and spherical from thermally induced structural changes). The front surface of each FPM is reflective and is topped with either a patterned dielectric coating with a small dimple (HLC) or a similarly sized raised spot (SPC) that interferes the central portion of the stellar point spread function (PSF) with the rest. As the beam propagates to successive pupil images, this transforms invisible phase variations into measurable intensity changes. A 128 nm-wide filter ($\lambda_c$ = 575 nm) and lenses create a pupil image on the LOCAM EMCCD. Differences in the pupil intensity pattern over time are translated into low-order wavefront error change measurements, and correction commands are then sent to the FSM, FCM, and DM1.

**Table 1. Roman CGI bandpass specifications[1]**

| Bandpass Name | Central Wavelength $\lambda_c$ | FWHM[2] Bandwidth $\Delta\lambda/\lambda_c$ |
|---|---|---|
| 1 | 575 nm | 10.1 % |
| 1a | 556 nm | 3.5 % |
| 1b | 575 nm | 3.3 % |
| 1c | 594 nm | 3.2 % |
| 2 | 660 nm | 17.0 % |
| 2a | 615 nm | 3.6 % |
| 2b | 638 nm | 2.8 % |
| 2c | 656 nm | 1.0 % |
| 3 | 730 nm | 16.7 % |
| 3a | 681 nm | 3.5 % |
| 3b | 704 nm | 3.4 % |
| 3c | 727 nm | 2.8 % |
| 3g | 752 nm | 3.4 % |
| 3d | 754 nm | 1.0 % |
| 3e | 778 nm | 3.5 % |
| 4 | 825 nm | 11.4 % |
| 4a | 792 nm | 3.5 % |
| 4b | 825 nm | 3.6 % |
| 4c | 857 nm | 3.5 % |
| lowfs | 575 nm | 22.5% |

[1]Design specifications
[2]Full-width-at-half-maximum (FWHM) transmission of filter relative to $\lambda_c$

## 2.2 Bandpasses

There are four broad filters, Bands 1 – 4. Band 1 ($\Delta\lambda/\lambda_c$ = 10%, $\lambda_c$ = 575 nm) is used by the HLC and is the primary technical demonstration and exoplanet imaging bandpass. Being the shortest bandpass, it offers the smallest IWA on the sky. Band 2 (17%, $\lambda_c$ = 660 nm) and Band 3 (17%, $\lambda_c$ = 730 nm), which have considerable overlap, are for use with the slit and prism in the SPC spectroscopic mode. Band 4 (11%, $\lambda_c$ = 825 nm) is intended for disk imaging in the wide-field SPC mode (though HLC can be used for disk imaging and the Wide-field-of-view SPC (SPC-WFOV) for exoplanets). Sampling the spans of these bandpasses are narrower filters, 3 each for Bands 1 and 4, and 5 for Bands 2 and 3. These "calibration" filters are used while digging the dark hole to measure the chromatic wavefront errors to derive a broadband DM solution. Some narrowband filters (2c, 3d) are used for spectroscopic calibration. The bandpasses are listed in

Table 1. Neutral density filters are also available to prevent detector saturation during bright star target acquisition.

## 3    Coronagraphs

There are three baseline CGI coronagraphic modes that we describe here: HLC, spectroscopy SPC (SPC-Spec), and wide field of view SPC. Of these, only the HLC is necessary to meet the technology demonstration requirements and will be tested fully before launch. Besides these, there is an assortment of other FPMs and Lyot stops contributed by the NASA Exoplanet Exploration Program[25]. These are not tested or calibrated by the CGI project on the ground and so will require on-orbit calibration, plus potentially some additional software development. They include HLC and SPC masks that cover additional field orientations and bandpasses, different field sizes, high contrast imaging around binary stars[31], and EXCAM Zernike wavefront sensing[32]. Only the baseline coronagraphs are discussed here.

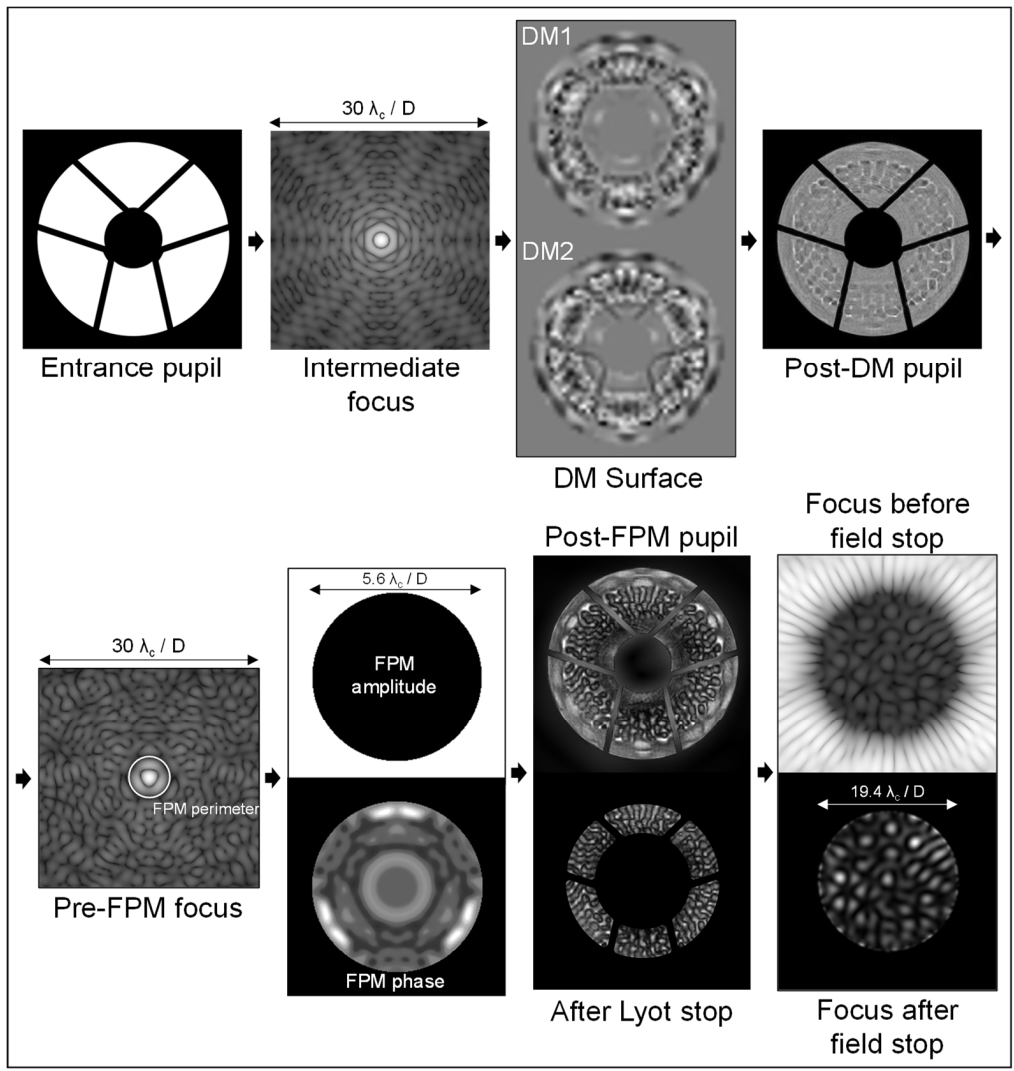

**Figure 2. Hybrid Lyot Coronagraph principal planes and masks. Images are at 575 nm in an unaberrated system. Grayscales are intensities except in panels labeled phase or surface.**

### 3.1 Hybrid Lyot coronagraph (HLC)

The HLC is designed for imaging extrasolar planets, and it is the mode that will be used to formally meet the CGI technology demonstration requirements. Optimized for Band 1 (10% bandpass at $\lambda_c = 575$ nm), it provides a 360° field of view around the star over an annulus of $r = 2.8 - 9.7 \, \lambda_c/D$ (the inner radius is defined by the FPM, the outer by the specified DM control region and field stop). It has an FPM and Lyot stop like a classical Lyot coronagraph, but it also uses the DMs to compensate for diffraction by the telescope obscurations.

The DMs, FPM, and Lyot stop are optimized together using modeling techniques and iterative wavefront control algorithms like those described later to create a dark hole with suppressed diffracted light in an unaberrated system. The flight design was generated using the FALCO (Fast Linearized Coronagraph Optimizer)[33,34] software, which is available for Matlab and Python. Early designs showed that it is not sufficient to simply dig the darkest dark hole possible, as doing so can degrade other important properties, such as the PSF sharpness or aberration tolerance. Hence, the coronagraphic performance is gauged against stellar light suppression, exoplanet throughput, and tolerance of low-order aberrations (including pointing errors), all combined to predict the signal-to-noise ratio of an exoplanet[35].

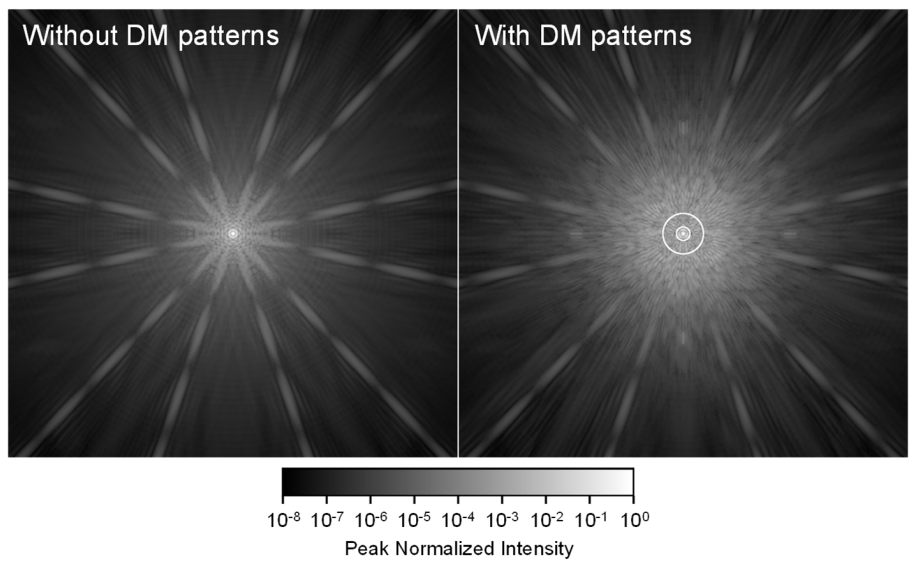

**Figure 3. The large-angle Band 1 field at the plane of the HLC FPM in the unaberrated system (left) without and (right) with the DM patterns used to compensate for the obscurations. Note that the diffraction spikes fade into a field of speckles close to the star, a result of the wavefront modification by the DMs. The dark hole region of $r = 3$ & $9 \, \lambda_c/D$ is noted by circles on the right image. Each image is 200 $\lambda_c/D$ (10") on a side and displayed as log(intensity).**

The DMs create a highly structured wavefront pattern, intentionally introducing ~76 nm RMS (root mean square) of WFE with ~184 nm peak-to-valley actuator stroke, before adding corrections for aberrations. Most of the modulation occurs in the annulus of the pupil that is not later masked by the Lyot stop; the masked areas have a much weaker effect, with their impact indirectly felt due to subsequent diffraction by the FPM. As shown in Figure 2, the obscurations remain prominent and fairly sharp in the post-DM pupil, unlike other methods that use analytically computed patterns on the DMs to perform pupil remapping to apodize the obscurations. At focus, these patterns result in a PSF (Figure 3) that has a halo of speckles within the region of influence of the DMs ($r \approx 24$ $\lambda/D$), inside which the diffraction spikes are obfuscated. Only 47% of the starlight falls inside the FPM perimeter – without the DMs patterns it would be 82%, so there is a steep price to pay in PSF sharpness for compensating for the obscurations.

The FPM[36] is a partially transmissive (1.6% amplitude) nickel-on-titanium spot 2.8 $\lambda_c/D$ in radius. It is topped with a patterned dielectric (PMGI) coating that modulates the phase of the PSF core in transmission (the combination of amplitude and phase modulation constitutes the 'hybrid' part of the coronagraph). The reflection off the front side of this pattern produces a phase-contrast pupil image in LOCAM. Early versions of the FPM used fully circularly-symmetric dielectric patterns (rings), but the final Phase C pattern is asymmetric, which modeling shows provides higher-contrast performance but at the expense of degraded LOCAM sensing.

The DMs and FPM create a complicated pupil distribution in the plane of the Lyot stop, with significant high-spatial-frequency ripples (Figure 2). This is unlike a classical Lyot coronagraph in which the remaining starlight is concentrated in and around the obscurations. The stop reduces the geometric clear area of the pupil, and consequently the transmission of field sources, by 58%. About 15% of the starlight passes through the Lyot stop in the form of high-spatial-frequency ripples that end up in the bright ring around the dark hole (only ~$10^{-8}$ of this flux ends up inside the hole). The $r = 9.7$ $\lambda_c/D$ field stop prevents scatter by the remaining optics and saturation of the detector. The Lyot and field stops are open apertures in silicon-on-insulator (SoI) wafers with aluminum coatings for opaqueness[37].

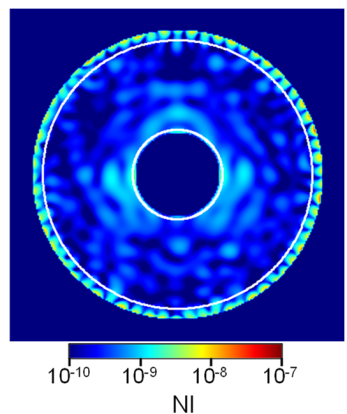

**Figure 4. HLC Band 1 normalized intensity map for the unaberrated system using the design DM patterns. The superposed circles indicate the evaluation region between $r = 3$ & $9$ $\lambda_c/D$. The mean NI within that region is $3 \times 10^{-10}$.**

### 3.1.1 Predicted HLC performance

**The as-designed HLC achieves a mean dark hole NI over $r = 3 – 9$ $\lambda_c/D$ of $3 \times 10^{-10}$ in Band 1 in the unaberrated model (**

Figure 4), which may seem superfluous given the CGI goal of ~$10^{-9}$ and the technology demonstration requirement above that. However, this margin provides a low base upon which aberrations, pointing errors, and fabrication defects will be added. After surface and polarization-dependent aberrations are included the mean contrast climbs to $5 \times 10^{-4}$, then WFC is used to dig

a dark hole, reaching $6 \times 10^{-10}$ for the mean polarization (additional effects such as misalignments and mismatches between the system and control models will raise the final value, as described later). Unless otherwise noted, the performance results presented in this section for the aberrated system represent the best possible, using the known and predicted system properties with no mask fabrication errors, misalignments, imperfect DM actuators, or modeling uncertainty factors.

### 3.1.2 HLC aberration sensitivities

Pointing errors, wavefront drift due to thermal changes, and polarization-dependent aberrations primarily manifest as low-order wavefront errors that impact dark hole static contrast and stability. The coronagraph is not uniformly sensitive to all such aberrations, so WFE drift will alter the resulting speckle pattern by different magnitudes depending on the form of the change. Each coronagraph has different sensitivities; the HLC's are shown in Figure 5. These are derived by introducing a fixed amount of a particular aberration and computing the change in intensity relative to the unperturbed image. Note that the RMS of the differences between images is shown.

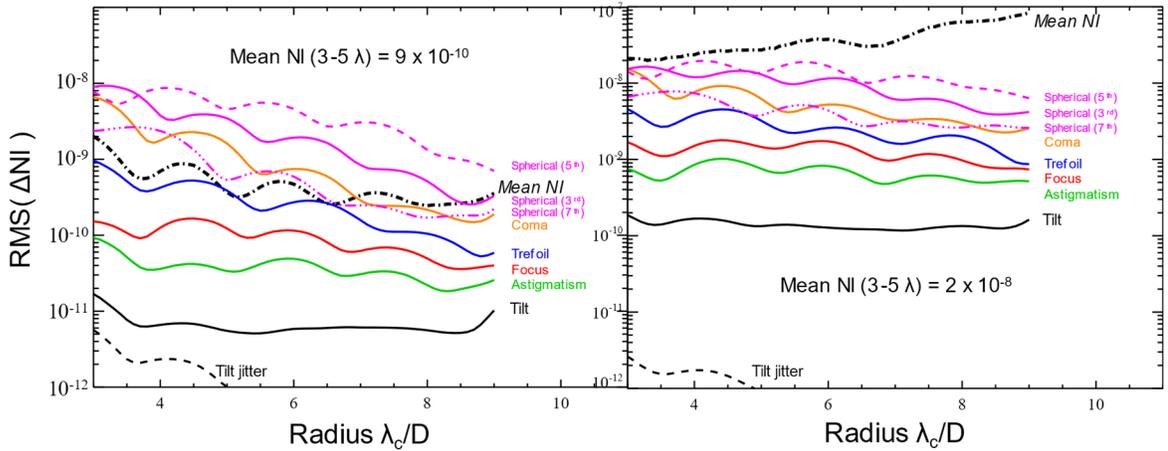

**Figure 5. HLC (Band 1) contrast changes measured in $\lambda_c/D$-wide annuli versus field angle for 100 pm RMS of low-order aberration with unperturbed dark hole fields of $9\times10^{-10}$ (left) and $2\times10^{-8}$ (right) mean NI over $3 - 5$ $\lambda_c/D$. *Tilt jitter* uses the mean of two images with -100 pm and +100 pm RMS of tilt.**

The HLC, partly by design[38], has low relative sensitivities to defocus, astigmatism (the dominant polarization-dependent aberration), and tip/tilt (introduced by uncorrected pointing errors and finite stellar diameter). It has higher sensitivities to coma and varieties of spherical aberration, which will be important later when discussing the thermal and structural modeling results. These sensitivities are proportional not just to the square of the aberration change but also to the level of the ambient field. The perturbed instantaneous electric field at time $t$ in the image plane as a function of the distribution in the entrance pupil can be described as:

$$E(t) = C\left(Ae^{i(\varphi + \delta(t))}\right) \qquad (1)$$

where $A$ is the pupil amplitude, $\varphi$ is the static aberration (e.g., surface and polarization-dependent aberrations), $\delta$ is the dynamic aberration (e.g., pointing error or thermally-induced wavefront change), and $C()$ is the function that propagates the pupil field through the coronagraph. The static, $E_0$, and dynamic $\delta E(t)$, fields can be separated:

$$E_0 = C\left(Ae^{i\varphi}\right) \qquad (2)$$

$$\delta E(t) = C\left(Ae^{i\varphi}\left[e^{i\delta(t)} - 1\right]\right) \qquad (3)$$

The perturbed field intensity can be written as:

$$I(t) = |E_0 + \delta E(t)|^2 = |E_0|^2 + 2Re[E_0^*\delta E(t)] + |\delta E(t)|^2 \qquad (4)$$

where $Re()$ takes the real part and $E_0^*$ is the complex conjugate of the static field. The static field in the cross-term determines the magnitude of the dynamic component, an effect sometimes called *speckle pinning*[39]. If $|E_0|$ is large compared to $|\delta E|$, then the cross-term dominates, and the intensity change is linear with $|\delta E|$, otherwise it is quadratic, scaling as $|\delta E|^2$. As shown in Figure 5, the sensitivities computed with a $2\times10^{-8}$ mean NI dark hole solution (an arbitrary illustrative case) are about an order of magnitude greater than those for at $9\times10^{-10}$. Figure 6 plots the sensitivities versus magnitude for a $9\times10^{-10}$ mean NI field and shows where the linear-to-quadratic transition occurs for each aberration at that contrast.

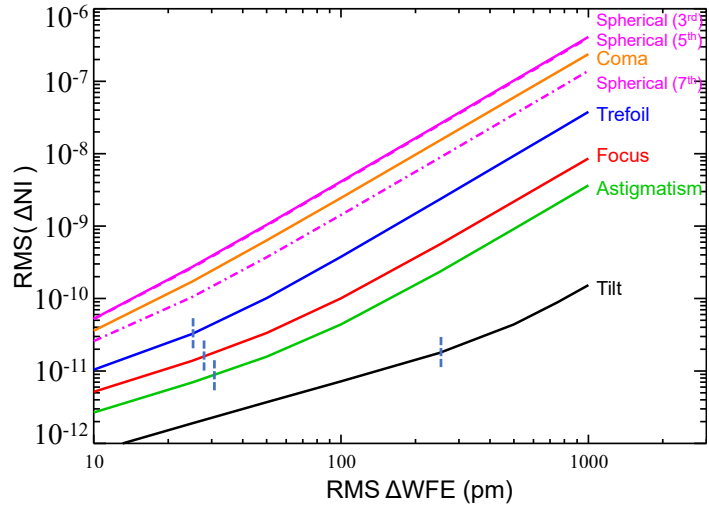

**Figure 6.** HLC (Band 1) contrast changes relative to low-order aberration measured over a $r = 3 - 5\ \lambda_c/D$ annulus. The default image was obtained after running WFC on the aberrated system model and includes polarization but no jitter and has a dark hole mean NI of $9\times10^{-10}$ over $3 - 5\ \lambda_c/D$. Vertical dashed lines indicate the approximate aberration sensitivity transitions from linear (leftwards) to quadratic (rightwards). The spherical and coma transitions occur <10 pm RMS.

The cross-term can be positive or negative, which greatly reduces the effect of aberrations that occur in equal amounts but with opposite signs over time (e.g., pointing jitter) or extent (finite stellar diameter). When the instantaneous intensities are added together over time or space, the cross-terms largely subtract out, leaving just the $|\delta E|^2$ terms that are independent of the static field.

This is demonstrated by the "tilt jitter" profiles in Figure 5, produced using the mean of images with -100 pm and +100 pm RMS of tilt.

The effects of pointing jitter together with a finite diameter star in the HLC are shown in Figure 7 and 8. How jitter is included in the models will be described in Section 5.7. The jitter range shown is what is expected as residuals from the post-FSM pointing corrections. The current dynamical predictions suggest that large jitter (>0.8 mas RMS, Gaussian distribution per axis) should only occur for a small fraction of the observing time, and most of the time the jitter is <0.4 mas RMS. Most stars likely to be observed will also be <1 mas in angular diameter; as a rule-of-thumb, a star of angular diameter $\theta$ will have a very similar dark hole impact as ¼$\theta$ of RMS jitter.

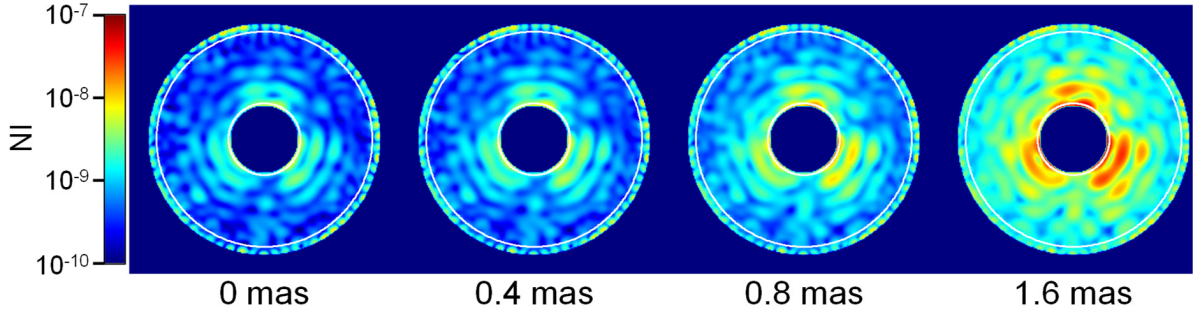

**Figure 7. HLC (Band 1) dark hole normalized intensity maps with different amounts of pointing jitter (values are RMS Gaussian jitter per axis), and all also include a 1 mas-diameter star. The default image was obtained after running WFC on the aberrated system model and includes all polarization components.**

### 3.1.3  HLC off-axis PSF

While the WFE injected by the DMs is beneficial for suppressing the diffraction from the obscurations, it significantly degrades the off-axis PSF by scattering light from the core into the wings (Figure 9). With flat DMs, the telescope would produce a PSF at the plane of the FPM that concentrates 80% of its total energy within $r = 2\ \lambda_c/D$; with the HLC DM pattern included, the same amount is contained within $r = 17.5$ $\lambda_c/D$ (Figure 10). The large amount of light in the wings is especially problematic for imaging extended circumstellar disks[40] as it creates confusion as to where flux seen at any location actually originated. The Lyot stop further reduces throughput and broadens the field PSF, pushing down its peak-pixel flux to 18% of the telescope's PSF with flat DMs and just 10% with the HLC DM patterns included.

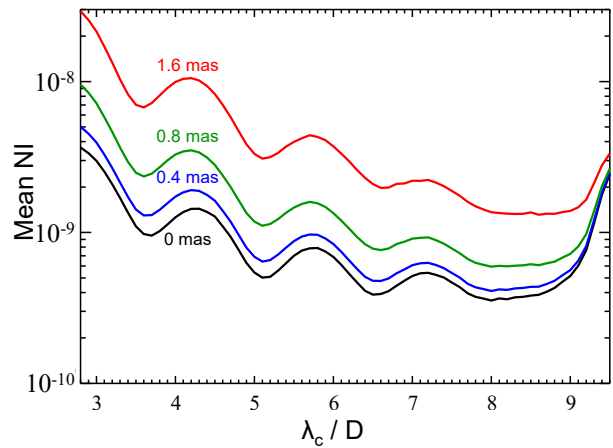

**Figure 8. Azimuthal mean contrast plots corresponding to the maps in Figure 7.**

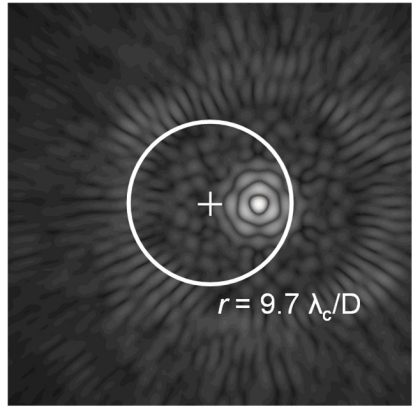

**Figure 9. A point source at the final focus in HLC Band 1 offset by 6 $\lambda c/D$ (0.3") from the FPM center, which is marked by a cross. The perimeter of the circular field stop (which is not applied here) is superposed. Displayed as log(intensity). Note that the shadow of the FPM is not seen because the field at the FPM is subsequently filtered by the Lyot stop.**

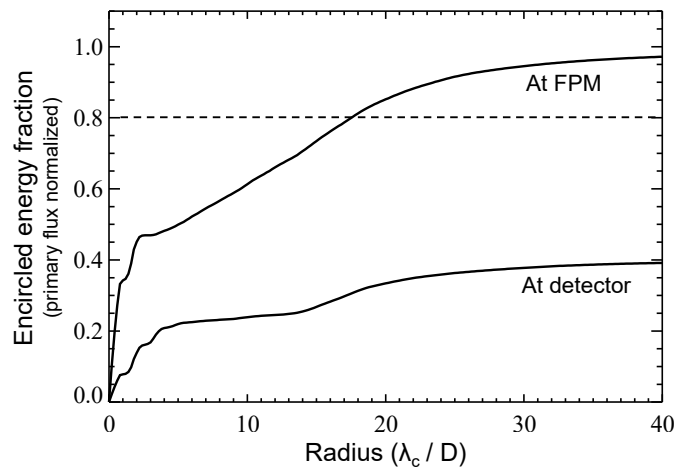

**Figure 10. Encircled energy (fraction of light within a given radius) of the Band 1 HLC PSF without the FPM or field stop. The HLC DM pattern is included. The curves are primary flux normalized. The "At detector" curve includes the 42% transmission of the Lyot stop and omits the field stop. The radius is relative to the center of the PSF, not the FPM. The dashed line indicates 80% encircled energy.**

Because most of the light in the wings is effectively lost in the detector noise, the amount in the core is of critical importance for exoplanet detection. As a measure of this we use the *core throughput* metric, the fraction of light from an exoplanet incident on the primary mirror that ends up in the PSF core (defined as all PSF pixels ≥50% of the maximum value, which is close in size to the optimal photometric aperture). The DM pattern is included, as are throughput losses from coronagraphic masks but not losses from reflections, filters, detector efficiency, etc. (*i.e.*, without the coronagraph masks, the total PSF flux would equal 1.0).

Core throughput for the HLC is plotted versus field angle in Figure 11, reaching a maximum of 4.2% (for the unaberrated *Roman* telescope without a coronagraph, it is 37%). It declines near the star due to truncation of the core by the FPM, with 50% of the maximum core throughput at $r$ = 3.0 $\lambda_c/D$. The area of the core region is 2300 mas$^2$ (1725 mas$^2$ without the coronagraph); the more compact the core area, the better, as it means there will be less background noise. At most, 38% of the light from a field source ends up on the detector. Note that the core throughput decreases beyond 6 $\lambda_c/D$ because the beam shear on DM2 from the off-axis source degrades the wavefront correction.

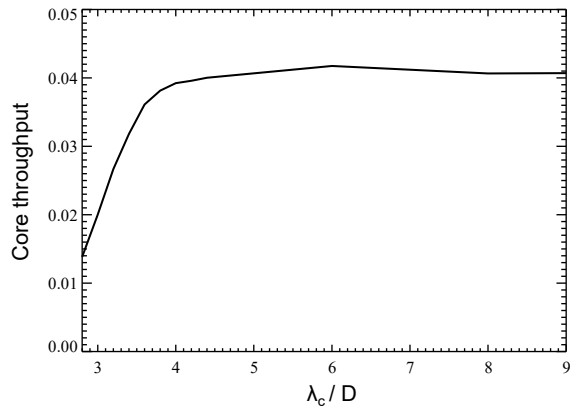

**Figure 11. HLC Band 1 field PSF core throughput fraction versus field position.**

A side effect of the truncation at the FPM (Figure 12) is distortion that causes the apparent position of the source to be different than its actual sky offset (Figure 13). In essence, the edge of the FPM acts somewhat like a pinhole, creating a new point source that the Lyot stop then redefines. An exoplanet at a sky offset of 3.0 $\lambda_c/D$ from the FPM center will appear at ~3.2 $\lambda_c/D$, a 10 mas difference. This effect will have to be included in astrometric calibrations.

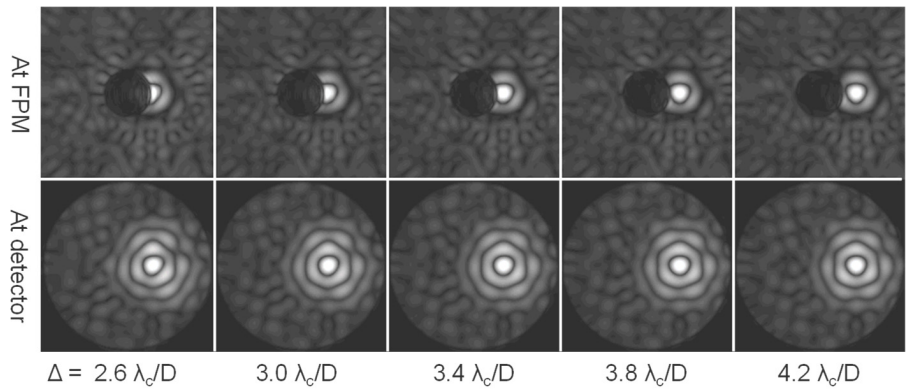

**Figure 12. A point source at different sky offsets from the FPM center seen (top) in the plane of the FPM and (bottom) at the detector (after the Lyot and field stops). Each image is individually scaled in intensity; in actuality, the intensity at the detector decreases for sources with small offsets.**

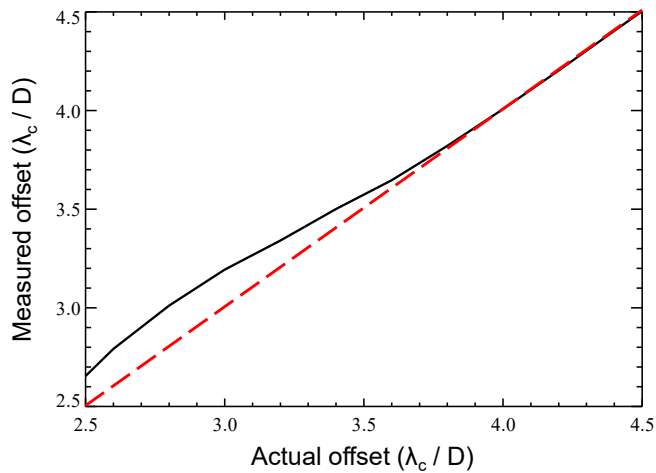

**Figure 13. The centroid-measured offset from the FPM center of the Band 1 HLC field PSF compared to its actual on-sky offset. The red dashed line plots equal offsets, for reference.**

### 3.2 Shaped pupil coronagraph

While the HLC primarily uses the DMs to compensate for the telescope obscurations, the SPC instead redefines the pupil with a complicated mix of apertures that mask them and create a tailored diffraction pattern with intrinsic dark zones. The first SPC proposed for CGI was "pure" in that only the pupil mask was responsible for diffraction suppression[41]. It produced a bow-tie-shaped, $10^{-8}$ mean contrast dark hole in a 10% bandpass that extended from $r = 4 – 22.5$ $\lambda_c$/D over two 60° wedges on opposite sides of the star (the small zones allow for a mask design with better contrast and throughput than would be possible for a larger region). The moderate contrast and low core throughput (2.7%) resulted in poor overall performance, though it was very insensitive to low-order aberrations, especially tip/tilt. Subsequent designs added bowtie FPMs and Lyot stops to offload some of the diffraction suppression workload from the pupil mask. These allow for better contrast, smaller inner working angle, broader bandpass, and higher throughput, but at the expense of additional complexity and increased aberration sensitivity.

The CGI SPCs have evolved considerably over the years, being constantly redesigned as the pupil pattern has changed and additional constraints on alignment tolerances have been imposed[42-,43,44]. This has spurred major innovations in apodizer optimization that have benefitted not just CGI but also coronagraphs for other obscured and/or segmented systems[45]. Figure 14 shows the SPC-Spec pupil masks over the various design cycles as the obscuration pattern has changed. Most challenging was accommodating the increase in the widths of the baffle plates on the secondary support struts that block scatter from insulation – the spider width has a major impact on SPC core throughput. Additional padding of the SPC mask along the spiders is necessary due to uncertainties in pupil alignment, distortion, and magnification, as well as along the outer edge to block the rollover on the primary mirror (described later).

The two SPC modes in CGI each have pupil masks, FPMs, and Lyot stops, but each has different aims. The spectroscopic SPC (SPC-Spec) is tailored to cover a ~17% bandpass (Band 2 or 3) and will be used with the slits and prism to produce exoplanet spectra on EXCAM (primarily the broad methane absorption feature at ~730 nm). It is assumed that the position of the exoplanet will have already been measured using the HLC, so only a limited field of view is needed (a bowtie-shaped region covering $r = 2.6 - 9.4$ $\lambda_c/D$ within two 65° wedges). The wide-field-of-view SPC (SPC-WFOV) is designed for imaging extended circumstellar disks in a 10% bandpass (Band 4) and covers a full 360° from $r = 5.6 - 20.4$ $\lambda_c/D$.

The SPC pupil masks are derived using algorithms different than those for HLC but optimized for the same metric, exoplanet signal-to-noise, with aberration sensitivities and alignment tolerances included. They are created on 1000-pixel-diameter grids representing the full pupil diameter (the outer 2% radius of which is masked to avoid the primary mirror edge rollover and absorb pupil magnification error). The designs are almost fully-binary, with <0.3% non-binary values. The actual fabricated masks are similarly pixelated, though the few non-binary pixels are each resolved into 3×3 subpixels to approximate a greyscale transmission; in the models, the corresponding non-binary values are used. The masks are matched to the 0.9909 geometric aspect ratio of the pupil as projected in the plane perpendicular to the chief ray, as predicted by ray tracing.

An SPC pupil mask is a pattern of low reflectivity ($R < 10^{-7}$) black silicon on a reflective substrate[42]. This avoids the potential of ghosts if using a transmissive substrate. The focal plane and Lyot stop masks are etched apertures in SoI wafers.

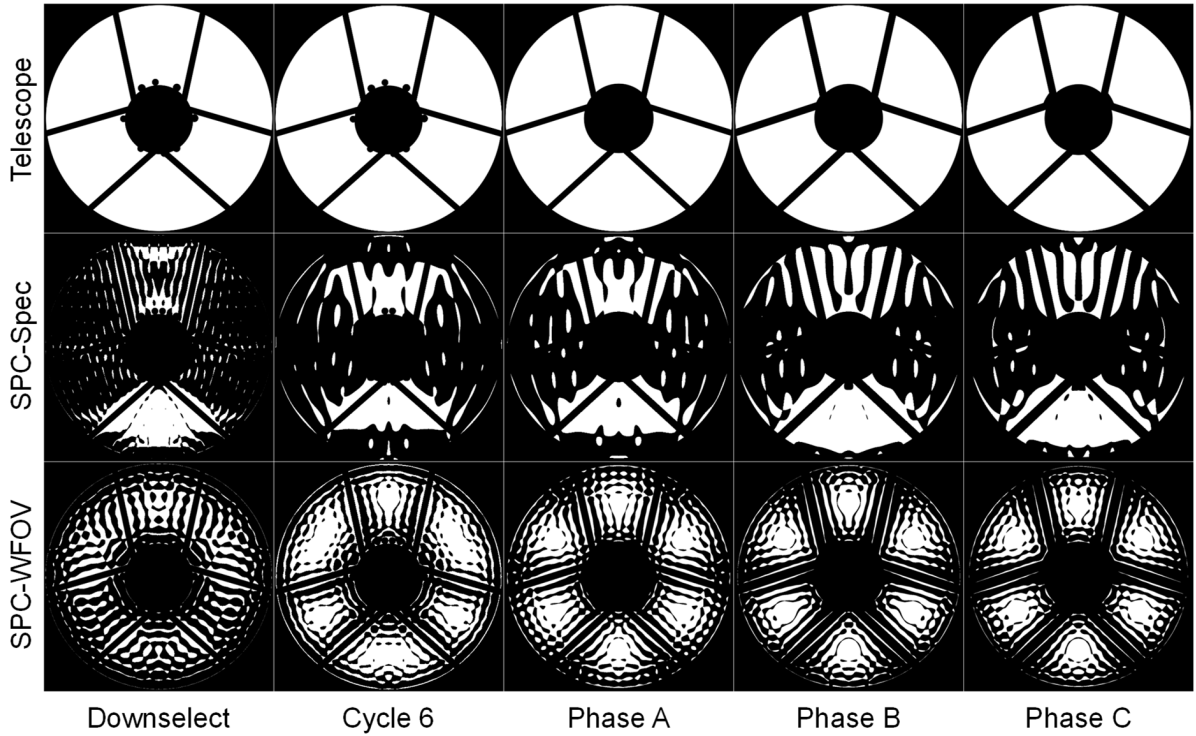

**Figure 14. The telescope obscuration pattern (top) and corresponding shaped pupil masks versus design cycle. The Downselect masks (2012) are "pure" SPC; the others are used with an FPM and Lyot stop.**

### 3.2.1 Spectroscopic SPC (SPC-Spec)

The SPC-Spec pupil mask (Figure 15) passes 34% of the incident light and creates a diffraction pattern at an intermediate focus with a mild-contrast ($4 \times 10^{-5}$), bowtie-shaped dark zone matching the FPM aperture (Figure 16). While the pupil mask and Lyot stop are the same for both Bands 2 and 3, separate FPMs are used, each scaled appropriately for the bandpass central wavelength. The bowtie-shaped Lyot stop is at the subsequent pupil image and reduces throughput of field sources by an additional 30%. Unlike the HLC, only a small fraction (~$10^{-7}$) of the original stellar flux passes through the Lyot stop. At the final image plane in an unaberrated system, no WFC, and a 17% bandpass, the mean NI is $1.1 \times 10^{-9}$ over $r = 3 - 9$ $\lambda_c$/D within the bowtie region (Figure 17).

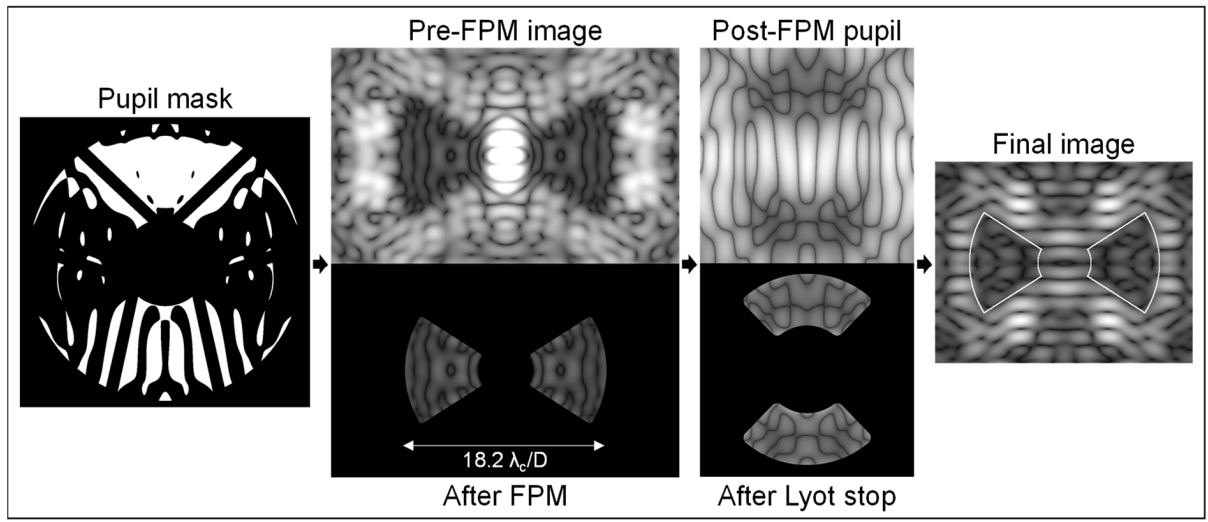

**Figure 15. SPC-Spec principal planes and masks. Images are monochromatic in an unaberrated system.**

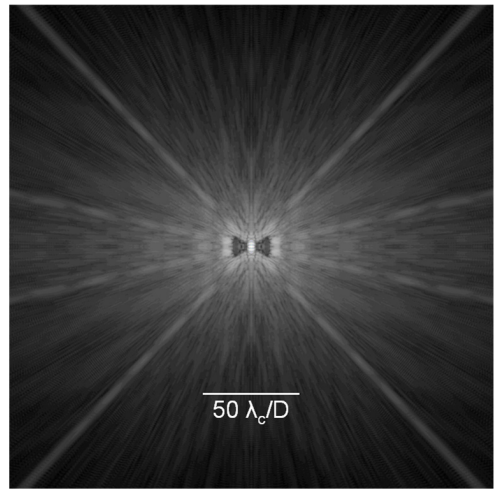

**Figure 16. The large-angle, unaberrated field for a broadband bandpass at the intermediate focus after the SPC-Spec pupil mask but before application of the FPM. Shown with a logarithmic stretch.**

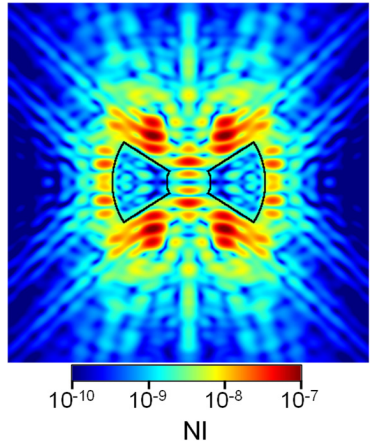

**Figure 17. SPC-Spec normalized intensity map for the unaberrated system in a 17% bandpass. The region of the bowtie FPM is superposed. Within this region between $r = 3$ & $9$ $\lambda_c/D$ the mean NI is $1 \times 10^{-9}$.**

The SPC-Spec low-order aberration sensitivities for Band 3 are shown in Figure 18. In general, it is less sensitive than the HLC to most aberrations, especially spherical. Notably, it is an order of magnitude less sensitive to pointing jitter. With aberrations (including polarization-dependent ones), and after WFC, the mean NI in Band 3 is $1.9 \times 10^{-9}$ (no jitter & 1 mas-diameter star) and $2.8 \times 10^{-9}$ (1.6 mas RMS jitter per-axis & 1 mas-diameter star); see Figure 19 and 20.

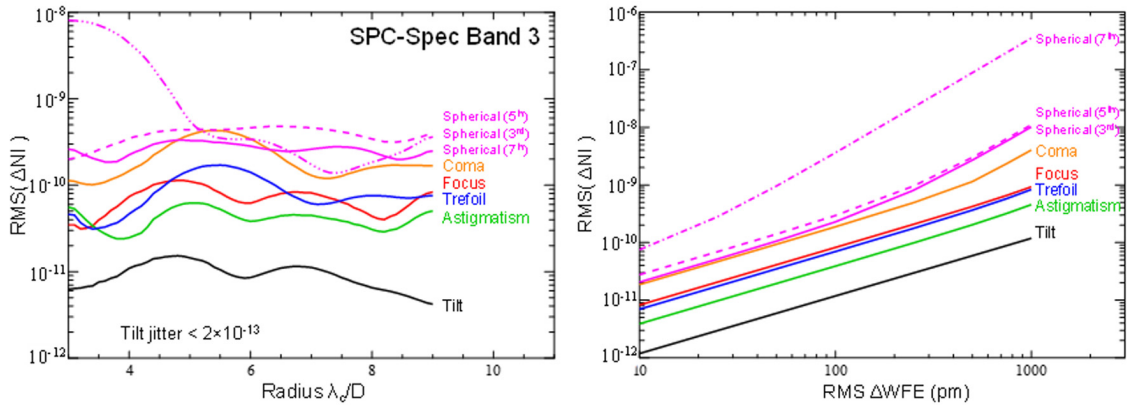

**Figure 18. SPC-Spec Band 3 sensitivities to low-order aberrations. (Left) RMS of the intensity changes versus field radius measured in $\lambda_c/D$-wide annuli within the bow-tie region due to the introduction of 100 picometer RMS of the specified aberration; (right) contrast change versus RMS WFE of each aberration measured over a $r = 3 - 5$ $\lambda_c/D$ annulus. For bidirectional aberrations (tip/tilt, astigmatism, coma, trefoil), the directions with the greater sensitivities are plotted. The default image was obtained after running WFC on the aberrated system model and includes polarization but no jitter. The mean field NI is $1.1 \times 10^{-9}$.**

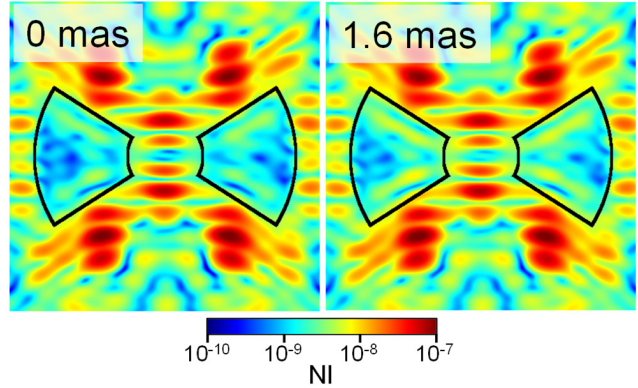

**Figure 19. SPC-Spec Band 3 dark hole normalized intensity maps with different amounts of pointing jitter (values are RMS Gaussian jitter per axis), and all also include a 1 mas-diameter star. The default image was obtained after running WFC on the aberrated system model and includes all polarization-dependent aberrations. The superposed bowtie aperture covers $r$ = 2.6 & 9.4 $\lambda_c$/D.**

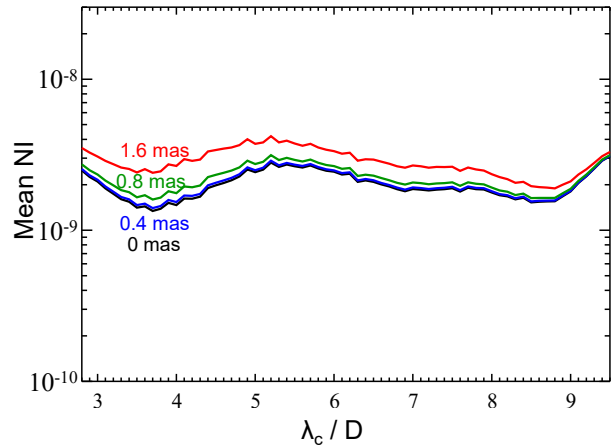

**Figure 20. Azimuthal mean contrast plots corresponding to the SPC-Spec Band 3 maps in Figure 19. These can be compared to the equivalent HLC values shown in Figure 8.**

The pupil mask creates a multi-lobed field PSF (Figure 21). The nearest sidelobes have maximum intensities >50% of the central peak, so they are included in the core throughput (Figure 22) and area metrics (0.051 and 9623 mas$^2$, respectively, in Band 3). If just the central core is included, then these values are 0.030 and 5366 mas$^2$. The spectroscopic slit is oriented along the vertical axis of the image in Figure 21, so each lobe will have its own spectrum in deep exposures. The broad SPC PSF suffers slightly more distortion near the FPM edge than does the narrower HLC's. For example, at 3.0 $\lambda_c$/D true sky offset, a source will appear at 3.35 $\lambda_c$/D on the detector (Figure 23).

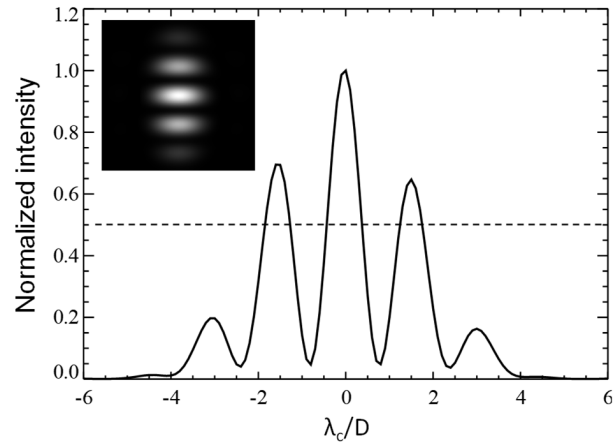

**Figure 21. Vertical cross-section through the center of the SPC-Spec field PSF (Band 3, with aberrations, after WFC). The PSF shown inset with a linear intensity stretch.**

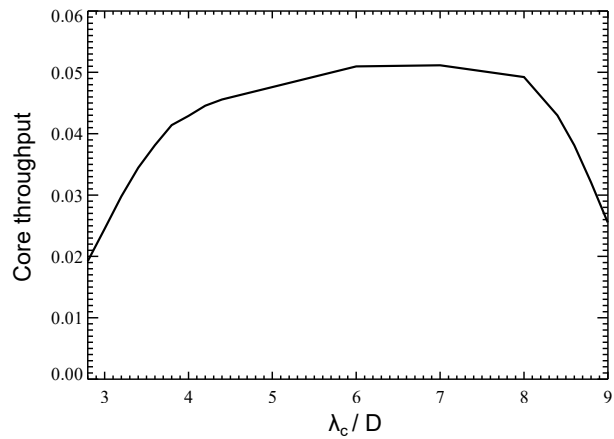

**Figure 22. SPC-Spec core throughput fraction versus field position. The reductions at the inner and outer bounds reflect the truncation of the core by the FPM.**

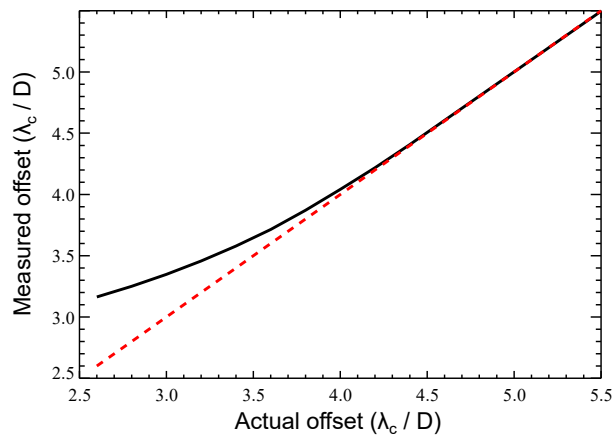

**Figure 23. The centroid-measured offset from the FPM center of the Band 3 SPC-Spec field PSF compared to its actual on-sky offset. The red dashed line plots equal offsets, for reference.**

### 3.2.2 Wide-field-of-view SPC (SPC-WFOV)

The SPC-WFOV gains a larger field of view, both radially and azimuthally, than SPC-Spec by sacrificing bandwidth (10%) and inner working angle (6 $\lambda_c$/D). The pupil mask (Figure 24) more closely resembles a binary approximation to a variable-transmission apodizer than does SPC-Spec's. It passes 34% of the incident light and creates a PSF at the FPM plane containing a ~$10^{-7}$ mean NI annular zone (Figure 25). The bulk of the flux in this plane is at large angles, with 80% within 45 $\lambda_c$/D (Figure 26). This is the PSF that applies to any field source located outside of the FPM annulus, so a significant fraction of light inside the dark hole may be from the wings of the PSF of a source located outside of it, which is not suppressed by the coronagraph. This creates potential issues when imaging extended sources such as circumstellar disks that may extend beyond 20 $\lambda_c$/D from the star, as it will be difficult to discern how much light originated at any given location. The Lyot stop reduces field source throughput by an additional 4.5%.

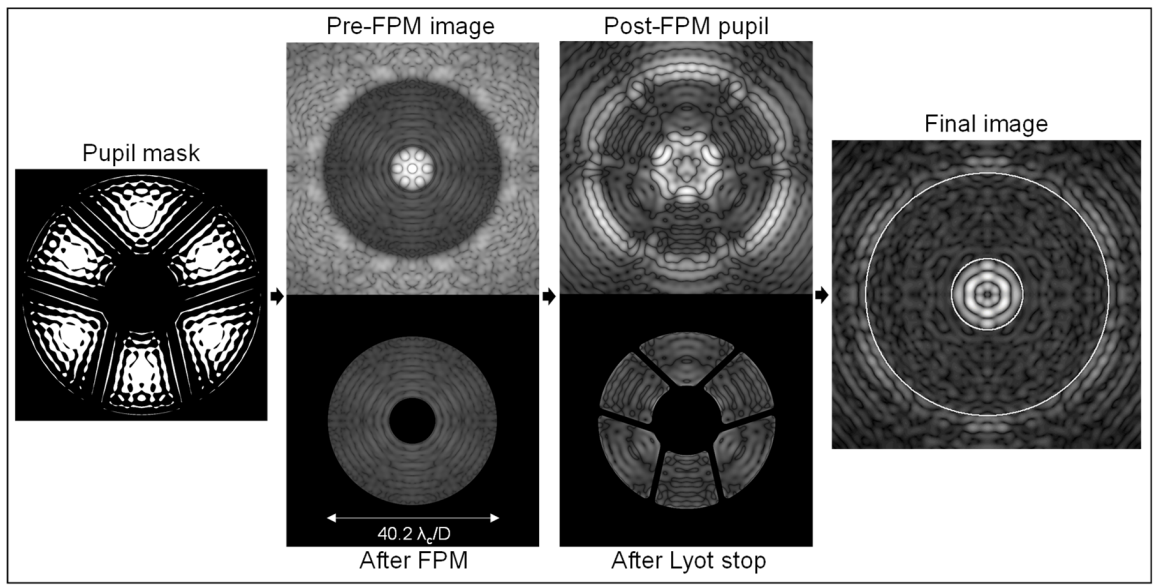

**Figure 24. SPC-WFOV principal planes and masks. Images are monochromatic in an unaberrated system. The projected outline of the FPM is superposed on the final image.**

In an unaberrated system with a 10% bandpass, the SPC-WFOV produces a $9 \times 10^{-10}$ mean NI dark hole from $r = 6 – 20$ $\lambda_c$/D (Figure 27). In the aberrated system with polarization and after WFC, the mean NI is $3.2 \times 10^{-9}$ with 1.6 mas RMS of jitter and a 1.0 mas diameter star. The large inner radius of the FPM suppresses most of the low-order aberrations, leading to sensitivities (Figure 28) orders of magnitude lower than those for HLC or SPC-Spec. The pointing jitter sensitivity is about two magnitudes less than HLC's, and there is no significant change in dark hole intensity over the expected jitter range (Figure 29).

The nearly threefold-symmetric SPC-WFOV pupil pattern produces a much more conventional field PSF than in SPC-Spec (Figure 30). The Band 4 maximum core throughput is 0.042 with 50% of its maximum at r = 6.0 $\lambda_c$/D (Figure 31), and the core area is 4614 mas$^2$. SPC-WFOV has the least amount of astrometric distortion near the IWA (Figure 32).

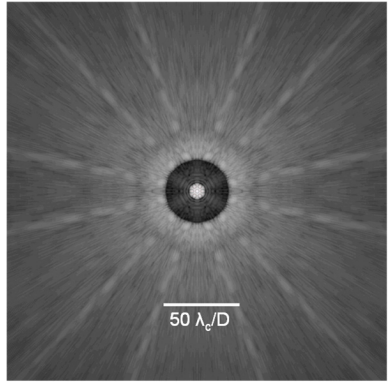

**Figure 25. The large-angle, unaberrated field for a 10% bandpass at the intermediate focus after the SPC-WFOV pupil mask but before application of the FPM. Shown with a logarithmic stretch.**

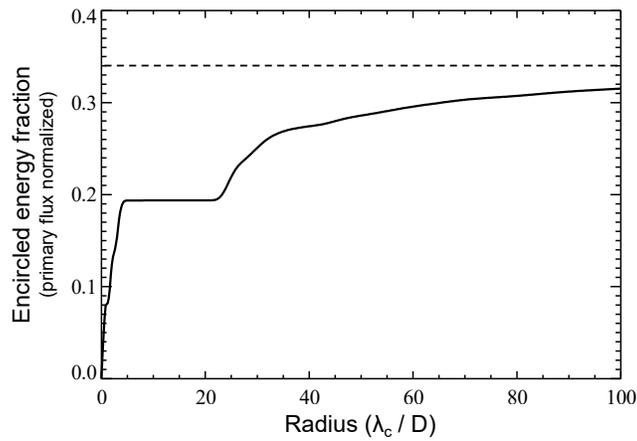

**Figure 26. Encircled energy for the SPC-WFOV Band 3 PSF *at the plane of the FPM* (before application of the FPM and prior to the Lyot stop and final image plane), primary flux normalized. The maximum value is 0.34. The plot at the final focus is very similar in shape and amplitude.**

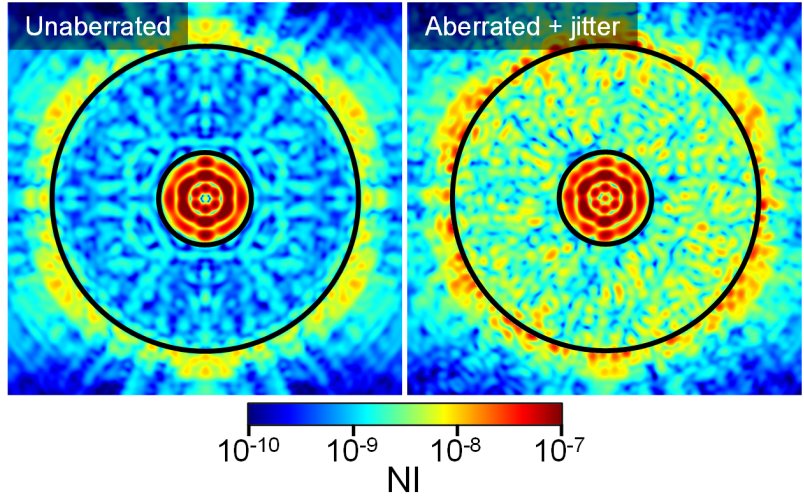

**Figure 27. SPC-WFOV Band 4 dark hole normalized intensity maps, (left) without aberrations or jitter and (right) with aberrations, including polarization-dependent ones, after WFC, including 1.6 mas RMS-per-axis jitter and a 1 mas-diameter star (residuals dominated by the mid-spatial-frequency optical aberrations). The superposed circles are $r$ = 6 & 20 $\lambda_c$/D.**

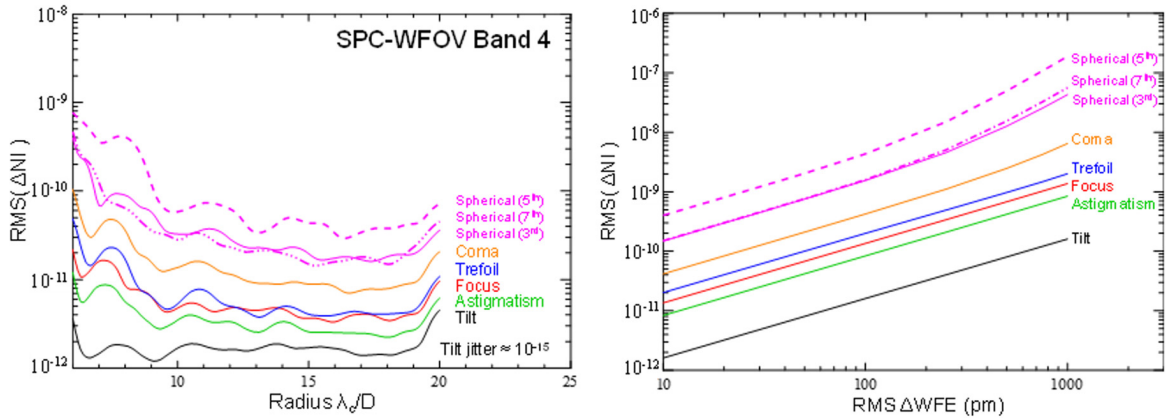

**Figure 28. SPC-WFOV Band 4 sensitivities to low-order aberrations. (Left) RMS of the intensity changes versus field radius measured in $\lambda_c$/D-wide annuli within the dark hole region due to the introduction of 100 picometer RMS of the specified aberration; (right) contrast change versus RMS WFE of each aberration measured over a $r$ = 6 – 8 $\lambda_c$/D annulus. For bidirectional aberrations (tip/tilt, astigmatism, coma, trefoil), the directions with the greater sensitivities are plotted. The mean field NI was $9 \times 10^{-10}$.**

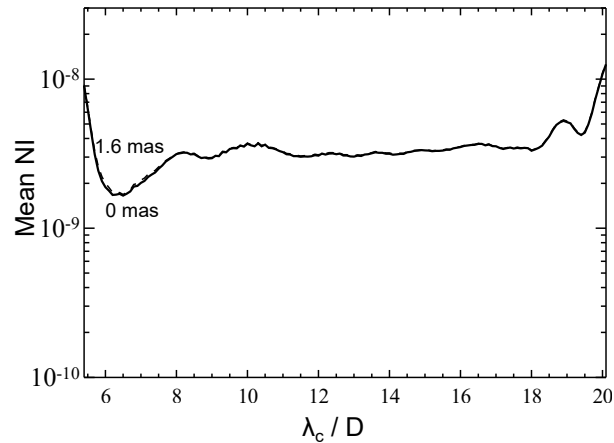

**Figure 29. Azimuthal mean contrast for SPC-WFOV Band 4 with varying amounts of pointing jitter (RMS per axis) and a 1 mas-diameter star in an aberrated system after WFC.**

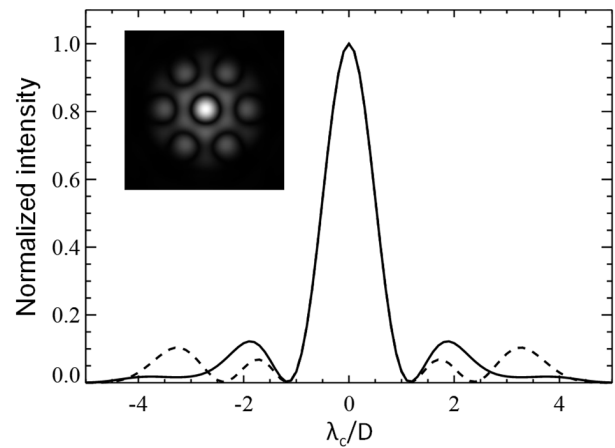

**Figure 30. SPC-WFOV Band 4 detector-plane PSF vertical (solid) and horizontal (dashed) cross-sections. Inset is the PSF (13 $\lambda_c$/D across) shown with a square-root intensity stretch.**

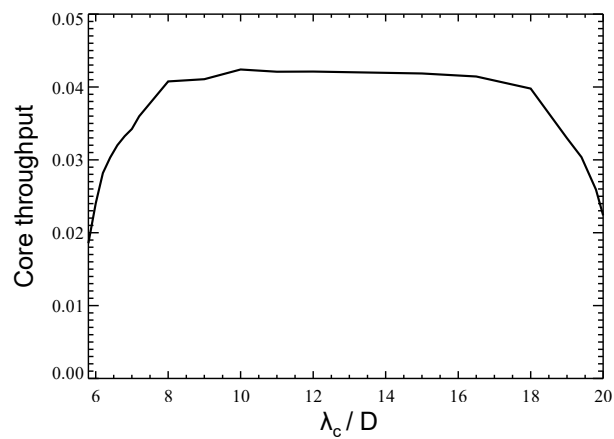

**Figure 31. SPC-WFOV core throughput fraction versus field position. The reductions at the inner and outer bounds reflect the truncation of the PSF core by the FPM.**

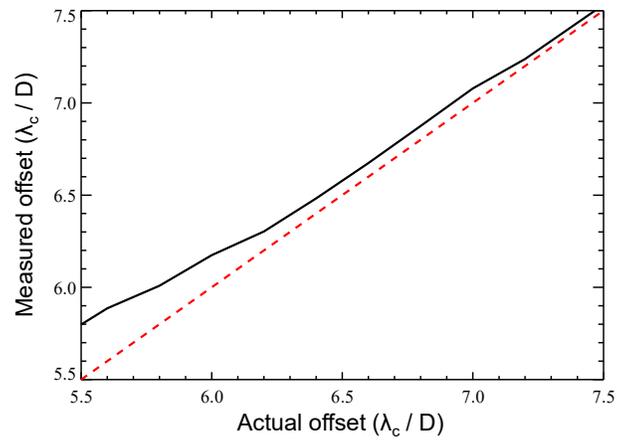

**Figure 32. The centroid-measured offset from the FPM center of the Band 4 SPC-WFOV field PSF compared to its actual on-sky offset. The red dashed line plots equal offsets, for reference.**

## 4   CGI model

The first step towards simulating the CGI system is to define the numerical model based on the prescribed optical layout, mainly the propagation distances between optics and their effective focal lengths. Static aberrations are then added to each optic representative of realistic polishing and figuring errors, and polarization-dependent aberrations and misalignments are introduced (dynamic disturbances are described later).

### 4.1  Unfolded model layout

The complex *Roman* CGI optical layout with its nearly 30 optical surfaces (mirrors, lenses, filters, masks) can be ray traced to derive the basic system performance in non-coronagraphic modes, but doing so does not account for its diffractive behavior. Some modeling packages exist that can compute diffraction through a complex system, but they are either too cumbersome, inaccurate, or slow for the needs of the CGI project (when generating time series simulations, tens of thousands of models must be computed). Instead, conventional Fourier-based angular spectrum and Fresnel algorithms are used. It is easier to run these on an unfolded model by rearranging the system into an on-axis, linear sequence of components with powered optics represented by thin lens approximations. The computations are limited to the paraxial regime, but that is typically not a significant issue for the coronagraph with its very small field of view (~3 arcseconds).

The approximation of non-parabolic and off-axis powered optics by ideal paraxial lenses in the unfolded layout does fail to accurately represent some facets of the true beam properties at particular locations in the system. For instance, ray tracing shows that there are multiple waves of coma and astigmatism at the Cassegrain focus located between the POMA fold and M3, but the unfolded model instead has a clean focus with no aberrations. However, there is nothing at this focus, and by the time the beam is at the FSM, the design aberration is just 7 nm RMS.

The beam shape changes significantly as it passes from the primary and through the TCA. It is skewed and highly elliptical at the POMA fold that is in a converging beam. For significantly inclined surfaces such as this and the FPM and the DMs, the surface errors (including DM actuators) are projected into planes that are perpendicular to the chief ray in the unfolded model (the planes we will refer to from hereon, unless otherwise noted). The pupil produced by the TCA at the FSM is slightly elliptical with an X/Y diameter ratio of 0.9909. The ray trace shows that this ratio is carried through the remainder of the system. This is represented in the model as an elliptical entrance pupil map imposed at the primary mirror that incorporates the projected secondary mirror and struts obscuration patterns.

### 4.2  Optical surface errors

Each mirror, filter, or lens has a unique set of errors ranging over low, mid, and high-order spatial frequencies that predominantly diffract light as speckles to small (<3 $\lambda$/D), medium (3 – 200 $\lambda$/D), and large (>200 $\lambda$/D) angles, respectively (these ranges are somewhat subjective but are what are assumed here). These can be phase or amplitude errors from surface figuring and polishing defects, misalignments, gravity strain release, differential polarization-dependent aberrations, coating nonuniformity and stresses, and material roughness. Static low-order errors such as coma or astigmatism are not of primary concern since they can be corrected by the DMs (except for incoherent polarization errors, described later). High-order errors scatter light to angles well outside the dark hole; an analysis by L3Harris indicates that these do not have a significant impact on contrast. The mid-spatial-frequency errors, mostly from polishing residuals[26], are the most important since they directly scatter light into the dark hole. The maximum dark hole radius ($r_{max}$) and maximum directly-correctable aberration spatial frequency ($\omega_{max}$) are limited by the number of DM actuators across the pupil (46.3, along the minor axis of the beam on the inclined DM), corresponding to $r_{max} = 46.3/2 \approx 23$ λ/D and $\omega_{max} \approx 23$ cycles/D.

The most critical optics in terms of surface quality are those prior to the FPM since they exist where the full amount of starlight is available to scatter. However, in the case of the HLC more than 80% of the light continues past the FPM to the Lyot stop, and 15% passes through the Lyot stop to the field stop (most of this light is concentrated just outside the dark hole region), so the optics between the FPM and the field stop must also be of high quality. Errors on optics after the field stop do not add light inside the hole but they do degrade the sharpness of the final image, including any planets, and introduce distortion. To maximize the suppression of the obscuration-produced diffraction pattern by the coronagraphic masks, the corrected wavefront incident on the FPM must be as close as possible to that expected for an unaberrated system, meaning that post-FPM errors cannot be directly controlled with the DM without also upsetting the pre-FPM solution. This demonstrates the need to represent each optic and include WFC to accurately model system performance. The results described here are produced by numerical models that include the interferometrically-measured surface error maps of the telescope and TCA flight optics and a mixture of measured and synthetic maps for the CGI optics (starting at the FSM).

As an aside, we note that Shaklan & Green[46] defined surface error requirements that depended on the contrast goal and location of the optic (i.e., near or far from a pupil). These were in the context of non-iterative, deterministic WFC accomplished with DMs in either a Michelson interferometer or in series; that is, a phase or amplitude error of a particular spatial frequency is corrected by the DMs with compensating patterns of the same frequency (the solution minimizes wavefront error). However, the iterative WFC algorithm used for CGI is not similarly deterministic, and by constraining the solution to minimize energy only within a given dark hole region it has additional degrees of freedom to deal with non-pupil optical aberrations. It is thus possible to tolerate optical errors larger than what would be expected from the Shaklan & Green specifications. The surface requirements are thus best verified by evaluating synthetic errors with models and then iterating to a final set of specifications. However, it may not be the static contrast goal that defines the surface requirements but rather the sensitivity of the speckle field stability to beam shear on non-pupil optics (from uncorrected pointing errors or structural deformations).

### 4.2.1 Measured OTA and TCA surface errors

The interferometric measurements of the uncoated OTA flight mirrors used in the models were obtained by the vendor, L3Harris, to which they added predicted low-order gravity release and thermal cool-down deformations. The primary, like most similar large mirrors, including *Hubble*'s, has a turned-over edge. The surface deviation increases from the mean by ~70 nm and then plummets nearly 200 nm at the very edge (Figure 33). The actual beam diameter is defined by a stop mounted just above the primary that happens to mask everything outside of the peak deviation. The remaining visible edge error is narrower than the projected size of a DM actuator and so cannot be well corrected with WFC. The inclusion of this feature in the models is important because it has a strong impact on the shaped pupil coronagraph, which explicitly accounts for it by blocking that region in its pupil mask. In the model, the primary map is stretched to match the 0.9909 aspect ratio of the pupil.

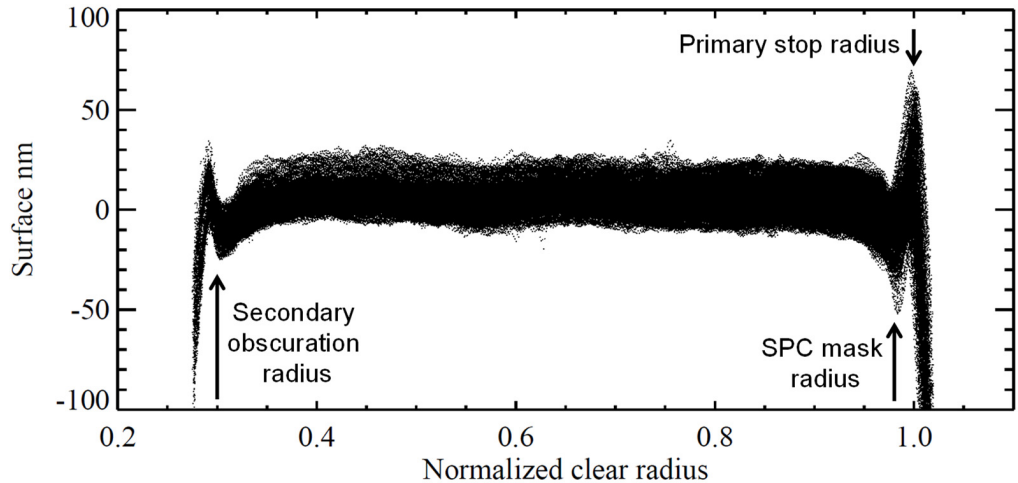

**Figure 33. Radial scatter plot of the measured primary mirror surface errors. The inner and outer rollovers are largely masked by the stop at the primary and the secondary mirror obscuration. The shaped pupil mask was designed to block the remaining portion of the edge rollover since it cannot be well corrected with the CGI deformable mirrors.**

Besides the turnover, the primary has two other prominent features. The ion figuring technique used to polish the final surface left two periodic, mid-order raster patterns with ~60° separation, producing two diffraction spikes. There is also print-through of the hexagonal cells that form the inner core of the mirror and generates a grid of spots in the image field. The simulations show that both effects are well-corrected by the DMs. The primary has WFE$_{>Z11}$ = 16.3 nm RMS. The secondary shows much lower raster and cell print-through effects and is dominated by other mid-order errors, with WFE$_{>Z11}$ = 6.6 nm RMS.

The TCA optics[26] were fabricated and measured by QED Technologies; at the time of writing, only measurements of the uncoated and unmounted components were available, with predicted low-order coating strain added. The combined OTA+TCA WFE (Table 2, Figure 34) of 35.7 nm RMS measured at the plane of the FSM is similar to the 39.4 nm RMS (including camera aberrations) for the *Hubble Space Telescope*[47] and well below the 76.4 nm rms interface requirement between the observatory optical train and CGI (the CGI optical system begins at the FSM, with the OTA+TCA elements the responsibility of the *Roman* project).

**Table 2. Wavefront phase error measured over the illuminated pupil at the FSM with contributions only from the specified optics, below and above 5th order spherical (Z22). Excludes alignment errors.**

| Optics | Z4-Z22 nm RMS WFE | >Z22 nm RMS WFE | Total nm RMS WFE |
|---|---|---|---|
| OTA & TCA | 31.7 | 16.5 | 35.7 |
| OTA only | 11.9 | 13.7 | 18.1 |
| TCA only | 27.2 | 9.2 | 28.7 |

Because the POMA fold is highly inclined (~45°) and measured face-on, its map was projected to a plane perpendicular to the axis of the chief ray for use in the models. The POMA fold has a WFE$_{>Z11}$ = 4.0 nm RMS and the TT fold is 5.4 nm RMS.

The maps for the aspheric M3, M4, and M5 mirrors show concentric ripples with spatial frequencies >20 cycles/$D_{CA}$ (Figure 35) that are residuals from polishing. Since their spatial frequencies scatter light mainly outside the dark hole, they do not appear to be performance-limiting features according to the simulations. Their imprint can be seen in both the system phase and amplitude patterns in Figure 34. These optics have WFE$_{>Z11}$ of 3.6 – 9.0 nm RMS, with M3 the worst.

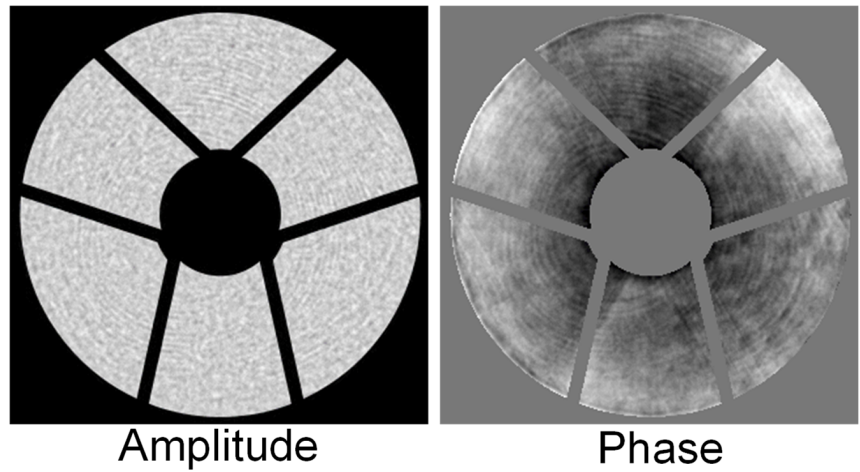

**Figure 34. The amplitude and phase at λ = 575 nm at the pupil plane coincident with the FSM. This represents the propagation of the measured OTA and TCA errors in the model (the artificial low-order errors used to meet the error budget are not included). The amplitude aberrations are dominated by the surface ripples in the TCA M3, M4, and M5 optics that are not at pupils, and so phase errors partially transform to amplitude errors during propagation. Coating non-uniformities are not included. The standard deviation relative to the normalized mean amplitude is 5%. The phase is displayed between ±130 nm WFE.**

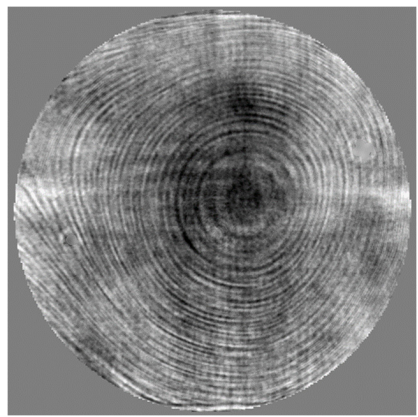

**Figure 35. Measured TCA M3 aspheric mirror displayed between ±40 nm of wavefront error (uncoated). The crosshatch pattern and rings are polishing artifacts. The optic has 12.6 nm RMS of WFE, 9.0 nm RMS after subtracting low-order errors up to Z11. The clear aperture is 149 mm, and the beam diameter is 137 mm.**

### 4.2.2 Synthetic surface errors for CGI optics

The CGI internal optics come from multiple sources, including international contributions. For example, OAPs 4 – 7 are contributed by the Centre National D'Etudes Spatiales (CNES) and fabricated by Laboratoire d'Astrophysique de Marseille (LAM) using a stressed-glass polishing technique[48]. The SPC mask substrates, polarizers, and lenses are contributed by the Japanese space agency (JAXA) and sourced from various Japanese vendors.

High-quality, interferometrically measured error maps of the FSM, FCM, and the post-OAP3 fold are used in the models. At the time of writing, the maps of the available CGI OAPs and lenses are unsuitable for direct use as they contain too many measurement artifacts (mostly rings that cannot be readily filtered out), so they are represented by synthetic error maps.

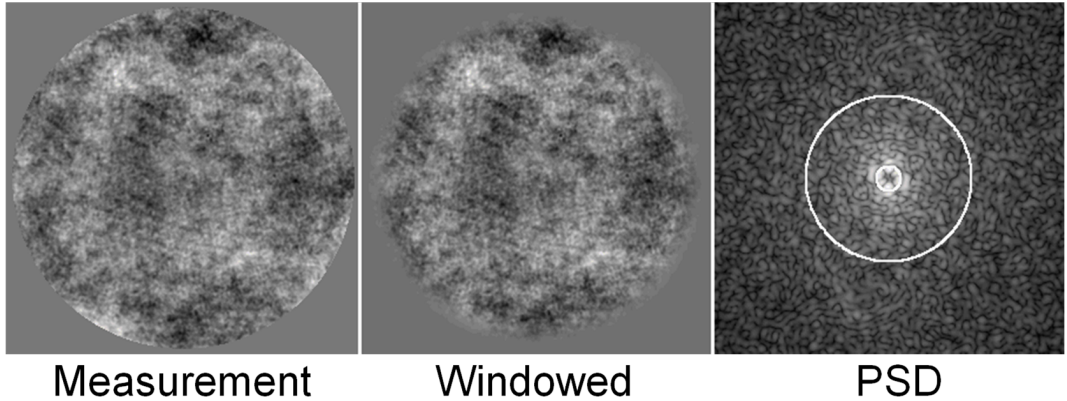

**Figure 36. (Left) OAP 7 prototype interferometrically measured wavefront error map after subtracting up to Z11, displayed between ±8 nm WFE; (Middle) Map after applying Tukey window function; (Right) modulus-square of the Fourier-transform of the windowed map (the two-dimensional PSD distribution). The superposed circles are $r$ = 3 & 20 cycles/D.**

The low and mid-order errors are synthesized separately, corresponding to how the actual errors are measured to produce a power spectral density (PSD) curve that describes how much aberration is present at each spatial frequency. The low-order errors, characterized by Zernike polynomials, are subtracted from the measured map (up to Z11 here, corresponding to spatial frequencies of up to 2-3 cycles/D). A windowing function is then applied to the remainder; we use a Tukey (cosine-tapered) window (Figure 36). The low-order subtraction and windowing are necessary to reduce ringing in the next step, computing the Fourier transform of the map. The modulus-square of the transform provides the two-dimensional PSD distribution, and the azimuthal average of that produces the PSD curve (Figure 37)).

The curve is approximated with a simple function:

$$PSD(f) = \frac{a}{1 + \left(\frac{f}{b}\right)^c} \qquad (5)$$

where $f$ is the spatial frequency, $a$ is the maximum aberration power, $b$ is the turnover spatial frequency, and $c$ is the high-frequency falloff power. This function is flat at frequencies below $b$, representing the suppression of the low-order errors, and then drops off at higher ones, a characteristic of the polishing process. We set the PSD turnover for all CGI optics at $b = 3$ cycles/$D_{CA}$, based on subtraction of aberrations $\leq$Z11. Our derivation of the PSD of a prototype OAP 7 optic shows $c = 2.5$, a common power-law for high-quality optics, and we assume it for OAPs 1 – 8. We also analyzed measurements of Gemini Planet Imager optics, which were fabricated by a variety of vendors, from which we derived $c = 3.0$ for the PSD of flats and lenses (these are of similar quality to those fabricated for CGI).

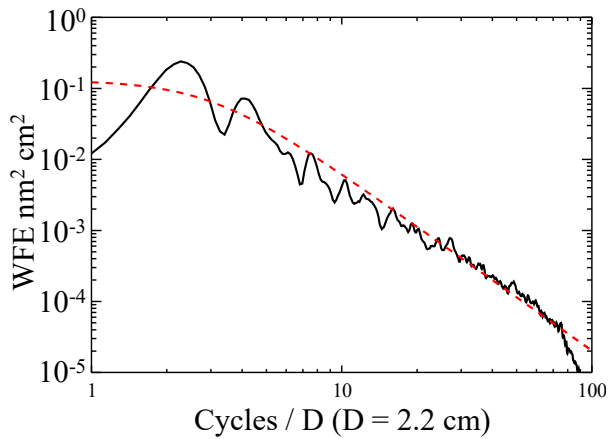

**Figure 37. The measured OAP7 prototype PSD curve. The dashed line is the PSD function with $b = 3$ cycles/D and $c = 2.5$.**

To create the mid-order error map, a two-dimensional PSD distribution is generated and then the square-root is taken to convert from aberration power to amplitude. To allow for more accurate interpolation of the map, the higher spatial frequencies are suppressed by applying to this distribution a cosine-tapered window with a maximum extent of 80 cycles/$D_{beam}$ (this corresponds to field radii well outside of the dark hole and is used to reduce interpolation errors caused by high-frequency features when resizing the map to match the wavefront sampling). The amplitude function is then multiplied by a random array of complex phases and the Fourier transform taken. The result is normalized to a specified RMS wavefront error. This represents an isotropic map of mid-order errors (Figure 38). There are no spatially correlated features such as polishing marks in these synthetic maps since the phases used to generate them were random (no such features are seen in the recently measured FSM-to-OAP8 wavefront error map). The statistics of the synthetic maps are given in Table 3.

The low-order aberrations in CGI include surface figure errors, misalignments, design errors, and polarization-dependent aberrations. Except for polarization, these have almost no impact on the models and coronagraphic performance when they occur prior to the FPM as the DMs can easily compensate for them. A random distribution of Zernike aberrations from Z5 – Z11 are generated for each optic and normalized to a total RMS WFE that matches the wavefront error budget. These are higher than the low-order errors measured in the available optics as they also include margins for misalignments and deformations from mountings. The measured CGI OAP maps show WFE$_{\leq Z11}$ of 1.4 – 4.6 nm RMS and WFE$_{>Z11}$ of 2.1 – 3.1 nm RMS (the latter range is impacted by the ringing and limited height resolution). The synthetic maps have a larger WFE$_{\leq Z11}$ range of 2.9 – 7.0 nm RMS, which include both surface error and potential misalignments, in keeping with the CGI wavefront error budget, and WFE$_{>Z11}$ of 2.9 – 3.9 nm RMS (these slightly higher values than the measured optics reflect uncertainty in the quality of the final flight optics).

**Table 3. Wavefront phase error parameters for the synthetic CGI optics (measured over optic clear aperture)**

| Optics | ≤Z11 Fabrication RMS WFE | ≤Z11 Alignment RMS WFE | >Z11 Fabrication RMS WFE | PSD Power $c$ |
|---|---|---|---|---|
| OAPs 1-3, 8 | 7 nm | 2 nm | 3 nm | 2.5 |
| OAPs 4-7 | 5 nm | 2 nm | 2 nm | 2.5 |
| DMs | 6 nm | 0 nm | 3 nm | 3 |
| Pupil mask | 10 nm | 0 nm | 4 nm | 3 |
| Folds | 5 nm | 0 nm | 1.5 nm | 3 |
| Filters | 2 nm | 0 nm | 1.5 nm | 3 |
| Lenses | 4 nm | 2 nm | 2 nm | 3 |

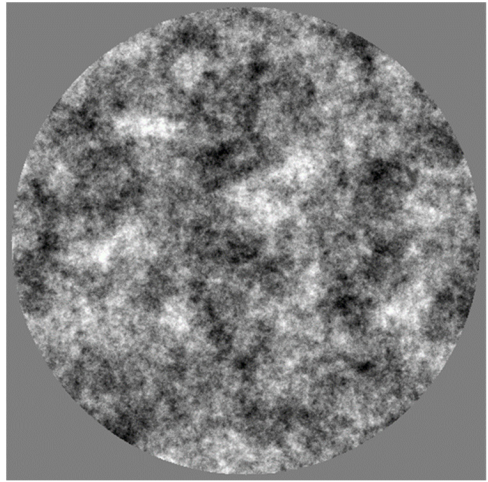

**Figure 38. Synthetic OAP7 mid-order error map shown between ±8 nm WFE.**

Late in the writing of this document, an interferometric measurement was obtained of the assembled CGI optics, from the FSM to final fold mirror but with high-quality flats substituted for the DMs. The measured total WFE (>Z3) of 17.0 nm RMS compares favorably to the model's 15.5 nm RMS.

### 4.3 Amplitude errors

Both phase and amplitude wavefront errors can create speckles in the final image, though with different chromatic behaviors that complicate wavefront control over a broad bandpass. In CGI, optical surface residuals from figuring and polishing are the largest source of phase errors, with polarization-dependent aberrations and misalignments being minor contributors. Surface errors actually constitute the main source of amplitude errors, rather than non-uniformities in mirror reflectances as one might assume. As the beam propagates, phase errors transform into a mixture of phase and amplitude errors and vice versa due to the *Talbot effect*[49].

A phase error of period $P$ will cycle between being a phase and amplitude error as it propagates over a *Talbot length*, $L_T = 2P^2/\lambda$, measured in a collimated beam or the equivalent effective propagation distance for a diverging or converging one. At propagation distances of ¼ $L_T$ and ¾ $L_T$, the phase aberration will transform into an almost pure amplitude error, and at ½ $L_T$ it is a phase error with opposite sign; in-between it is a mixture of phase and amplitude. The cycle repeats over each $L_T$ distance of propagation. Phase errors from optics at pupil planes will create phase, but not amplitude, speckles at conjugate pupil and image planes. However, phase errors from optics not at a pupil will generate both phase and amplitude speckles.

The Talbot effect is the basis for the sequential DM wavefront control used in CGI. DM2, which is 1.0 meter from the pupil located at DM1, is used to compensate for amplitude errors seen in projection on the pupil. DM1 controls phase errors, including those generated by DM2 when it corrects amplitude. Phase errors at a pupil can be well corrected for all wavelengths with DM1 up to a maximum spatial frequency limited by the number of actuators, $N$, across the pupil (nominally $f_{max} = N/2$ cycles, though actually less due to the finite widths of the actuators; for CGI, $N \approx 46$).

Low-order errors will transform more slowly than higher-order ones as they propagate. In the CGI collimated DM space with beam diameter D=46.3 mm and $\lambda = 575$ nm, $L_T$ for a 3 cycle/D error is 828.5 m while for a 20 cycle/D one it is 18.6 m. The long $L_T$ of the low-order error means that it largely remains a phase aberration and is not a significant contributor to amplitude WFE; it can be well corrected over a broad bandpass by DM1. The 20 cycle/D aberration reaches a maximum conversion to amplitude after ¼ $L_T$ = 4.7 m of propagation, so it can create non-negligible amplitude errors. The DM1-DM2 separation of 1.0 m corresponds to ¼ $L_T$ for ~42 cycles/D; this is where DM2 is most efficient at correcting amplitude errors, in terms of stroke required and minimal introduction of additional phase errors. At other spatial frequencies DM2 must use more stroke to affect amplitude due to the incomplete transformation of phase-to-amplitude over the propagation distance to the pupil at DM1. This also introduces additional phase error that DM1 must compensate.

Figure 34 shows the computed amplitude at the pupil plane located at the FSM using the measured surface error maps of the telescope and TCA optics. The variations are mostly caused by surface errors on the TCA optics that are not at pupils (as evidenced by the high-order ripples from the finishing used on the TCA powered mirrors). The standard deviation of these amplitude variations is σ = 5% with a P-V of 43%, both relative to the mean. When low-pass filtered to ≤20 cycles/D, corresponding to field angles more relevant to CGI performance, σ = 0.8% and P-V = 8%. Note that the primary mirror surface aberrations do not create amplitude errors because it is at a pupil, and the secondary mirror is effectively close enough to the primary that it is not a major source. The contributions of selected out-of-pupil optics, in isolation, to the system amplitude error are given in Table 4.

The vendor-accepted requirement on reflectivity uniformity is a peak-to-valley (P-V) = 0.5% per optic, which corresponds to a 0.25% P-V amplitude error. If we assume an 8.6× conversion from amplitude σ to P-V (derived from the 43%/5% ratio above), then we estimate a σ = 0.03% amplitude variation due to coating nonuniformity. Using this, the root-sum-square for the 7 OTA and TCA optics nonuniformity is then σ = 0.08%, which is 63 times less than the amplitude nonuniformity from the phase-to-amplitude errors. Nonuniformity is thus not a significant performance limitation (plus, it is largely wavelength-independent over a bandpass, while phase-to-amplitude conversions are not).

**Table 4. Amplitude non-uniformity standard deviation at the CGI exit pupil due to phase errors on each out-of-pupil optic in isolation ($\lambda = 575$ nm)**

| Optic | Amplitude $\sigma$ |
|---|---|
| Secondary | 0.1% |
| POMA fold | 0.3% |
| M3 | 2.9% |
| M4 | 1.5% |
| M5 | 2.1% |
| TT fold | 1.4% |
| OAP 1 | 1.5% |
| FCM | 0.7% |

*4.4 Polarization-dependent aberrations*

A wavefront incident on an inclined surface will undergo phase and amplitude modulation depending on its polarization, whether in reflection or transmission. Essentially, it sees a different local surface inclination depending on the orientation of the electric field vector. Integrated over the beam, this results in different low-order wavefront errors depending on polarization, with larger inclinations resulting in greater aberration differences[50] (i.e., inclined flats or highly curved optics). Coatings and/or surface materials will further increase or decrease these effects.

An unpolarized input plane wave can be decomposed into two separate wavefronts with orthogonal polarizations that add incoherently. In the absence of any surfaces that introduce polarization-dependent aberrations, these wavefronts will behave identically as they propagate through the system. In real systems, however, any inclined optics (e.g., flats or curved mirrors) and birefringent materials will introduce significant differential aberrations between polarizations that vary by wavelength. After propagation, the output can be described as four separate, incoherent wavefronts representing orthogonal linear polarization states: two for the direct components (e.g., $0°_{in}$, $0°_{out}$ & $90°_{in}$, $90°_{out}$) and two for the cross-terms from induced leakage into the other polarization (e.g., $90°_{in}$, $0°_{out}$ & $0°_{in}$, $90°_{out}$); note that the input coordinate system may be rotated relative to the output's. These are equivalent to the propagation of the Jones pupil matrices. Each component can have quite different phase and amplitude aberrations, and it is impossible to fully correct all components simultaneously with conventional WFC (i.e., deformable mirrors). Without a polarizer, one can correct for the mean aberration of the direct components – the cross-terms have relatively miniscule power in most cases, so the remaining difference between the direct terms dominates (but can be further suppressed by the low-order aberration rejection properties of the coronagraph).

When a polarizer is used, the single direct term can be fully corrected (at least monochromatically) but the uncorrected cross-term remains, unless polarizers are used both at the input and output (which is not an option in CGI). For most astronomical observations, the polarization-dependent aberrations are too small to be of importance, but they are very much so at the CGI contrast levels.

The CGI polarization aberrations were predicted by ray tracing the Code V prescription, including known and estimated coating properties (the primary and secondary coatings are proprietary), from the primary mirror to the FPAM (polarization-dependent aberrations from optics afterwards are minor and do not contribute significantly to speckle generation). The outputs were the electric fields at the FPAM entrance pupil for each polarization component at wavelengths spanning the CGI bandpasses. The input fields were defined at polarization angles of ±45° and the outputs at 0° and 90°. Using rotated input axes produces four output fields of equal intensity that have smooth phase and amplitude patterns that are easily characterized by Zernike aberrations (Figure 39); this allows such errors to be easily represented at any spatial sampling and interpolated in wavelength to match the modeling needs.

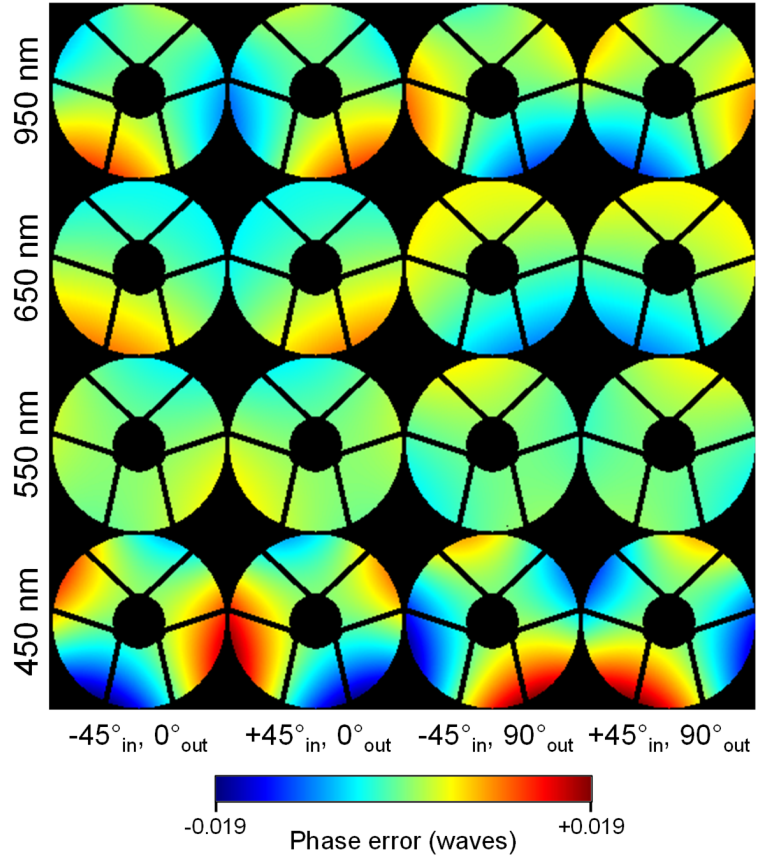

**Figure 39. Predicted polarization-dependent phase error versus wavelength at the FPM entrance pupil for the rotated (±45°) input polarizations. The mean phase at each wavelength has been subtracted.**

These can be readily converted to aligned (direct and cross-term; Figure 40 and Figure 41) E-fields by the addition or subtraction of the two E-fields for a given output polarization:

$$E(0°_{in}, 0°_{out}) = [E(+45°_{in}, 0°_{out}) + E(-45°_{in}, 0°_{out})] / \sqrt{2} \qquad (6)$$

$$E(90°_{in}, 90°_{out}) = [E(+45°_{in}, 90°_{out}) + E(-45°_{in}, 90°_{out})] / \sqrt{2} \qquad (7)$$

$$E(90°_{in}, 0°_{out}) = [E(+45°_{in}, 0°_{out}) - E(-45°_{in}, 0°_{out})] / \sqrt{2} \qquad (8)$$

$$E(0°_{in}, 90°_{out}) = [E(+45°_{in}, 90°_{out}) - E(-45°_{in}, 90°_{out})] / \sqrt{2} \qquad (9)$$

Note that the signs may change depending on how the orientations are defined in the optical prescription. These equations can also be used to convert final image plane E-fields for rotated-input coordinates into input-aligned ones. Notice that the cross-term pupil phase aberration pattern has sharp features and discontinuities where the corresponding amplitude is zero, which makes interpolation inaccurate (and is not readily solved by phase unwrapping), hence the benefit of using the rotated input polarizations instead. The diffraction model is run separately for each set of polarization-dependent aberrations and the intensity images are then added together – four for normal imaging (no polarizer) or two for a single polarization channel.

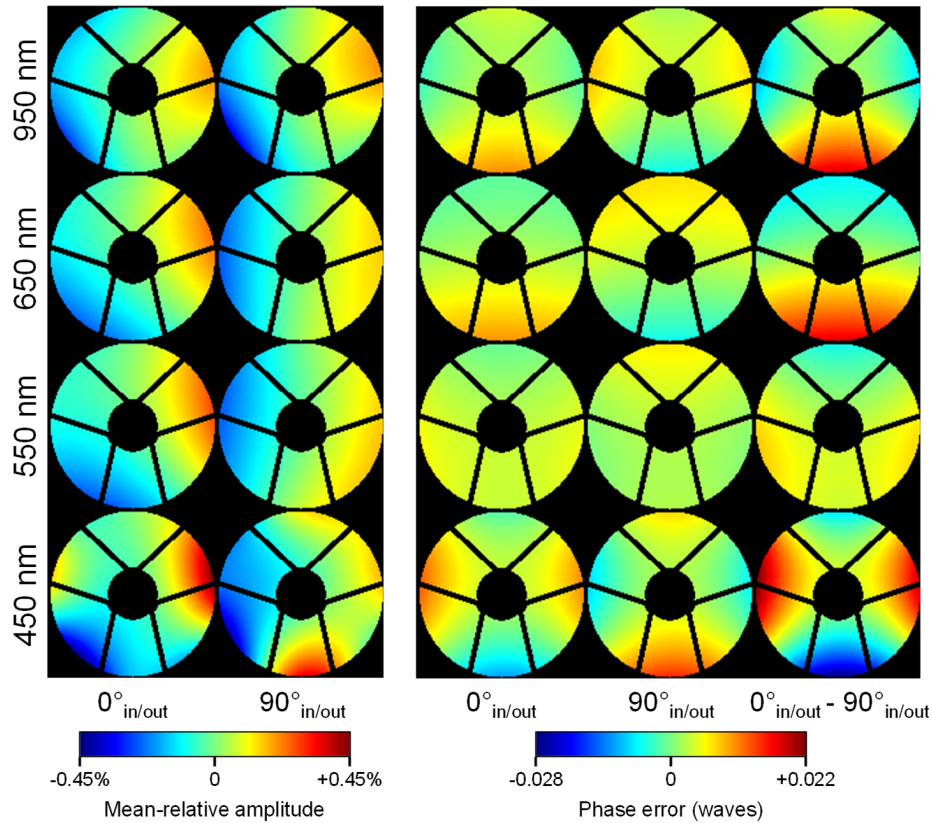

**Figure 40. (Left) Polarization-dependent direct-component amplitude errors versus wavelength relative to the uniform mean value across the pupil. (Right) Phase errors versus wavelength relative to the mean of the 0° and 90° phase aberrations. Shown at the FPM entrance pupil.**

The polarization-dependent aberrations consist almost entirely of astigmatism and tilt. The fast primary mirror and off-axis location of the CGI field in the OTA focal plane account for most of this. As mentioned before, the inclination of the pickoff mirror was initially a concern in regard to generating large cross-terms, but our analyses show that is not the case, at least at the CGI contrast levels.

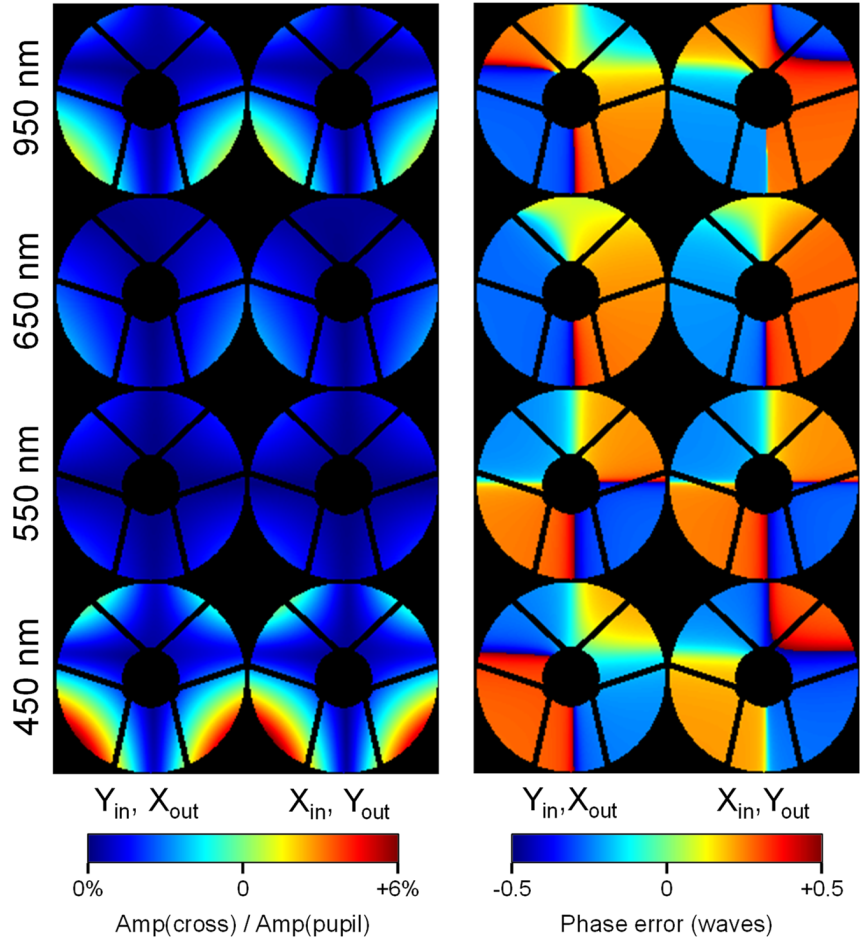

**Figure 41. Polarization-dependent cross-term components versus wavelength. (Left) Amplitude variations, shown relative to the total pupil amplitude, and (right) phase errors (some phase wrapping is present). Shown at the FPM entrance pupil.**

Without a polarizer, the wavefront sensing and control algorithm measures and corrects the mean aberration of the direct terms:

$$E_{mean} = (E_{0°} + E_{90°})/2 \qquad (10)$$

so that the residual polarization aberration present in each incoherent direct term is one-half the difference between the two:

$$E'_{0°} = E_{0°} - E_{mean} = (E_{0°} - E_{90°})/2 \qquad (11)$$
$$E'_{90°} = E_{90°} - E_{mean} = (E_{90°} - E_{0°})/2 \qquad (12)$$

As shown in Figure 42a, the direct-term aberration difference varies with wavelength (dependent on the coatings) between $1 - 8$ nm RMS in tilt and astigmatism, with a minimum conveniently at $\lambda \approx 525$ nm, at the short wavelength end of Band 1. Fortunately, the HLC and SPC are least sensitive to these particular aberrations, so the impact of the residual errors is not as significant as it would be if other forms dominated. Figure 42b shows that the cross-terms contain a very small fraction of the total intensity incident on the FPM, with a minimum of ~0.002%, again conveniently within Band 1, and peaking in Band 4 at ~0.05%. These low intensities reduce the effect of having very large phase errors.

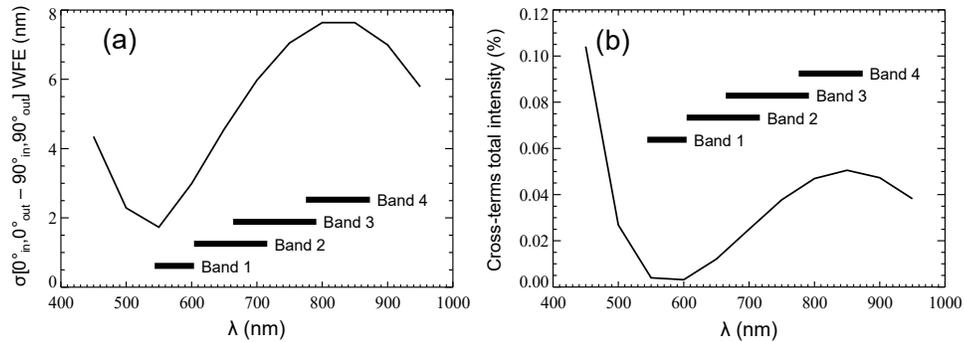

**Figure 42. (a) Standard deviation versus wavelength of the difference between the direct polarization term phase errors. (b) Fractional intensity versus wavelength of the polarization cross-term components prior to the FPM.**

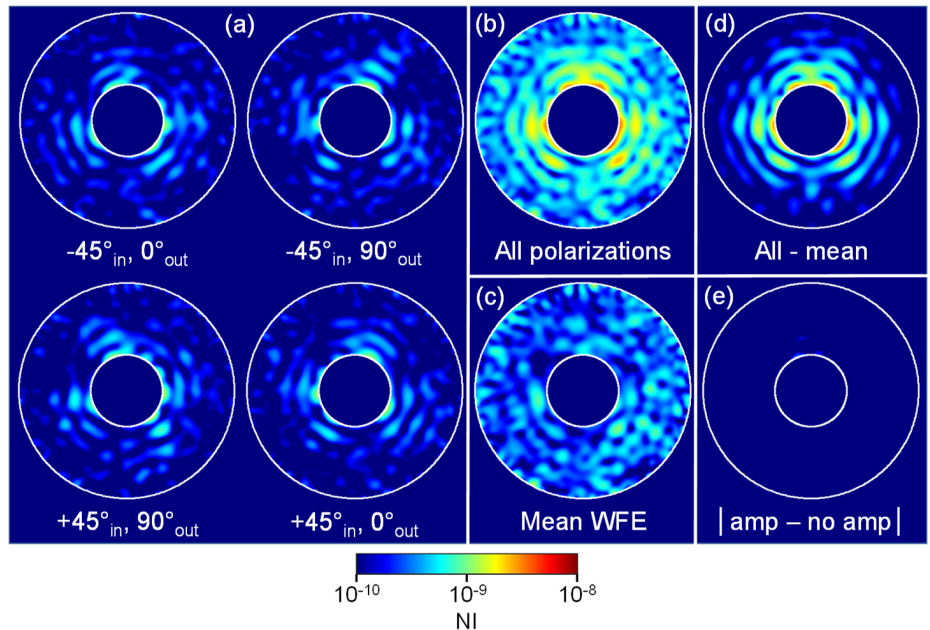

**Figure 43. HLC (Band 1) dark hole normalized intensity maps obtained after running WFC on the aberrated system model, including all polarization-dependent aberrations. The same DM solution is used for all images. The images are normalized relative to the full system illumination (the four polarization component images should be added together, not averaged). (a) Images for ±45° input and 0° & 90° output polarization combinations; (b) The incoherent sum of the four polarization components that are shown in (a) representing the image seen without polarizers; (c) The coherent (common) component including the mean wavefront error, which is what the wavefront sensing algorithm sees; (d) The difference between the all-polarizations (b) and mean (c) images showing the incoherent polarization contribution; (e) The absolute difference between images generated with and without polarization-dependent amplitude aberrations, showing their insignificant impact. The superposed circles are $r = 3$ & $9$ $\lambda_c$/D.**

Figure 43 shows the contrast maps for the HLC in Band 1 for the ±45° input polarization components and generated using the DM solution for the mean polarization-dependent aberrations. The incoherent sum of the four components (Figure 43b) represents the image that would be observed without a polarizer. The difference between Figure 43b and Figure 43c, shown in Figure 43d, reveals the total incoherent contributions from the polarization cross-terms and the uncorrectable portion of the direct-terms. Using the same DM solution and converting the input WFEs to 0° and 90° conventions, the direct and cross-term contrast maps are shown in Figure 44. While the cross-term components comprise only ~0.002% of the flux entering the coronagraph in this bandpass, they contribute 16% of the light within the HLC dark hole at the final image.

As was shown in Figure 39 and Figure 40, the amplitude wavefront errors due to polarization are miniscule compared to those that arise from phase-to-amplitude transformation of optical surface errors. When the DM solution derived with both phase and amplitude polarization-dependent aberrations is used and the polarization amplitude errors then omitted, the change in contrast is <$10^{-10}$ (Figure 43e).

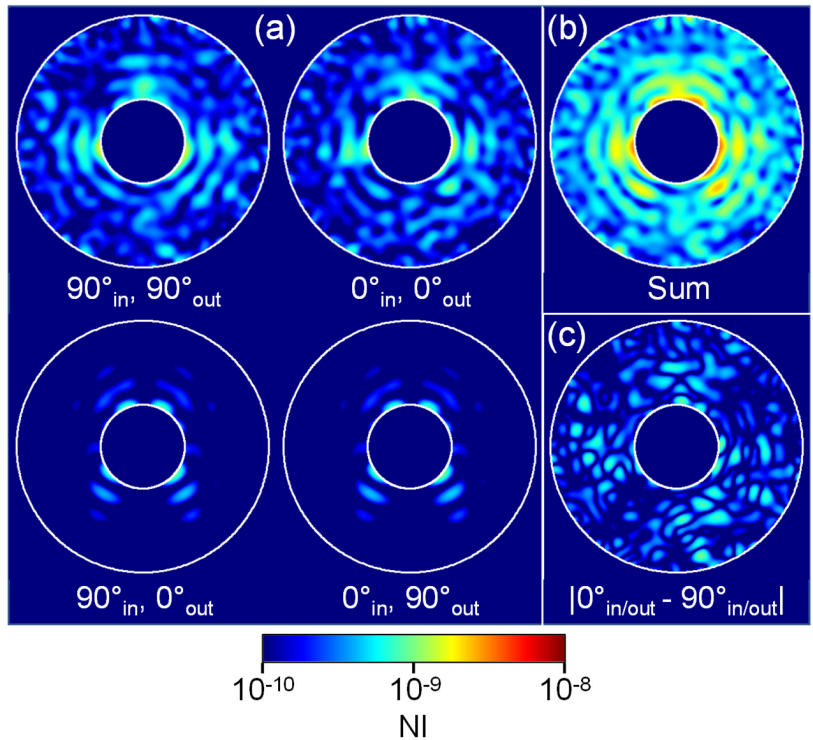

**Figure 44. HLC (Band 1) dark hole normalized intensity maps, equivalent to those in Figure 43 and using the same DM solution but with 0° and 90° input and output polarizations. Images in (a) are normalized relative to the full system illumination (the four polarization component images are added together, not averaged, to create the sum). (a) Direct and cross-term polarization component images; (b) The intensity sum of the polarization component images shown in (a), representing what would be seen without a polarizer; (c) The absolute difference between the 0° and 90° outputs, highlighting the differences between fields that would be seen with orthogonal polarizers. The superposed circles are $r$ = 3 & 9 $\lambda_c$/D.**

The very low aberration sensitivity of the SPC-WFOV coronagraph means it is largely unaffected by polarization-dependent WFE (Figure 45). In Band 4, the dark hole dug in an aberrated system including polarization differs in mean NI by only $3\times10^{-11}$ from a hole generated using the same DM solution but without polarization; the maximum differences are $<10^{-9}$, confined near the IWA. The differences seen in the dark hole between the two orthogonal output polarizations (Figure 44c and Figure 45c) indicate how suitable polarization differential imaging[51,52] (PDI) might be for observations of some circumstellar dust disks. PDI is a post-processing technique that has been used with data from ground-based telescopes to distinguish instrumental artifacts (speckles) from intrinsically polarized circumstellar disks. The assumption is that the speckle field observed simultaneously in two polarizations will appear the same and share the same time-dependent responses to wavefront changes, while a disk that has a strong polarization signature (dependent on the properties of its dust grains) will differ. By subtracting the two images, the speckles will subtract out, revealing the polarization differential image of the disk. This has the benefit of avoiding speckle changes between non-contemporaneous images that would create residuals after subtraction. The models suggest that PDI may be applicable to imaging with SPC-WFOV and the polarizers since the speckles do not have a strong polarization dependence.

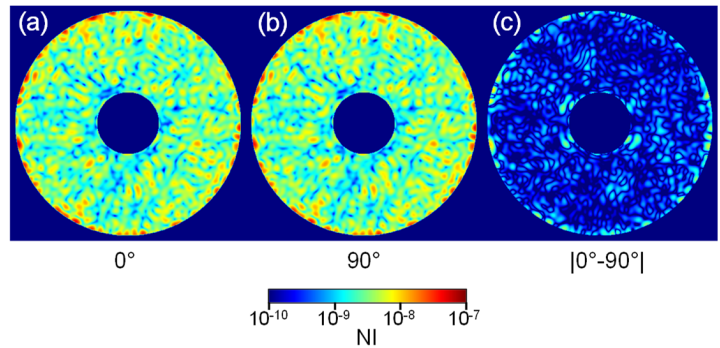

**Figure 45. The SPC-WFOV (Band 4) dark hole at 0° (a) and 90° (b) polarizations in the aberrated system. The hole was dug using WFC without polarizers. The small absolute differences between the two channels (c) demonstrate the minimal impact that polarization-dependent aberrations have on the SPC-WFOV. The hole is shown between 6 and 20 λc/D (λc = 825 nm).**

### 4.5 Low-order pre-CGI aberrations

The OTA and TCA error maps include the measured low order figuring errors as well as predicted gravity release and cool-down aberrations. Besides these, there are also low-order alignment and polarization-dependent errors, as previously discussed. The dominant low-order surface errors tend to be Z4 – Z11. Defocus (Z4) is largely removed by alignment during system integration, though at the expense of a magnification error; on-orbit, the secondary mirror and FCM can be used to adjust focus. Astigmatism (Z5 & Z6) and coma (Z7 & Z8) may be partially compensated by alignment as well, but not trefoil (Z9 & Z10) or spherical (Z11). No attempt is made in our models to align out any low-order surface errors; as it is, the total measured errors are significantly below the requirements. The estimated WFE at the FSM is 35.7 nm RMS (Table 2) while the *Roman* requirements specify a conservative WFE of 76 nm RMS entering the CGI over all spatial frequencies, excluding Z4, which can be corrected in isolation by the FCM. That leaves a healthy WFE margin of ~67 nm RMS, practically all of which is low order. This margin could absorb misalignments and other unexpected deformations. To make up some of this surplus in the model, a random collection of low-order WFE errors totaling 60 nm RMS is added at the FSM (these are not included in the errors shown in Table 3). As these errors are easily compensated by the DMs, the only real utility of including them is to estimate the maximum DM stroke.

### 4.6 Pupil non-planarity and offset

One significant characteristic not explicitly present in the unfolded layout is the non-planarity of the telescope's pupil image within the CGI[53]. The TCA optics correct the off-axis aberrations but also create a three-dimensional, distorted pupil. The interface requirements between the telescope and CGI specified that the pupil would be located at the CGI entrance (FSM) to within some tolerance. Technically, that is what has occurred, but only along the chief ray. It was an oversight that there was no requirement for the planarity of the pupil image, which was not part of the TCA design optimization. The result is that the surface on which the pupil image is in-focus is curved and inclined relative to the internal CGI pupil (which explains some of the pupil aspect ratio), and it is only coincident with it at a point or along a chord across the beam, depending on the optic. This property was discovered after fabrication of the TCA, so a redesign was not possible.

This distortion of the pupil cannot be practically represented in the diffraction model employed (Section 5) as it is due to non-paraxial aberrations. Instead, the *entire* pupil in the model is displaced along the optical axis by the maximum offset from the FSM that is in the actual pupil. This is done by increasing the separation in the collimated beam between the TT Fold and the FSM, which places the in-focus pupil ~33 mm in front of the FSM. This then offsets the pupil 6 mm in front of the SPC pupil mask, which is where the sensitivity to the error is greatest. This represents a worse-case condition where the entire pupil is defocused by the maximum amount rather than just one portion.

### 4.7 Scattered and stray light

The diffractive propagation model used to compute the dark hole is not suitable for predicting the additional scattered and stray light caused by far-off-axis astronomical sources, dust on the optics, surface microroughness, incomplete baffling, and radiation-induced luminescence from the transmissive optics (e.g., imaging lens). These are instead computed separately, including using scattering software such as FRED®. Analysis shows that the expected contribution from these sources is equivalent to a $<10^{-11}$ contrast increase in the dark hole. This factor is therefore not included in the simulations described in this paper.

## 5    Wavefront propagation

It is critical to capture the change in the wavefront as it propagates from optic to optic, given that surface errors on out-of-pupil optics can morph into a combination of phase and amplitude errors at focus, and beamwalk on out-of-pupil optics caused by source offsets can change the system WFE. Such effects cannot be captured using simple Fraunhofer (far-field) diffraction calculations that use single Fourier transforms to propagate directly between the pupil and focal planes. Angular spectrum and Fresnel algorithms can be used for efficient propagations over arbitrary distances between optics, with the constraint that the results are limited to the paraxial regime (a valid assumption for coronagraphs with small fields, like CGI)[54].

There are a variety of programs that can be used compute diffraction between surfaces. Most incorporate versions of propagators based on Fourier transforms. Some use alternatives, such as the beam propagation method that is a hybrid of ray tracing and analytical diffraction propagation of Gaussian beams; these tend to be difficult and costly to use at the precision required for coronagraphs due to the large number of rays required and the need to frequently remap the rays at surfaces (CGI has over two dozen optical surfaces to propagate between at multiple wavelengths and polarization states). When generating a time series of CGI images for a days-long observing scenario, tens of thousands of runs through the model are required. It is therefore necessary that the computations be done both accurately and efficiently.

### 5.1  PROPER

We use the PROPER[55,56] optical propagation software library for most of our numerical simulations. PROPER is a free, open source set of wavefront propagation routines developed primarily for modeling coronagraphs and is available in IDL, Matlab, and Python versions. While not optimized to be the fastest possible code (it does not utilize speedups available with graphics processing units, for instance), it provides cross-platform compatibility and ease of use. It has been verified against more rigorous diffraction algorithms (e.g., S-Huygens[57], and transitively, Rayleigh-Sommerfeld) for a simple coronagraph[58]. The CGI model has been translated into those three languages and consists mostly of optical property definitions and calls to PROPER routines. An advantage seen for CGI modeling is that an analyst can run the model within an environment they are most comfortable. Versions of these models, identical to those used by the CGI project except for the export-controlled primary and secondary mirror error maps, are publicly available (Section 5.9).

PROPER provides functions for propagating a wavefront with automatic selection of Fast Fourier Transform (FFT) based algorithms (near & far field), as well as representing realistic optical surfaces. Powered optics are replaced with ideal thin lens approximations. Aberrations can be added to each surface as specified by Zernike polynomials, PSDs, and error maps. Deformable mirrors are modeled with an actuator influence function measured from a DM like those used in CGI. Functions are provided for drawing obscurations composed of ellipses, rectangles, and polygons, all antialiased. Note that PROPER does not include any wavefront control algorithms.

PROPER chooses the propagator depending on whether the beam is converging/diverging or collimated (or very nearly so). In either case, the geometric beam diameter is kept constant within the wavefront array, so in a converging or diverging beam the sampling will change with propagation distance while it remains constant in a collimated one (which also includes the region at and very close to focus, or more specifically, locations within the Rayleigh distance of the beam waist).

The user specifies the beam diameter in both physical units and pixels ($N_{pupil}$) at the entrance pupil (the primary mirror aperture scraper, in the case of CGI) and the diameter in pixels of the wavefront array ($N_{array}$). Since FFTs are used in the propagators, sufficient array zero-padding must be provided to minimize aliasing and wrap-around artifacts (the beam diameter should be no more than half the wavefront array diameter). The sampling at focus is $\Delta = (N_{pupil} / N_{array}) \lambda/D$ per pixel, and it can be made finer by making $N_{pupil}$ smaller. The extent of the field is expanded by making $N_{array}$ larger.

The choices for sampling and extent are typically a compromise between accuracy and execution speed. The minimum pupil diameter used in the CGI model is 309 pixels in the HLC mode. This is set by the need to sufficiently sample the obscurations and the DM actuators (approximately 7 pixels per actuator). The SPC uses a diameter of 1000 pixels to sample the many edges of the pupil mask. The nearest factor-of-two array widths (because conventional FFTs are used for propagation) that are at least twice these beam diameters are 1024 (HLC) and 2048 (SPC) pixels. These provide focal plane samplings of $\Delta = 309/1024 = 0.3 \lambda/D$ (HLC) and $1000/2048 = 0.49 \lambda/D$ (SPC).

Because all the CGI modes use hard-edged FPMs, it is critical to sample the masks at high resolution to accurately capture their effects. To produce such high sampling using the default PROPER routines would require an impractically large $N_{array}$ and relatively small $N_{pupil}$. As will be discussed in the next section, an alternative algorithm is used to propagate the beam to/from the FPM at high resolution.

In the PROPER model the user can specify the final image sampling, which is obtained by using damped sinc (Lanczos) interpolation of the complex-valued electric field (we interpolate the E-field rather than intensity because the E-field representation contains less high spatial frequency content, and so suffers less due to the finite sampling resulting from the trade between precision and efficiency). The CGI detector has 13 μm pixels, corresponding to 0.5 $\lambda/D$ at $\lambda = 500$ nm. In cases where the area of the detector pixel may impact the results (e.g., studying the ability to dig a dark hole with large pixels that undersample the field), it is necessary to integrate the intensity over each pixel. This is done by interpolating the E-field to 9× finer sampling than the detector pixel, converting the result to intensity, and then 9×9 block averaging to detector sampling. Our studies have shown that integration over the pixel is unnecessary for evaluating performance at the detector sampling in a comparative sense (e.g., comparing WFC algorithm parameters), but we use it when simulating science observations.

### 5.2 Modeling the HLC

Most of the HLC can be represented in the diffraction model using the default routines in our chosen software, PROPER, including the DMs, propagation between the DMs, and propagation to/from the masks (FPM, Lyot stop, field stop). The designer provides antialiased representations of the *Roman* pupil and the Lyot stop (both assume 309 pixels across the geometric beam diameter) and maps of the DM actuator strokes for the unaberrated dark hole solution. The FPM is specified as two-dimensional thickness maps for the nickel, PMGI, and titanium coatings, 209 pixels across (0.027 $\lambda_c/D$ per pixel), along with tables of assumed indices of refraction versus wavelength for each material. Standard thin-film equations are used to compute the complex reflective and transmissive modulations. The mask is on an anti-reflection (AR) coated glass substrate, though the AR layers are not included in the models (the actual coating recipe is proprietary and thus cannot be accurately represented, but modeling using a common AR coating shows that including it would not have any significant impact on the model fidelity).

Special consideration is required to account for the small size and hard edge of the FPM. Since the simulations are used to evaluate fabrication and aberration tolerances, the finest practical sampling is needed to ensure the best accuracy. Our 309-pixel diameter HLC pupil representation and 1024-pixel wavefront array diameter used with PROPER's propagators provides a sampling of 0.3 $\lambda/D$ at focus. The FPM thickness maps are sampled at 0.027 $\lambda/D$ per pixel, almost 10× finer. To match this, $N_{array}$ would need to be over 11,000 pixels wide and would increase computation times excessively. To avoid this, some modelers resort to resizing the FPM maps to coarser samplings by interpolating either in thickness or complex transmission, allowing for more practical array sizes but inevitably introducing some artifacts. Doing so may result in an apparently suitable representation, especially with wavefront control applied to compensate for any numerical errors. However, the resulting performance estimates have been shown from our experiments to be inconsistent and may not match those of the actual FPM, depending on the interpolation method used.

Instead of scaling the FPM, we instead scale the wavefront by upsampling the E-field at the FPM using a Matrix Fourier Transform (MFT)[59]. This is a variety of the Discrete Fourier Transform that allows user-specified output sampling, unlike an FFT. Temporarily detouring from the usual PROPER routines, the field at the FPAM, before the FPM is applied, is propagated via an FFT to a virtual pupil plane, where two copies are produced, $E_0$ and $E_1$. $E_1$ is propagated back to focus using an MFT at a sampling equal to the thickness map's, but only within the region of the FPM, producing $E_{focus}$. The field multiplied by the FPM is then:

$$E_{FPM} = E_{focus} M (FPM - T_{clear}) \qquad (13)$$

where $M$ is set to 1.0 where the FPM thickness is non-zero and 0.0 elsewhere, $FPM$ is the complex-valued focal plane mask transmission function where $M$ is 1.0, and $T_{clear}$ is the complex-valued transmission where the mask is transparent ($M$ is 0.0) and includes the antireflection coating. $E_{FPM}$ is propagated back to the virtual pupil with the inverse MFT and then added to $E_0$. This effectively subtracts the original E-field within the FPM area and then adds in the same region multiplied by the complex FPM pattern – the field outside the FPM is unmodified (this process is an application of Babinet's principle). The result is Fourier-transformed to focus at normal sampling, and then the usual propagators are used to go through the rest of the system.

### 5.3 Modeling the SPC

Modeling the SPC is easier than the HLC since the DMs are not part of the diffraction suppression and the FPMs are simple apertures. For both modes the designer provides the *Roman* pupil, pupil mask, and Lyot stop patterns at 1000 pixels/D resolution and the FPM at 0.05 $\lambda_c/D$ (SPC-Spec) or 0.1 $\lambda_c/D$ (SPC-WFOV) sampling. The PROPER model uses 2048 × 2048 wavefront array grids for the SPC modes, providing 1000/2048 ≈ 0.49 $\lambda/D$ sampling at focus. The pupil and Lyot masks can be applied directly, but as with the HLC, the FPMs are hard-edged and not well represented at the default model resolution, so MFTs are also used to propagate at high sampling. Because the SPC FPMs have limited outer extents, unlike for HLC, it is not necessary to create copies of the virtual pupil field and subtract out the region inside the FPM – the high-resolution field can be directly multiplied by the FPM.

### 5.4 Representing thick lenses and doublets

Powered reflective optics in CGI are represented by ideal thin lenses in the PROPER model. This provides a reasonable approximation to the actual system in terms of resulting beam sizes and effective focal lengths. However, CGI has a variety of actual lenses in the back end: the imaging and pupil imaging lenses (both doublets) and four defocus lenses (singlets). The beam sizes and effective focal lengths change during propagation through these in ways that cannot be replicated with a single ideal thin lens approximation. In the CGI model a thick lens representation is adopted for these elements using the thick lens equation; note that this capability is not yet provided by PROPER but is implemented explicitly in the CGI model because of the limited conditions under which this representation is appropriate. This approximation reproduces both the sizes and focal lengths of the beams exiting the lenses. Using just thin lenses, in the normal imaging mode the exiting beam diameter is 3% different than that predicted from ray tracing when the focal length is forced to be the same. With the thick lens model, the beam size error is only 0.5% and the focal length error is 0.1%, using the prescribed lens properties. Including the indices of refraction also means that the system exhibits the expected chromaticity.

### 5.5 Deformable mirror model

The deformable mirror is represented as an array of actuator influence functions. The function describes the deformation of the DM facesheet when a single actuator is pushed or pulled while all other actuators are kept at a uniform height. The deformation can extend over the span of multiple actuators and depends on the thickness and stiffness of the facesheet. By default, PROPER uses a function that spans 9×9 actuators sampled at 10 points per actuator in each dimension. It was derived from measurements of an AOA Xinetics 32×32 DM, like the type used on CGI. A similarly sampled, sparse grid of delta functions representing actuator pistons is convolved by the influence function and the result is interpolated to the wavefront sampling. The interpolation can include offsets and rotations of the DM to map it to an inclined surface. With 9.65° tilts of the CGI DMs relative to the chief ray, there are 46.3 actuators spanning one direction and 47.0 actuators over the other when projected into the pupil. Note that the simulations presented later in Section 8 use the recently-measured functions of the flight DMs, described in Section 7.1.3, while all other simulations presented herein use the default PROPER function.

The actuator pistons in PROPER are specified in meters of stroke, but volts are sent to the real DM, with a maximum 100 V. This involves a digital-to-analog converter (DAC) and a calibration of the gain (volts-to-stroke) for each actuator. The DAC is modeled as a 15-bit system to conservatively represent the actual, nonideal 16-bit one used in CGI. In the results shown prior to Section 7, where a revised DM model will be introduced, it is assumed the mean maximum actuator stroke is 500 nm (5 nm/V gain) and there are gain variations of 5% among the actuators. The quantized mean stroke resolution is $500/2^{15} = 15.3$ pm. Recall that the effect on the wavefront phase is twice the surface displacement.

Neighbor rules are applied to DM strokes. These are intended to provide a wide margin against inelastic deformations in the bonds between actuators and the facesheet if the height difference between adjacent actuators is too great. The flight limits are 50V between horizontally or vertically adjacent actuators and 75V between diagonal ones. To enforce compliance with the neighbor rule, if a desired DM shape would violate it then the model instead substitutes a DM shape in which the neighboring actuator heights are brought closer to one another enough to comply. In practice, we rarely encounter such violations in our models except in cases where there are defects or misalignments that require large, relatively high-spatial-frequency DM corrections (such as dead DM actuators).

Out of 2304 actuators on each DM, 1640 are individually addressable, in an annular pattern (Figure 47). Models were used to determine which actuators are useful for wavefront control. These were identified by applying piston to each actuator individually, propagating the disturbance through the model, and measuring the change in the dark hole total intensity. Thresholds were applied to these "strength" maps (Figure 46) for each coronagraph iteratively to determine the minimum number of actuators required to dig a dark hole, including tolerances for alignment errors. Ineffective actuators do not need to be individually addressed, avoiding expensive additional driver electronics and their associated mass and power needs. As one might expect, an actuator that is not within the illuminated region of the pupil has no significance, unless its influence function extends inside of it. Those within the clear regions of the Lyot stops have the greatest strength, but even actuators in the blocked areas still have some effect (and typically have more chromatic behaviors, useful for broadband control). Ineffective actuators in each corner and the center are electrically tied together so that the full DM surface can be pistoned to a specified bias.

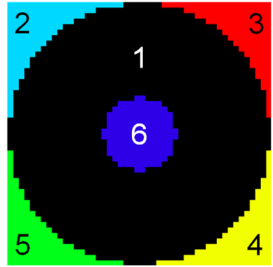

**Figure 47. Actuator addressing map for the CGI DMs as used in the models. Actuators in region 1 are individually addressable. In the other regions, actuators are electrically tied together within each region. In reality, each corner region is split into two subregions.**

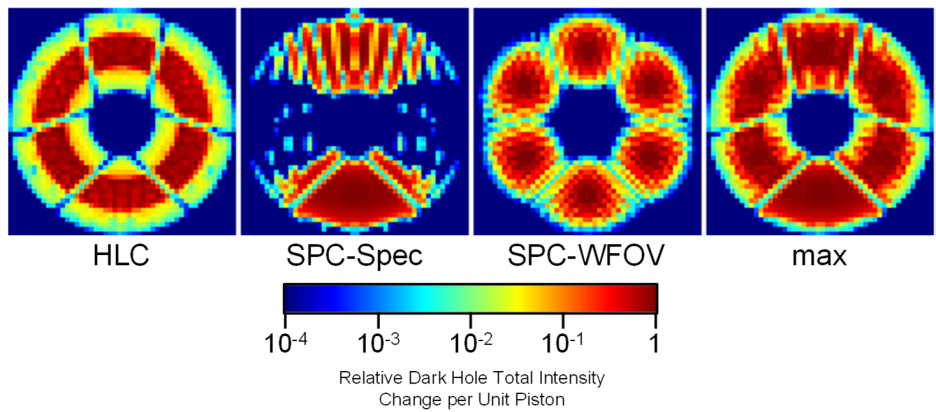

**Figure 46. The "strength" of each DM actuator for each baseline coronagraphic mode. These maps were derived by separately pistoning each actuator by an equal amount and measuring the total change in simulated dark hole intensity. The maximum value for each actuator is shown in the "max" map, which is used to determine which actuators need to be individually controlled. Each map is 48 × 48 actuators.**

Modeling has also been used to evaluate the effects of anomalous actuators, those classified as "dead", "tied", or "weak". A dead actuator either has no electrical connection or is shorted to a ground. In these cases, the facesheet is fixed in height since the actuator cannot move. Modeling shows that it is necessary to set the mean DM stroke to a level that minimizes the difference between the dead actuator and the adjacent ones while still allowing for a good dark hole. The dead actuator can also be imperfectly compensated by the matching actuators on the other DM, though there will always be an uncorrectable error that will cause some contrast degradation. The current chosen flight DMs each have a dead actuator, and their impact will be discussed later. The orientation of the electrostrictive actuation is such that increasing voltages cause the facesheet to contract toward the interior of the DM, *i.e.*, pull down, and so dead actuators (at 0 V) protrude relative to their neighbors at nonzero voltages.

During testing of the flight DMs, several defects were identified, including dead/weak actuators and deformations of the DMs. These are addressed by a revised DM model as described in Section 7.

*5.6  Representing misalignments*

Certain components may become misaligned due to thermally induced structural changes or launch vibrations, and the model must be able to represent these. The most critical, position-sensitive ones are the pupil masks (shaped pupils and Lyot stops) and the deformable mirrors, as they are diffractive elements. The orientations of the DMs can be easily changed in the model as described before. Masks displacements require Fourier-transform-based shifting and/or rotation as a means of interpolation. The transforms induce some ringing in sharp-edge masks, but experiments show these are small enough to have a negligible impact, with contrast errors of $<6 \times 10^{-12}$ over the range of expected displacements. Besides mask shifts, the entire CGI can be displaced by applying the same techniques to the field at the FSM; this replicates a bulk motion of the CGI due to thermally induced structural changes at the interface with the instrument carrier (IC), leading to identical shears at each pupil inside the instrument.

It is important to distinguish between the misalignments that take place prior to a coronagraphic observation in orbit (e.g., during launch) and alignment drifts during such an observation (e.g., due to thermoelastic drifts). For the former, we can correct them in many cases during the process of setting up an observation, using the degrees of freedom afforded by the CGI mechanisms and deformable mirrors. For the latter, we can only measure and correct in real time the low-order wavefront error terms using LOWFSC. This subject is discussed in more detail in Section 6.5 in the context of open-loop and closed-loop sensitivities.

*5.7  Pointing jitter*

In a conventional imaging system, the effect of jitter can be easily represented by convolving the model image with an appropriate blurring kernel. This is not possible with the coronagraph because the star is occulted by the FPM, and any motion of the star will alter the dark hole. Instead, multiple source offsets sampling the jitter distribution must be individually propagated through the system model and then added in intensity.

For just a single image, computing multiple stellar offsets may be onerous, but they only have to be done once. In the case of simulating jitter in an observing scenario with hundreds or thousands of images, computing a multitude of offsets at each timestep becomes impractical, especially given the time it takes to generate a single broadband image through the complicated CGI system.

A practical shortcut is to precompute the E-field *changes* for a predefined set of offsets, (*x*,*y*), add these to the E-field computed at each timestep, and then combine the fields incoherently via weighted averaging. We use an irregularly sampled distribution of 125 offsets (Figure 48a) extending out to *r* = 6.4 mas. For instance, offsets of ≤0.6 mas are spaced by 0.15 mas to well-sample small jitters and the stellar disc. Larger offsets are spaced further apart, up to 1.6 mas. Each offset represents a given area, *A*(*x*,*y*), of the potential jitter distribution (Figure 48b).

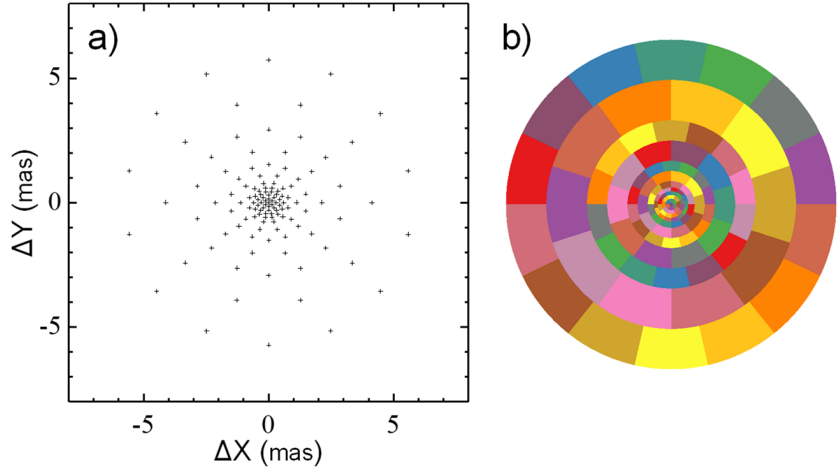

Figure 48. (a) Star offsets used for computing jitter. (b) Area of the jitter distribution that each
    pointing offset represents.

Each offset is converted to its equivalent wavefront tip/tilt at the primary mirror and then propagated through the system using the dark hole DM solution used at the beginning of the observing scenario. This is done at each wavelength spanning the bandpass, *λ*, and each polarization component, *p*, producing an E-field, $E_0(x,y,λ,p)$. The star-centered E-field, $E_0(0,0,λ,p)$, is subtracted, and the difference, $ΔE_0(x,y,λ,p) = E_0(x,y,λ,p) - E_0(0,0,λ,p)$, is stored.

At each timestep, *i*, in the observing scenario, a star-centered E-field, $E_i(0,0,λ,p)$, is generated that includes the system drifts predicted from finite-element models, with LOCAM-derived corrections. Each pre-computed $ΔE_0(x,y,λ,p)$ is added to $E_i(0,0,λ,p)$ and converted to intensity:

$$I_i(x, y, λ, p) = |E_i(0,0, λ, p) + ΔE_0(x, y, λ, p)|^2 \qquad (14)$$

and then the ensemble of images over wavelength are averaged together with weighting appropriate for the system throughput and stellar spectrum to produce the broadband image, $I_i(x,y,p)$.

A weighting function is defined on a uniform grid, (*X*,*Y*), spaced by 0.05 $λ_c$/D with an elliptical Gaussian distribution matching the predicted RMS jitter values, $σ_X$ and $σ_Y$:

$$W_{jit}(X, Y) = exp\left(-0.5\left[X^2\big/σ_X^2 + Y^2\big/σ_Y^2\right]\right) \qquad (15)$$

This is then convolved with a uniform circular top-hat function representing offsets corresponding to the stellar disc. This distribution is then resampled to match the available offsets, (*x*,*y*), and normalized to a total of 1.0. $A(x,y)$, which is also total-normalized, is then applied. The weighted offset images are summed to produce the jittered image for each polarization component, $I_i(p)$:

$$I_i(p) = \sum_{x,y} I_i(x,y,p) W_{jit+disc}(x,y) A(x,y) \qquad (16)$$

W·A effectively represent the amount of time the source spends within a sector during an exposure. The separate polarization component images are then averaged together to create the final image incident on the detector.

This shortcut is an approximation given that it uses E-field changes derived at the initial, unperturbed timestep and not at each subsequent, perturbed step. To assess its accuracy, jittered HLC images were generated via explicit modeling of offsets in a perturbed system and the results compared with those created using the shortcut. The entire CGI was shifted perpendicular to the optical axis by 0.7 μm at the CGI-IC interface (the pupil at the FSM); this is at the high end of the expected shear and represents the largest contributor to speckle instability, next to jitter itself. Over a range of jitter spanning 0.5 – 1.5 mas RMS, the maximum error at the IWA was ≤0.5%, decreasing at larger field angles. The time savings amounts to more than two orders of magnitude.

### 5.8 The compact and control models

The full CGI model represents each optic and mask, from the primary mirror to the final focus. Given that most of the optics are not in pupil or focal planes, and multiple DMs are used for wavefront control, Fresnel propagators are used to go from surface to surface. This allows the model to capture phase-to-amplitude transformations, the effects of beamwalk relative to surface aberrations, and displacements of masks from the fundamental planes (pupils and foci). Propagation through the entire model is computationally expensive given the number of surfaces and the array sizes required. When high absolute accuracy is not needed and execution time is more important, a simplified representation of the system can be used, the *compact model*.

The compact model is mainly useful for rapidly predicting the dark hole E-field *change*, $\Delta E$, rather than the absolute field, $E$, in response to alterations of the DM patterns during wavefront control. The model simplifications that allow for the increased execution speed may result in significant errors (order of magnitude, depending on what non-ideal properties are in the full model) in $E$ (e.g., the speckle patterns may not appear the same as those from the full model). However, when differenced relative to a similarly computed, unperturbed E-field, the numerical errors from the approximations tend to subtract out, leaving a representation of $\Delta E$ of sufficient accuracy for wavefront control.

In the compact model the entrance pupil is at DM1 rather than the primary mirror. The pupil amplitude and phase there can be defined by extracting them from a propagation through the full model (which includes the summation of the system errors, though only for a particular alignment state). Once CGI is in space, phase retrieval-derived phase and amplitude measurements may be used instead.

After DM1, the beam is propagated via the angular spectrum method to DM2 and then back to the pupil. If the SPC is being modeled, the pupil mask is applied. The wavefront is then propagated directly to focus with a single FFT, and the FPM is applied. While the same MFT-based propagation method used in the full model can be used here to allow high-resolution representation of the FPM, we have found that it is not necessary when computing $\Delta E$ with the compact model. Rather, the FPM can be resampled via MFTs to match the default pixel scale.

From the FPM another FFT takes the wavefront to the Lyot stop, followed by another to the final focus. The field stop is ignored as there are no further aberrated optics in the compact model to scatter light from outside the dark hole. If the back-end optics contain sufficiently high aberrations to distort the final image, then those can be included (they could be derived in reality using phase retrieval on images taken with a pinhole in the FPAM).

The analytic treatment associated with diffractive propagation through the full model simplifies to that of the compact model in the case with no wavefront modifications aside from the thin lens treatment (i.e., no additional surface errors) at the individual optic surfaces that are not explicitly visited in the compact model (at OAPs, flat mirrors, lenses). No propagation errors are introduced by this approach, but as mentioned, the computational savings available to the compact model can only be realized by neglecting the details of the surfaces that are not modeled beyond their thin lens effects.

A version of the compact model, the *control model*, allows for rapid computation of the DM response matrix, the *Jacobian*. The Jacobian predicts how the electric field in the dark hole changes in response to an actuator piston, for each active actuator on each DM and in each sensing bandpass; for the SPC spectroscopic mode, this is 1640 active actuators per DM × 2 DMs × 5 sensing bandpasses = 16400 cases. The model is run separately for each case, poking an actuator by a small amount and recording the E-field change at focus.

The time required to generate the Jacobian could be hours using the compact model as described, even if running separate pokes in parallel. A massive improvement in speed can be realized by recognizing that most of the wavefront remains essentially unchanged when modification by an actuator poke is made at a pupil (DM1, SPC mask, Lyot stop) or near it (DM2) – only the region within and immediately surrounding the poked actuator changes significantly. Instead of propagating the entire beam with FFTs, just the region around the actuator (e.g., 5 × 5 actuators in the models examined here) is propagated using MFTs, both for angular spectrum (between DMs) and Fraunhofer (pupil-to-focus and vice-versa) diffraction (note that the array size after the first MFT used in the angular propagator must be large enough to contain all of the significant spatial frequencies). The unperturbed field is subtracted from the perturbed one at the DM so that only the disturbance, not the full field amplitude, is propagated. Using this method, parallel processing can be used to generate a Jacobian in the span of minutes, not hours.

The control model is used to generate the WFC Jacobian whether in simulation or in reality (the flight control model uses dedicated, Python-based code rather than PROPER). Mismatches between it and the full model or actual system can significantly degrade how deep a dark hole can be produced. While the simplifications in the control model do not appear to be limiting factors, differences between the assumed and actual system properties can be (e.g. unknown misalignments or FPM fabrication errors). Such mismatches will be discussed in later sections. A summary of the different models is provided in Table 5.

## 5.9 *Publicly available modeling software from the CGI project*

The PROPER-based Phase C CGI propagation model is publicly available[60] for IDL, Matlab, and Python. It is largely unmodified compared to the version used here, except that it replaces the export-controlled primary and secondary mirror surface measurements with synthetic maps that approximate them. The measurements of the other optics have no restrictions and are thus provided. At the time of writing, it also does not include the revised DM model described in Section 7. Representative DM solutions that produce good dark holes are provided for the three baseline coronagraphs (HLC Band 1, SPC-Spec Band 3, and SPC-WFOV Band 4). The model includes masks for the other coronagraphs, but no DM solutions for the aberrated system are provided.

**Table 5. Summaries of model types used for simulation and wavefront control**

| Model | Purpose | Propagates | Speed | Initial Plane | Features |
|-------|---------|-----------|-------|---------------|----------|
| Full | Dark hole evaluation | Full wavefront | Slow | Primary mirror | Propagation between each optic/mask, including aberrations; thick lenses |
| Compact | Rapid estimation of sensitivities, image change prediction for WFC | Full wavefront | Medium | DM1 pupil (Full-model-derived WFE up to DM1) | Far field diffraction (pupil-focus, no lenses), except between DMs |
| Control | Generate Jacobian | Single actuator region | Fastest | DM1 pupil (Full-model-or phase-retrieval-derived WFE up to DM1) | Far field diffraction (pupil-focus, no lenses), except between DMs using subarray transforms |

The PROPER model only represents the system and does not include wavefront sensing and control algorithms. Those must be provided by the user, or the FALCO[33,34] package may be used.

Another program, CGISim, is a Python-based wrapper around the PROPER model. It allows for easier simulation of dark hole images by including databases of the filters, system throughputs, and stellar spectra. Given DM settings, it can generate an image at CCD sampling with the appropriate flux rates for a given combination of filter, stellar spectral type, and brightness. Since it runs the PROPER model, it can also utilize its optional parameters (e.g., mask displacements). For spectroscopic simulations, CGISim will return a datacube of well-sampled frames over 31 wavelengths – it is up to the user to apply a slit and compute the dispersion onto the detector. CGISim[60] has been used by the project primarily to generate images for testing phase retrieval.

A spectrographic observation simulator will soon be available, developed by the Goddard Space Flight Center. It will take the multi-wavelength datacube produced by CGISim or the observing scenario time series, map a slit on it, and then disperse it over a detector. It will be available at the CGI distribution site[61].

A Python-based EMCCD simulator, emccd_detect (also used by CGISim) and a photon-counting calibration[62] routine, PhotonCount, are also available[61]. We note that the detector itself does not perform photon counting, which requires post-acquisition processing.

A summary of public CGI simulation tools (both diffraction and astronomical image generation) is provided by Douglas et al.[63]. Recently, the Phase B CGI PROPER model was converted[64] to the POPPY[65] propagation package.

## 6 Wavefront sensing and control

There are three stages of wavefront sensing and control (WFSC) used in CGI:

1. Phase retrieval (PR): derive the total system aberrations and the alignments of masks and DMs using in-focus, defocused and pupil images, then correct gross ($\geq$5 nm RMS) wavefront errors with the DMs.
2. Low-order wavefront sensing and control (LOWFS): derive the changes in low-order aberrations (Z2-Z11) over time and compensate for them using the FSM, FCM, and DM1 to maintain a stable dark hole.
3. High-order wavefront sensing and control (HOWFS): iteratively derive the electric field at the final focus and use the DMs to dig a deep dark hole.

Testbed demonstrations provide real-world proof of concept for coronagraph and WFSC effectiveness[15,19]. The CGI performance testbed at JPL, an operational duplicate of the flight instrument, was dedicated to demonstrating high and low-order wavefront sensing and control to TRL-5. WFSC will also be demonstrated with the flight instrument hardware and software during the upcoming validation and verification (V&V) campaign during pre-launch tests. The testbeds are, however, resource intensive (time, labor, budget) and cannot reproduce all facets of on-orbit conditions, such as the telescope's aberrations or measured defects in flight masks (which would not be risked being used in a testbed). At the same time, the models described in this paper have played key roles in predicting wavefront control performance prior to V&V, evaluating requirement flow-down and assessment of any lower-level requirement incompliance, developing CGI observing scenarios, and assessing end-to-end Roman observatory performance in orbit.

We note that there is a plethora of WFSC methods that could be used with CGI. The technology demonstration task, however, is specifically limited to the methods described here that have been used for years in the JPL testbeds. After successful completion of the demonstration, NASA may choose to allow the use of alternative algorithms, contingent on resources.

### 6.1 Phase retrieval

The first step of wavefront control is to measure the gross system aberrations and compensate for them using the DMs. This requires deriving the wavefront over the full pupil from images with some known diversity. For this purpose, in addition to the imaging (in-focus) lens, there are four selectable lenses that introduce -5 to +49 waves (at $\lambda$ = 575 nm) of peak-to-valley defocus. Another lens produces an image of the pupil, for a total of six images. The focal plane, Lyot stop, and field stop masks are left out of the beam to provide a full view of the pupil. Images of an isolated, bright star are taken in a filter through each of these lenses, from which the low and mid-order aberrations can be derived using phase retrieval.

The PR procedure used by CGI is a mixture of techniques and uses a simplified propagation model. Propagations are calculated from the pupil image backward through the PIL to a plane at the front surface of the lens, then forward again to the camera after the different lens. Differences between the simplified propagation model in the PR procedure and the full propagation model in CGISim include strictly monochromatic fields, and paraxial thin lenses. The pupil image serves as the reference plane, and the phase retrieval essentially estimates the phase at this plane.

The first technique estimates low-order aberrations (Z4-Z36) at the reference plane and individual image centrations. The parameters are iterated via a least-squares solver until the real

– simulated image differences are minimized. The focus differences between images are held constant to the measured powers of the lenses, and just the overall focus offset is derived (Z4). This technique, constrained to a limited set of parameters, does not allow for derivation of the mid-spatial-frequency errors, such as the polishing errors or DM actuator displacements.

A non-parametric, iterative algorithm (a variant of Gerchberg-Saxton[66,67]) is used to derive the pixel-to-pixel WFE in the pupil domain. The low-order errors measured in the prior modeling-fitting stage are used as the initial estimate of the phase errors in the pupil plane, and the measured pupil image provides the amplitude errors (square-root of the measured image). This combination is propagated backward and forward, with the appropriate focus and decentering (Z2, Z3) terms added, to produce an image for a given defocus lens. The field amplitude is replaced with the measured amplitude and propagated back to the pupil image, subtracting the defocus and decentering terms. This process is repeated for each lens and iterated until convergence, producing a well-sampled WFE map of the pupil aberrations.

The well-sampled WFE map is then used as the initial estimate, and the whole procedure of parametric modeling fitting followed by GS iterations is repeated. This outer loop is repeated until convergence. A similar combination was used to characterize the *Hubble Space Telescope* wavefront[68], though in that case the focus diversity was accomplished by moving the telescope's secondary mirror.

As discussed in Section 4.2, the aberrations in the back-end optics (post-FPM) are not as critical to speckle creation as those prior. Optimally, then, one would want to start the WFC process by correcting just the front-end WFE. During pre-launch ground tests, the back end WFE map will be derived with PR using a pinhole in the FPAM to simulate a star, filtering out the upstream aberrations. This will then be subtracted from the full-system WFE measured on-orbit to determine the front-end aberrations. The pinhole cannot be used on-orbit because the pointing system relies on the starlight reflected from the front of an FPM with a phase dimple, which would not be present. No significant changes in the back-end optics are expected during the ground-to-space transit (this is based on the analysis and component level test results, and will be further verified by comparing optical alignment and phase retrieval test results before and after CGI dynamic testing). A pinhole is included as an option in the PROPER model.

Experiments using PR on CGISim-simulated images are used to determine estimation accuracy and identify error sources. These models allow comparison of the derived wavefront to the "truth" (the computed field), something not possible in testbeds or on-orbit. The simulations (Figure 49) can include errors unknown to the PR routine: different aberrations for each of the lenses, uncertainties in the assumed focal lengths, detector noise, and incoherent polarization-dependent aberrations. Comparisons are made to the computed pupil E-field without lens aberrations included. In the absence of any of the above uncertainties, the derived PR error is 2.8 nm RMS. Adding lens aberrations – only one wavefront is derived from six images, with each having a different lens WFE (Figure 50) – increases the estimation error to 4.4 nm RMS; folding in measured lens errors into the solver could reduce the error, but it already meets CGI's needs. Adding uncertainties of up to 0.5% in the assumed lens focal lengths raises the error to 11.5 nm rms. The other error sources have no significant impact.

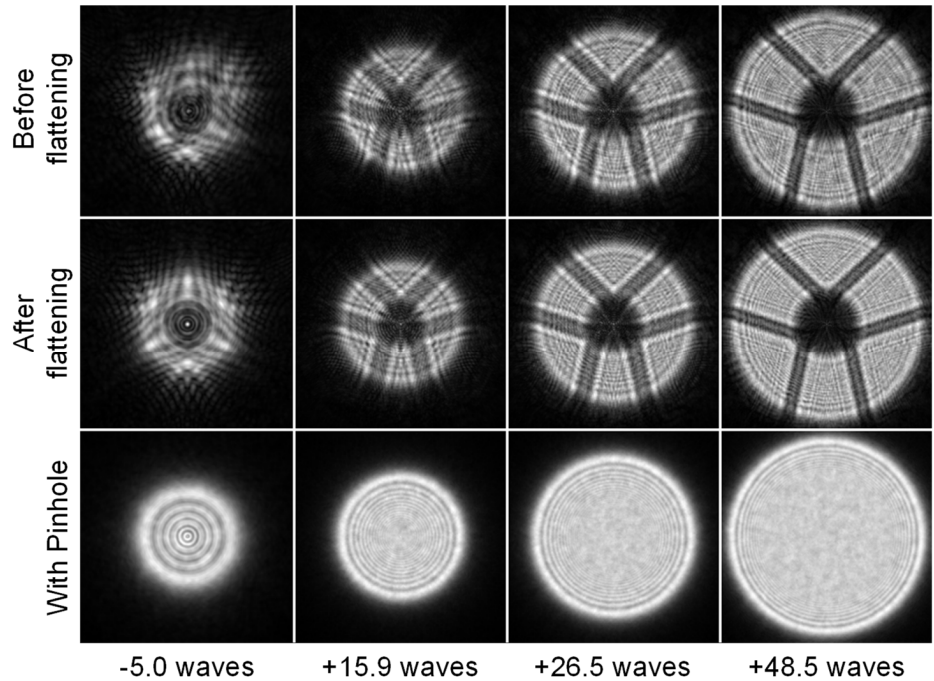

**Figure 49. Images in Band 1b produced by CGISim using the four defocus lenses (no detector noise). On top are images taken through the full system, with no masks inserted and flat DMs. Below are the images after flattening the wavefront. On the bottom are images taken with a 9 μm pinhole in the FPAM, so that aberrations prior to the FPM are not visible; these are used to derive the back-end aberrations. The images are displayed with the same spatial sampling. The peak-to-valley defocus is specified in waves at 575 nm.**

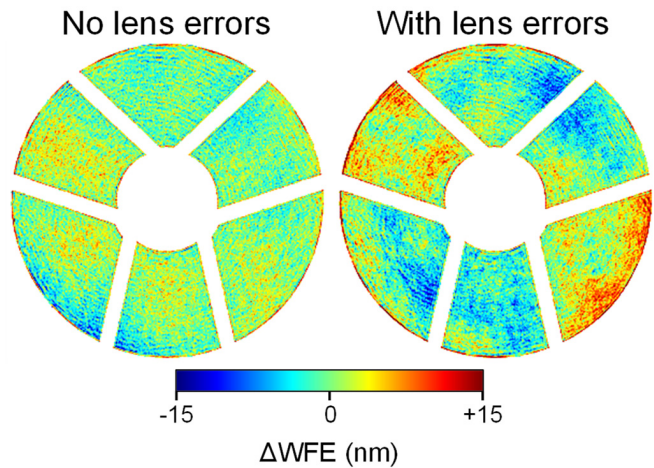

**Figure 50. Differences between the phase-retrieval-derived and actual (model) wavefront error distributions (Band 1b) based on CGISim simulations, shown in the plane of the pupil lens image. Lens (defocus and pupil imaging) aberrations were included in simulations used for the bottom map. Note that DM defects are not included in these simulations.**

In flight, the DMs will initially be set to ground-derived voltage maps that produced suitably flat phase measurements during CGI ground testing. With these, phase retrieval will produce wavefront estimates, including the OTA+TCA WFE, that are subsequently iterated to "flatten" the on-orbit wavefront (Figure 51). The CGI flight procedure is to compensate aberrations above Z11 (spherical) using DM1, while Z5 – Z11 are split between the two DMs to reduce total stroke required on just one (the low-order aberrations do not transform significantly due to the Talbot effect between the DMs; see Section 4.3). Z4 is corrected by the coarse stage of the FCM. This is an important stage as it corrects the low-order wavefront phase errors that high-order wavefront control does not optimally sense (because the FPM blocks the region where they are most apparent). It also initializes the system to operate in a more linear wavefront control regime that agrees more closely with the models. The retrieved WFE map, along with the obscuration pattern derived from the pupil image, represents the ensemble of aberrations and misalignments present in the system that is used in the control (compact) model.

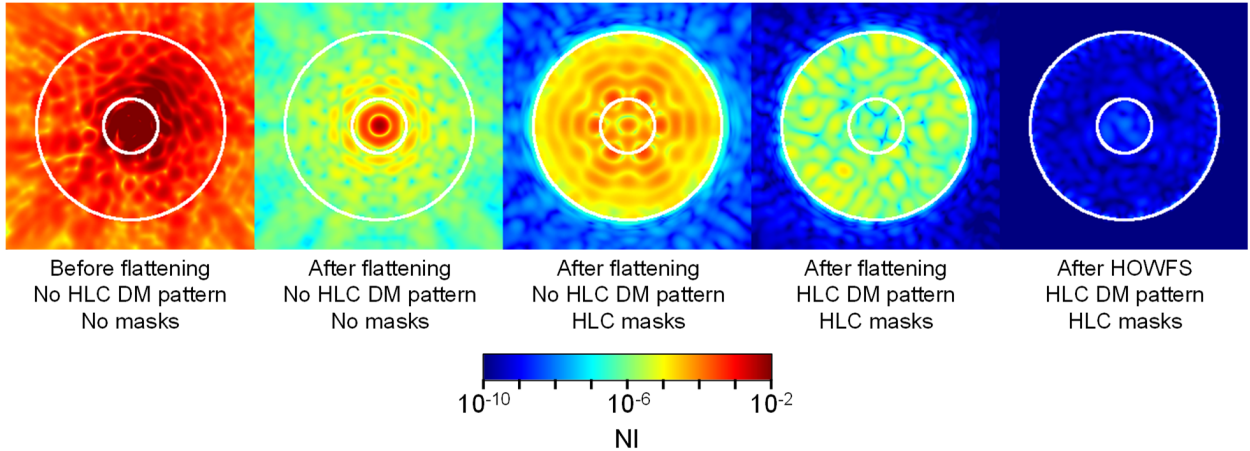

**Figure 51. Progression of the intensity within the dark hole region during WFC, starting with (left) flat DMs and no coronagraphic masks to (right) the HLC masks after HOWFS. HLC Band 1, no incoherent polarization contributions. Circles are r = 2.8 & 9.7 $\lambda_c$/D.**

### 6.2 High-order wavefront sensing

The next stage is the iterative HOWFS loop in which the dark hole region's E-field, $E_0$, is obtained and a DM solution, $\delta$, is computed that reduces its intensity. While $E_0$ is known in simulations, for a real coronagraph only intensities are measurable, so the complex representation ($\tilde{E}_0$) can be obtained only through an estimation algorithm. Because it is chromatic, the field must be obtained in multiple bandpasses across the span of the science filter to produce a broadband correction.

The *pair-wise probing* technique[69] for E-field estimation has been baselined for CGI. Pairs of positive and negative patterns ($\pm\Delta\delta_i$) on DM1 are used to perturb $E_0$, and the resulting intensities are recorded. The patterns are usually a combination of sinc functions that introduce a generally uniform offset over the dark hole in the real or imaginary E-field components. The imaginary perturbations are generally uniform across the region. The real ones are Hermitian, with the changes having opposite signs on opposite sides of the field. The resulting discontinuity along the centerline leads to undefined values in the E-field solution, so another set of images is obtained with the probe pattern rotated to solve for the real components over 360°. In total, three pairs of probes are needed to fully sample an HLC or SPC-WFOV dark hole; the SPC-Spec requires only two pairs since the centerline is masked by the bowtie FPM, though at the time of writing the project WFC strategy uses three pairs for it as well. Besides the probes, an unperturbed image is also obtained.

The amplitude of the E-field change, $\Delta E_i$, caused by the probe patterns $\pm\Delta\delta_i$ is obtained from the measured probed ($I_i^\pm$) and unprobed ($I_0$) intensities:

$$|\Delta E_i| = \sqrt{\frac{I_i^+ + I_i^-}{2} - I_0} \qquad (17)$$

The corresponding phase change, $\Delta\varphi_i$, is estimated from the numerical coronagraph (compact) model, $C()$:

$$\Delta\varphi_i = \varphi(\Delta E_i) = \varphi(C[\Delta\delta_i]) \qquad (18)$$

The two are combined to approximate the complex-valued $\Delta E_i$, which with the probe intensity differences, $\Delta I_i = (I_i^+ - I_i^-)/2$, form a set of linear equations that can be solved, pixel-by-pixel, for $\tilde{E}_0$ in each calibration filter:

$$\begin{bmatrix} \Delta I_1 \\ \Delta I_2 \\ \Delta I_3 \end{bmatrix} = 2 \begin{bmatrix} -\text{Im}\{\Delta E_1\} & \text{Re}\{\Delta E_1\} \\ -\text{Im}\{\Delta E_2\} & \text{Re}\{\Delta E_2\} \\ -\text{Im}\{\Delta E_3\} & \text{Re}\{\Delta E_3\} \end{bmatrix} \begin{bmatrix} \text{Re}\{\tilde{E}_0\} \\ \text{Im}\{\tilde{E}_0\} \end{bmatrix} \qquad (19)$$

Probing is the first step where the propagation model is used for WFC, whether in simulation or reality. The accuracy of the predicted phase change is strongly dependent on the representation of the DM in the model (alignment, assumed influence function shape, gain calibration, current DM shapes). However, because the model is only used to estimate the phase change caused by the probe, and measured values are used for everything else, its absolute accuracy does not have to be extreme (i.e., it does not reproduce the actual, ambient dark hole field to better than $10^{-8}$ contrast or worse, but because the ambient field subtracts out, it doe not need to).

The probe amplitude can be adjusted in proportion to the dark hole contrast. Because $\Delta E_i$ heterodynes with $E_0$, the probed image intensity can be significantly brighter than the unprobed one, an advantage when there is a deep dark hole and competing detector noise. There is no official formula, but the modeling presented here uses $NI_{probe} = (10^{-5} \times NI_{darkhole})^{1/2}$; this provides $10^{-5}$ mean NI probes for a $10^{-5}$ dark hole, and $10^{-7}$ probes for a $10^{-9}$ hole. The relation between probe brightness and measurement error has been studied by Groff et al.[70]

In order to set exposure parameters (gain and frame time), the compact model, $C$, can also be used to predict the dark hole intensities in the current iteration, $i$, that will result from applying a DM probe pattern, $\delta_i$, to the current DM solution, $\varphi_i$. The model is not accurate enough on its own to predict absolute brightness well, so a correction is derived based on the previous iteration's probe-measured ($\tilde{E}_{i-1}$) and predicted ($\acute{E}_{i-1}$) E-fields:

$$\acute{E}_{i-1} = C(\varphi_{i-1}) \qquad (20)$$

$$\Delta_{i-1} = \tilde{E}_{i-1} - \acute{E}_{i-1} \qquad (21)$$

$$E_{probe_i} = C(\delta_i + \varphi_i) + \Delta_{i-1} \qquad (22)$$

The resulting predicted mean probe brightness is typically good to within a factor of two in our simulations, sufficient to set the exposure. The same method can also be used to predict the brightness of the new unprobed image after a DM solution has been determined.

Probing is used in simulations when digging the dark hole in the presence of incoherent intensity effects (estimation error[71], finite bandpasses, pointing jitter, detector noise, scatter, and all polarization components). In practice, most of these effects subtract out when $\Delta I_i$ is computed, and only the coherent component of the E-field is estimated (e.g., the derived E-field represents only the mean polarization aberration). The field is assumed to correspond to the central wavelength of the calibration filter. Probing is mainly used in the simulations to determine the impact of detector noise (Figure 52). For example, in the very simple case of additive Gaussian noise (ignoring the higher-order details of a Poisson distribution due to shot noise, for example), the simulations show that a signal-to-noise ratio (SNR) of $\geq 5$ per pixel is needed under these conditions to achieve a good dark hole (SNR = mean probe intensity / noise). This means there is sufficient margin in the detector's dynamic range to capture those speckles above the mean without saturation. Testbed experiments with realistic low-light fluxes have been successful[72].

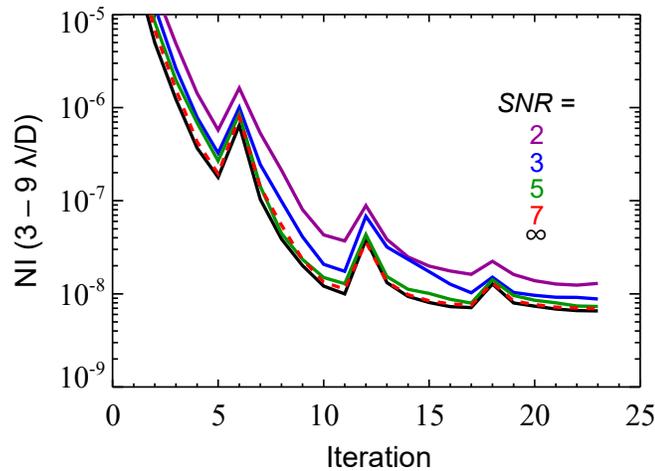

**Figure 52. Dark hole intensity versus WFC iteration using probes with the specified SNR (mean brightness/Gaussian noise). The same amount of noise was added to the fainter, unprobed image. These results suggest that a minimum SNR of 5 is needed (for normally-distributed noise).**

In real-world use probing is expensive, time-wise: in the case of HLC or SPC-WFOV, each probe is imaged in 3 calibration filters, so a total of $3 \times (6 + 1) = 21$ images are needed at each WFC iteration. SPC-Spec has 5 calibration filters spanning its larger bandpass, so it needs 35 images. Probing in simulations is computationally expensive. The image from each calibration filter is represented as the sum of $3 - 5$ monochromatic images; using the same wavelength for the end of one band and the beginning of the next reduces the total number needed. At a minimum for HLC or SPC-WFOV, a total of 7 monochromatic images are used for 3 calibration filters, so $7 \times (6 + 1) = 49$ such images are generated with the full model at each WFC iteration just for probing.

If polarization components are included (necessary when fully evaluating detector noise impacts on HLC WFC), then this is multiplied by four.

When performing investigative modeling of the system behavior, where detector noise is not important, it is most efficient to skip probing and directly use the multiple monochromatic E-fields computed by the full model, providing perfect knowledge. Typically, 7 – 9 wavelengths evenly spanning the 11% - 16% bandpasses are used. In this case, both the sensing and control wavelengths are the same. This provides a somewhat optimistic number of samples, given that in real conditions only 3 or 5 field estimates are available across the broad bandpass. This overabundance produces the best broadband dark hole, but it can lead to overoptimistic results. However, if fewer wavelengths are used, the potential exists for overly pessimistic results.

We have found empirically that a good compromise is to average the real and imaginary E-field components separately for 3 – 5 wavelengths that sample a calibration filter's bandpass to form an estimate of the E-field at the central wavelength, which we call *simulated probing*. This is technically not correct, as fields from different wavelengths should be averaged incoherently (in intensity). However, this appears to capture much of the effect of a finite bandpass while providing a single estimate of the field for that filter[73]. The dark hole results are typically the same as those obtained with noiseless true probing (10% difference at worst), but they are computed in significantly less time.

### 6.3 High-order wavefront control

The next step is to use the derived E-field to determine the DM settings that will improve the dark hole. The high-order WFC algorithm used for CGI is Electric Field Conjugation (EFC)[74]. Changing actuator heights at the DMs has a phase-only effect on the E-field reflected from the DM surface, nonlinearly changing the complex E-field via $\exp(i\varphi)$ (with $\varphi$ linearly related to actuator height). EFC approximates this as a linear relationship, appropriate for small phase changes, so the eventual DM solution must be arrived at iteratively using multiple steps of a sense-and-control loop. EFC solves for $x$ in the linear equation:

$$Gx = -\tilde{E}_0 \qquad (23)$$

where $\tilde{E}_0$ is a column vector containing the derived dark hole E-field values, sized $2N_{pix}N_\lambda$, and $x$ is the column vector of DM actuator pistons, sized $N_{act}N_{DM}$ ($N_{pix}$ is the number of pixels in the dark hole E-field, with the real and imaginary components, $N_\lambda$ is the number of sensing bandpasses, $N_{act}$ is the number of useful actuators per DM, and $N_{DM}$ is the number of DMs). $G$ is the Jacobian, the matrix that relates DM actuator pistons to changes in the image-plane E-field. As described in Section 5.8, it is built using the control model. Each useful actuator on each DM is pushed by a specified amount (typically a nanometer or so) and the model-predicted change in the E-field is recorded at each sensing wavelength. This forms a matrix that is sized $2N_{pix}N_\lambda$ rows by $N_{act}N_{DM}$ columns. Such matrices can reach hundreds of megabytes in size (the matrix for SPC-WFOV is about a gigabyte, using single-precision values).

The ideal goal when digging the dark hole is to improve its contrast, not intensity. The distinction is that contrast is the speckle field intensity divided by the unocculted source's PSF peak intensity. It is possible for EFC to dig a very low intensity dark hole but at the expense of reducing the PSF sharpness, typically by using high-spatial-frequency DM patterns that scatter light from the core into the wings. One wants to simultaneously decrease field intensity while minimizing PSF degradation. This can be done by modifying the cost function associated with the Jacobian[75]. In practice, optimizing contrast rather than intensity is primarily useful when generating the HLC DM solutions from scratch, starting with flat DMs. Once the patterns have been derived and then inserted into an aberrated system, there are no large differences in the quality of the subsequent post-EFC dark hole either way.

Field-dependent weights can be applied simultaneously to both $G$ and $\tilde{E}_0$ to guide the algorithm to a solution that creates a darker region near the IWA where more exoplanets can be seen, and sometimes this also results in a darker hole overall compared to using uniform weights. All the modeling presented herein applies 2× weighting to pixels from the radius of the FPM to IWA+2 $\lambda_c$/D. Appendix A illustrates the impact of weighting on the principal component modes that form the least-squares solution for the dark hole. The weights can also be set to zero for bad pixels (bad detector pixels or poor probing solutions). Likewise, dead DM actuators can be accommodated by zeroing-out the corresponding elements in $G$ or omitting them altogether.

### 6.3.1 Regularization

Due to the nonlinear response of the dark hole to actuator adjustments, it is necessary to dampen the changes to the DMs, otherwise the solution will not converge. The DMs cannot fully negate all light within the dark hole region due to chromatic effects and the limitations imposed by the influence functions. Trying to solve Eqn. 23 using the direct inversion of the Jacobian would therefore lead to an ill-posed condition, resulting in large DM strokes that would degrade, not improve, the dark hole. The necessary constraint is provided by regularization, and it is the most important WFC parameter. Within the CGI project, Tikhonov regularization[70] is used with factors applied using a $N_{act}N_{DM} \times N_{act}N_{DM}$ diagonal matrix appended to the bottom of the Jacobian; if singular value decomposition is used to solve for $x$, similar regularization can be achieved as a weighting function applied to the singular values. The optimal regularization values are typically determined through trial-and-error. In general, a larger value will result in greater damping and slower convergence. Conversely, wavefront control with excessive damping may never reach satisfactory levels of contrast. Note that convergence is not guaranteed with EFC, especially with insufficient damping or an inaccurate control model.

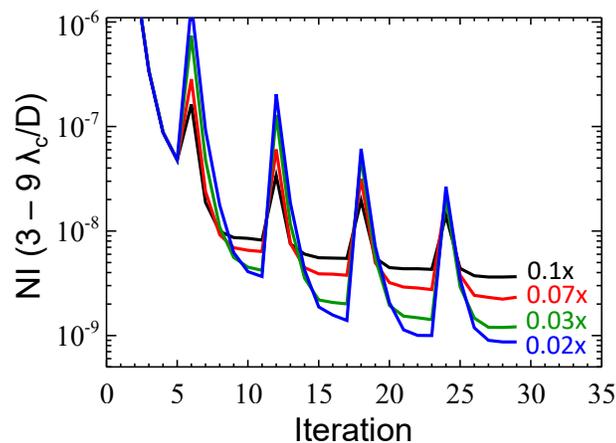

**Figure 53. Dark hole intensity in HLC Band 1 versus WFC iteration using different values for the regularization β bumps, specified as relative to the same default regularization. The Jacobian was recalculated at each iteration (see Section 6.3.3).**

The use of constant regularization over all WFC iterations will often lead to stagnation in the convergence, resulting in a poor dark hole solution. Significant improvement is achieved by periodically reducing the damping, a technique[76] called "β bumping" or "β kicking" ("β" is how the CGI project defines the regularization parameter). This behavior can be conceptually described by equating regularization to weights applied to the principal component modes used in the least-squares solution (*i.e.*, applied to the singular values; see Appendix A). In general, the default regularization suppresses modes corresponding to high spatial frequency features in the dark hole, those most likely to be highly chromatic and most sensitive to control model errors (these modes have low singular values). Occasional and momentary reductions in regularization (e.g., once every few iterations; Figure 53) allow higher-order modes into the solution, but they require relatively large actuator motions to have an effect. When this is done, modest errors in the model or calibration alter the lower-order (larger singular values) modes, and since small changes in them result in large E-field changes, the result is a significantly increased dark hole NI after the β kick. The impact can be compensated with a few subsequent iterations using the default regularization. The overall solution eventually converges to a better result compared to a single regularization.

The choice of when and by how much to change the regularization is defined by the *regularization* or *β schedule*. A variety of schedules have been explored in simulation and on testbeds: alternating every iteration; one regularization for a few iterations, then the other for a few more, then repeat; a few iterations, then a single β bump iteration, then repeat. Most simulations (including those here) and testbed experiments have ended up using the latter. The CGI EFC implementation allows any schedule.

### 6.3.2 Importance of the control model and known/unknown errors

As previously discussed, the control model is a speed-optimized version of the compact model, which itself is a simplified representation of the full optical system. It is intended to predict the *change* in the dark hole E-field caused by modification of the DM patterns rather than the field itself. The model is limited to representations of the entrance pupil, DMs, and critical masks (i.e., FPM, Lyot stop, shaped pupil, image plane sampling). Where available, measured properties are used (FPM sizes, DM and mask alignments, etc.). Instead of including aberrations on each optical surface, the entrance pupil contains a map of the system's total WFE (or at least the front-end map), derived via phase retrieval or, as is done in the simulations presented herein, from propagation through the full model.

The EFC convergence rate and solution quality are dependent on the accuracy of the control model[77]. Mismatches between the actual system (or its full model reproduction) and the control model induce errors in the predicted E-field changes that may divert the control to an unsuitable solution. This may be partially accommodated by increasing the damping via regularization, slowing convergence. The final solution may be limited by these inaccuracies to an unacceptable contrast.

There are two classes of system errors that impact control model accuracy: known and unknown[78]. Known errors are measured deviations from the design, and they may degrade performance in ways that cannot be fully compensated using WFC. These may include FPM fabrication errors, DM misalignments, mask rotations, DM gains, the phase retrieved WFE, and focal lengths (which determine magnifications), among many others. These are included in both the control and full models.

Unknown errors are the differences between the imperfect measurements and the actual values, and they are only included in the full model. An example is the thickness profile deviation from design of the dielectric coating on the HLC FPM. This can be measured accurately using an atomic force microscope, but only on a representative FPM, as the measurement process could endanger the flight article. So, there could be a small but not insignificant uncertainty in the assumed typical fabrication error included as a known quantity in the control model. Another example is the alignment of the Lyot stop or shaped pupil mask; their measurement accuracies are limited by the resolution of the pupil imaging lens image (300 pixels/D). The simulations presented herein include a variety of known and unknown errors detailed in Section 8.5.

An interesting case of control model mismatch was identified while digging dark holes on a testbed using an earlier version of the SPC-WFOV design[79]. The contrast was stagnating at an unexpectedly poor level, and it was concluded that the DM influence function used in the control model was to blame. After replacing the function that was predicted by finite-element-analysis (FEA) with a measured one, a good dark hole was obtained. The FEA function was previously used with success with the HLC and SPC-SPEC coronagraphs, but their dark hole regions are <10 $\lambda$/D in radius. This limits their sensitivities to higher spatial frequency structure in the influence function. The SPC-WFOV, with its larger field, is less tolerant of such mismatches in the control model.

### 6.3.3 Jacobian relinearization

Related to the control model accuracy is the suitability of using a single Jacobian for all EFC iterations. Especially early on, the DM pattern changes may be large enough that the linear approximation provided by $G$ is not able to predict the system response with enough accuracy to achieve convergence to a good solution. It is thus necessary to recompute $G$ using the current DM settings (*relinearization*). An example of this is shown in Figure 54. In this case, the system begins with the as-designed, obscuration-compensating HLC DM patterns added to the "flat" wavefront solutions. In addition, there is a dead DM1 actuator, fixed to a stroke of +350 nm relative to a DM bias level. This is initially compensated by manually setting the corresponding actuator on DM2 to -350 nm relative to the bias. These are the conditions used to construct the initial Jacobian. After the first EFC iteration, the stroke on DM2 changes by ~30 nm at the compensating actuator. Without relinearization and with a regularization schedule having the same $\beta$ kick values, a poor solution is reached as $G$ becomes inaccurate after the first iteration, with a slow and somewhat chaotic convergence. Repeating the experiment, the kicks are instead reduced in later iterations to avoid the degradation caused by imperfect correction of higher-order modes. This improves the result by a factor of ~2. Using these same two schedules while also recomputing $G$ at each iteration provides a 4× improvement relative to the first case, along with much faster and more stable convergence.

Relinearization also allows for larger $\beta$ kicks (allowing for correction of higher-order modes), which would otherwise cause divergence. For example, the worst result plot in Figure 53 (0.1x, black line) is the same as the best result one in Figure 54 (dashed red line). Figure 53 shows that with relinearization, more aggressive $\beta$ kicks can produce even better results, while Figure 54 demonstrates that without relinearization, even weak kicks can lead to erratic solutions. If the schedule must be predefined at the beginning of the run, it may be safer to use a more moderate kick and accept a potentially non-optimal solution rather than risk divergence.

Relinearization is required when generating the HLC DM patterns during the design process, which involves running EFC starting with flat DMs. The stroke rapidly increases during early iterations to hundreds of nm, so a fresh $G$ every iteration is a necessity.

Generating a new $G$ on-orbit would tax the ability of current flight-qualified computers. An advantage to using ground-based WFC calculations rather than the now-descoped onboard computer is that nearly unlimited computing power is available to quickly relinearize $G$ each iteration, which is the default. The compute power available in space for future missions may still be lacking for such tasks, even a decade from now, and especially if DMs with more actuators are used. Alternative WFC algorithms are being studied that do not rely on a Jacobian to avoid these issues[80,81].

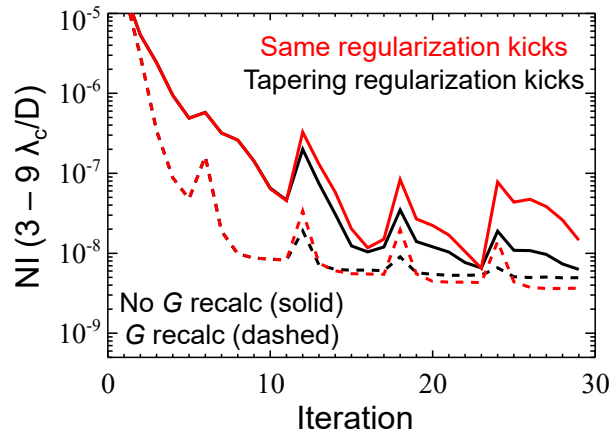

**Figure 54. Dark hole intensity in HLC Band 1 versus WFC iteration, with and without calculating the Jacobian at each iteration. The black lines represent the cases where the regularization bumps become less aggressive at later iterations to avoid divergence. The red lines show using the same regularization bump value (the initial one). The solid lines represent using the initial Jacobian at all iterations, and the dashed lines are for recalculation at each iteration.**

### 6.4  LOWFS with LOCAM

The sensitivities of the coronagraphs to pointing offsets and low-order aberration drifts as low as a few tens of picometers, such as shown in Figure 6, necessitate rapid sensing and control of wavefront changes to maintain stable dark holes. CGI's LOCAM[24], as described in Section 2.1, is a Zernike phase contrast system fed by reflection of starlight from the patterned front surface of an FPM and creates a small (~38 pixel diameter) pupil image on a dedicated camera. Images are taken at a rate of 1 kHz and are differenced to a reference image obtained at the beginning of the observation sequence – aberration changes, not absolute values, are measured. Zernike modes Z2 – Z11 are measured for each difference image. The 1 kHz tip and tilt (Z2, Z3) low-latency estimates are sent to a line-of-sight controller that operates with a control bandwidth of ~20 Hz, using the FSM in coordination with the spacecraft's pointing system. Z4 – Z11 measurements are averaged to reduce noise and sent at 0.1 Hz to controllers with 1.6 mHz bandwidths using the FCM (Z4) and DM1 (Z5 – Z11).

The difference images provide intensity imprints of the low-order phase changes and are compared, in a least-squares sense, to stored modes corresponding to Z2 – Z11, resulting in the required corrections. Early on, models were used to predict these mode patterns, but testbed experience showed that empirically derived patterns were better. This makes sense given that the system's response to modes can be directly measured by perturbing the actuators that correct those same modes (FSM, FCM, DM), avoiding any model/system mismatches. It can also measure pupil shear, but there are no corrections for it.

The rapid rate at which LOCAM acquires and measures frames means that when simulating it for a realistic observing scenario that is tens of hours long, tens or hundreds of millions of images are generated for a single run. The PROPER CGI model does not include LOCAM (PROPER is intended for more general use and cross-language compatibility and is not fully optimized for speed). A more targeted code was developed, LOWFSSim[82,83], that can compute the LOCAM image using a compact model and quickly apply EMCCD effects. It is based on the Prysm[84] propagation package for Python that is speed optimized, including GPU utilization. LOWFSSim is used to measure and provide low-order corrections in the observing scenario simulations described in Section 8.4.

### 6.5 Open and closed loop sensitivities, tolerancing, and error budgets

Throughout the past years on the CGI project, it has been surprising how well algorithms like EFC can deal with a wide variety of wavefront errors, including aberrations, DM defects (e.g., dead actuators), and mask fabrication errors and misalignments. As an example, consider a misalignment of the HLC Lyot stop by 55 μm (0.3% of the pupil diameter, or exactly one pixel in model sampling), a relatively large value. Prior to the shift, the dark hole contrast is $2 \times 10^{-9}$, which explodes to $4 \times 10^{-6}$ afterwards. However, running EFC then brings the contrast to within $0.5 \times 10^{-9}$ of its pre-shift value, even when using a control model that does not know about the shift.

This demonstrates that there can be significant differences between closed-loop (with WFC) and open-loop (no subsequent correction by WFC) sensitivities. The dark hole contrast immediately after running EFC is dominated by closed-loop sensitivities to errors such as static misalignments of masks and optics introduced during instrument integration and fabrication defects. Open-loop sensitivities include properties that can change over time and are not perfectly corrected by some mechanism, such as thermally induced changes in low-order aberrations or alignments and pointing jitter; these impact the stability of speckles in the dark hole.

Because the dark hole must maintain good contrast with an ideally constant speckle morphology to allow post-processing, wavefront stability is a critical characteristic that defines the primary metric used for CGI performance, the flux ratio noise (FRN)[85,86]. This is essentially the error in the measurement of the exoplanet/star brightness ratio derived from a set of post-processed observations, including speckle and detector noises. The CGI performance error budget tool is used to analytically predict the FRN, allowing quick evaluation of the impact of variations in many system parameters, both static and time-dependent, without having to run a full set of end-to-end simulations for a given observing scenario. The primary components of the error budget are sensitivities computed using the numerical model.

As shown in Equation 4, the instantaneous coherent dark hole intensity can be described as sum of the intensities of a constant E-field (closed loop), $E_0$, a field disturbance (open loop), $\delta E$, and their cross-term. Over a finite timespan (the time it takes to create an image), $E_0$ can be redefined to be the mean E-field, and $\delta E(t)$ is the variation over time relative to the mean. These can further be decomposed into E-field summations of individual contributors, $i$ (e.g., mask shift, focus change), $\bar{E}_i$ and $\delta E_i(t)$. It can be shown that over a finite timespan, uncorrelated E-field variations cause the cross-terms to drop out. Then, the mean dark hole intensity can be expressed as the sum of the mean contributor intensities added to the sum of the contributor field variances:

$$I_{av} \equiv \langle I(t) \rangle = \left\langle \sum_{i=0}^{n} |\bar{E}_i|^2 \right\rangle + \left\langle \sum_{i=0}^{n} |\delta E_i(t)|^2 \right\rangle \qquad (24)$$

$\bar{E}_i$ and $\delta E_i$ are obtained from the numerical model by differencing the ambient and perturbed E-fields for a variety of system variations (mask shifts, aberrations). Note that the *intensity* of each separate E-field is used. This allows the contrast sensitivity of each to be arbitrarily changed in the analytical model to accommodate uncertainty factors, something that cannot be readily done in the numerical model (the sensitivity of a specified FPM cannot be altered at will as it is a function of the physics of its design). The first sum giving the mean E-fields is simply the mean contrast over the timespan and can be arbitrarily set, guided by the contrast error budget that relates closed-loop sensitivities to dark hole intensity. Along with detector noise effects and the numerical-model-predicted field PSF properties (core throughput and area), the statistical variation over time of the dark hole can be estimated and the FRN rapidly derived.

The error budget has been validated[86] against the numerical model for a scenario that describes a timeline of observations of specified reference and target stars, including telescope orientations relative to the sun. As will be described in Section 8, thermal and structural modeling is used to predict the low-order WFE changes (including LOCAM-derived compensation) and mask shifts over a number of timesteps sampling the scenario. These are fed into both the numerical and analytical models to predict the statistics of the dark hole changes and FRN. The resulting RMS error between the error budget and numerical model over all dark hole annuli was 16%, well within the range needed to evaluate the impact of various parameters.

### 6.6 Numerical model validation

So many of CGI's performance predictions are dependent on the numerical model that it is critical to validate the simulations. Over the past couple of decades, experiments in high-contrast testbeds have shown that the models are primarily limited by the accuracies of the incorporated measured system characteristics (e.g., mask and DM alignments and scales, FPM fabrication errors) rather than the numerical application of the relevant physics. There are no expectations of being able to reproduce the testbed dark holes speckle-for-speckle in the models, so instead the goal has been comparing the post-EFC mean contrast levels and open-loop contrast sensitivities to selected induced disturbances.

The first CGI model validation tests were part of a Technology Milestone demonstration program[87,88]. These utilized the HLC and SPC-Spec coronagraph mask designs of that time, which were similar to the final flight designs described in Section 3. The testbed layout resembled flight from the FSM onwards – it had multiple pupils and foci for masks, DMs, filters, etc., but things like camera sampling were different from flight, and off-the-shelf parts were used. Initially, the input point source was provided by a compact, custom-built Cassegrain telescope optical simulator with a laser-drilled pinhole, but this produced too much chromatic pupil aberration, especially polarization-dependent WFE from the walls of the pinhole. A subsequent configuration replaced the simulator with a lithographically produced pinhole and OAPs, along with a polarizer and analyzer. A pupil mask was used to mimic the telescope obscurations. Perturbations were induced by moving a source/component or adding a low-order aberration pattern to a DM. The testbed was represented by a PROPER-based compact model with DMs, SPC masks, FPMs, and Lyot stops. The testbed total system phase and amplitude WFE maps were derived at multiple wavelengths using phase retrieval (PR) with flat DMs and no masks; these were then inserted at the entrance pupil (DM1) in the models, along with the PR-measured mask and DM alignments.

A "baseline" model was defined in which both the testbed and control models were essentially the same, with identical measured system properties and no unknown errors. This was then modified to create a Monte Carlo set of 10 testbed models (the control model remained the same), each with different values of unknown errors added to the measured ones[89]. The errors were randomly pulled from a distribution based on the estimated measurement errors. These represented 10 different instantiations of the observation. These provided an envelope of possible solutions, combinations of known and unknown errors within which the testbed results might fall. Given the considerable computation time involved for each Monte Carlo instance, we considered 10 runs to provide a reasonable estimate of the expected range of performance while being practical, given the milestone timeline.

The Milestone 9 SPC-Spec models showed excellent agreement to the testbed results[20,90], given the wide variety of potential mismatches due to unknowns (misalignments, aberrations, mask defects, etc.) and noises (detector, vibration-induced jitter). The mean contrasts of the Monte Carlo simulations and the testbed experiments agreed to within 30% and were an order of magnitude below the technology demonstration requirement. The errors in $Z2 – Z11$ sensitivities for 1 nm RMS perturbations ranged from 0% - 23%. The FPM and SPC pupil masks shear sensitivity errors were 20% and 33%, respectively.

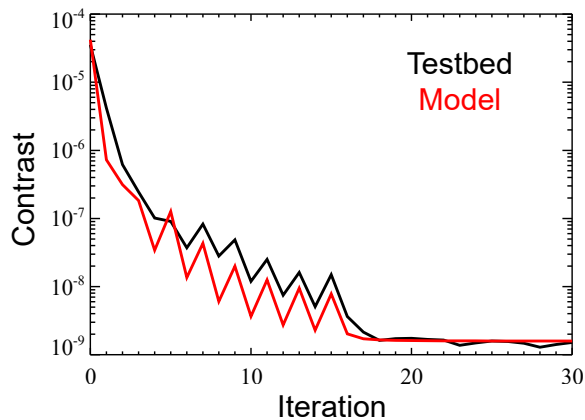

**Figure 55. HLC Band 1 mean dark hole contrast (modulated) measured in the testbed (black) and simulation (red). The regularization remained constant after iteration 15.**

The HLC tests in Milestone 9 were later found to be partially invalid[21]. Prior to running EFC, the HLC design DM patterns, which were optimized to reduce sensitivity to tip/tilt, were added to the flat-wavefront DM solution. Modeling, however, suggested that they were incorrectly added upside-down to the DMs on the testbed. This resulted in a substantial increase in the number of iterations required to reach convergence and unexpectedly higher tip/tilt sensitivities compared to the models. This was corrected in a later series of experiments meant to verify the flight wavefront control procedures using the HLC, providing a second opportunity for model validation. Besides fixing the pattern orientations, better calibrations of the DM actuator gains were available, and the models included the tied actuators present on the testbed. These HLC model and testbed results matched well, with the mean contrasts agreeing to within 5% and with generally similar convergence rates (Figure 55). The sensitivities to 1 nm RMS of tip/tilt agreed to within 40%, and the 0.1% pupil obscuration shear and 4 mrad pupil rotation sensitivities were within a factor of 2. Sensitivities to low-order aberrations above tip/tilt were not evaluated in these tests; in Milestone 9, they varied by $0\times - 2\times$ for Z4 – Z11 between model and testbed.

The testbed contrast values in these comparisons are the modulated (coherent) intensities derived from probing. The raw measurements include unmodulated (incoherent) contributors not included in the reported model results, with jitter being the most prominent for HLC, adding an estimated $\sim 10^{-9}$ in contrast.

## 7 Revised deformable mirror modeling

*Note: This section deals with models incorporating the measured properties of the flight DMs obtained late in the preparation of this document. This information was not available for the simulations presented in the prior sections, the results of which we believe reliably represent the stated coronagraphic and wavefront control behaviors. Due to limited resources, we are unable to rerun those simulations. The revised model is used in all simulations shown in later sections.*

The DMs are fundamental parts of CGI, and they are perhaps the most complex optical element in regard to the number of ways that they can impact performance. The DM model used for the results shown in the prior sections was a somewhat idealized representation of what was expected for flight. It used a mean stroke/voltage gain of 5 nm/V over all voltages that varied by 5% randomly among the actuators, with a mean maximum stroke of 500 nm at 100 V. The influence function used was the same for both DMs, based on a similar device used in the JPL testbed.

Characterizations of the two flight DMs were made in mid-2022 using an interferometer in a thermal vacuum chamber at JPL (the Vacuum Surface Gauge 2[91], VSG2). The surface of each was measured over weeks at different voltages and temperatures, allowing for the monitoring of long-term drifts. Some critical properties were measured:

1. The gain of each actuator was measured over the available voltage range, and the presence of a dead actuator on one DM (eventually assigned to DM1) and some weak, practically dead actuators at the edges of the other (DM2) were confirmed. These defects were suspected from earlier capacitance tests.

2. A large region on DM2 contains actuators that are electrically coupled to those in the row below – when one actuator is pistoned, the one below it moves in a proportionally smaller amount.

3. The mean actuator influence function of each DM was derived.

4. A large cylindrical surface deformation and a smaller amount of power (defocus) are seen on each DM, and both change once in vacuum or a dry-purge environment until it is fully desiccated.

5. Actuators continue to move over a period of days after voltage has been applied, leading to a slow change in the wavefront called *creep*.

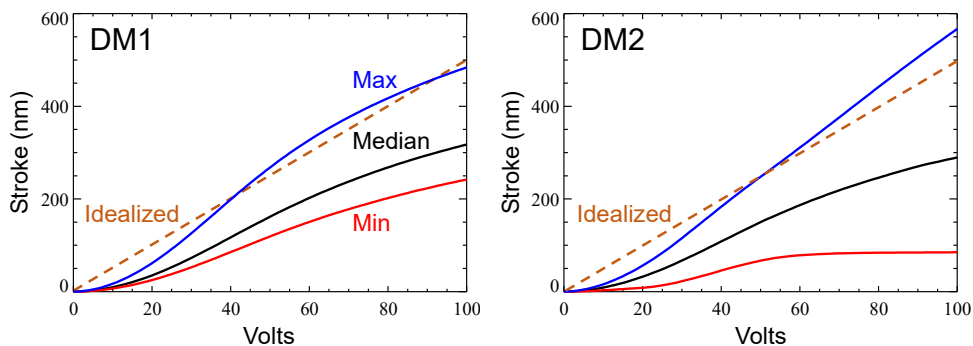

**Figure 56. Measured stroke (cumulative integral of gain) curves for the flight DMs for the weakest and strongest actuators and the median curve for all active actuators. Only a small number of actuators occur near the extrema (see Figure 57). The idealized DM curve used in prior sections is also shown.**

### 7.1.1 DM gains and dead actuators

The gain of each actuator was measured in VSG2 at 10 V increments. Differences in gains result in variations in maximum stroke (Figure 57). On DM1, excluding the dead actuator, the minimum, median, and maximum total strokes are 242, 318, and 484 nm, respectively; on DM2, ignoring the very weak actuators, they are 85, 290, and 567 nm. Linear fits to the median profiles shown in Figure 56 provide approximate gains of 3.6 (DM1) and 3.3 (DM2) nm/V. The mean dispersions in gain about the median are $\sigma = 9\%$ (DM1) and 20% (DM2); the higher DM2 value is due to a band of electrically coupled actuators (Section 7.1.2). Note that these values are for isolated actuator pistons, where the facesheet stress between poked actuator and adjacent unpoked actuators constrains the surface deformations – when a group of adjacent actuators is moved in unison, the mean surface displacement can be larger, up to 1.5× for DM1 and 1.7× for DM2. In the revised DM model the voltage is converted to stroke using the gain curve for each actuator. It is then sent to the PROPER DM model, which uses the measured influence function for each device. When a uniform voltage is applied, the variations in gain reproduce the patterns observed by VSG (Figure 62).

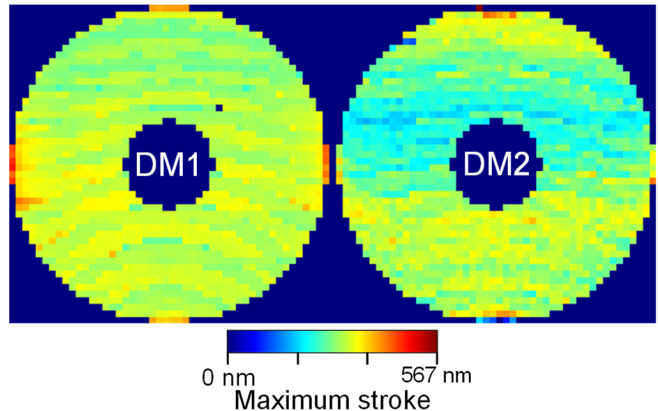

**Figure 57. Maps of the maximum stroke per actuator for the flight DMs based on pistons of single actuators in isolation. Note the dead pixel towards the upper right from center on DM1 and the weak/nearly-dead actuators along the top and bottom rows and in the upper left on DM2.**

The dead actuators are due to poor or failed electrical connections. The one on DM1 is electrically shorted to the common return line due to a metallization error. This defect was determined by the project to be acceptable and, owing to schedule constraints, a repair was not attempted. It was therefore necessary to determine if a suitable dark hole could be dug with them as they are, and since the flight DMs cannot be evaluated in a testbed, modeling was the only way to do this. A related issue was choosing where the DM with the single dead actuator should go (DM1 or DM2). Because the dead actuators were confirmed after electrical fitment, their orientations are fixed relative to the pupil and masks (Figure 58) – they cannot be rotated to put those bad actuators in more favorable locations (e.g., behind the shaped pupil mask). With the fixed orientations and the symmetries in the masks and pupil, the actuators end up in the same locations relative to the obscurations whether they are on DM1 or DM2.

The post-EFC model results showed that for the HLC there is no significant difference regarding where each DM is assigned. This is not surprising given that the dead actuator falls almost entirely within the shadow of the Lyot stop and has much less impact than it would within the clear region (Figure 46). The choice is also not significant for SPC-Spec, despite only a small portion of the actuator being blocked by the pupil mask. This might be due to its small field of view, which reduces its sensitivity to higher frequency errors. SPC-WFOV, however, is highly impacted by the assignment, with a dark hole more than an order of magnitude worse when the dead actuator is on DM2 instead of DM1. It is unclear why this is so. This highlights the need in the future to evaluate the sensitivity of coronagraph designs to dead actuators, especially those using patterned pupils for apodization. In the end, the DM with the dead actuator was placed at DM1. The nearly dead actuators on DM2 are not problematic as they are at the pupil's edge and are largely blocked by the SPC pupil masks and the HLC Lyot stop.

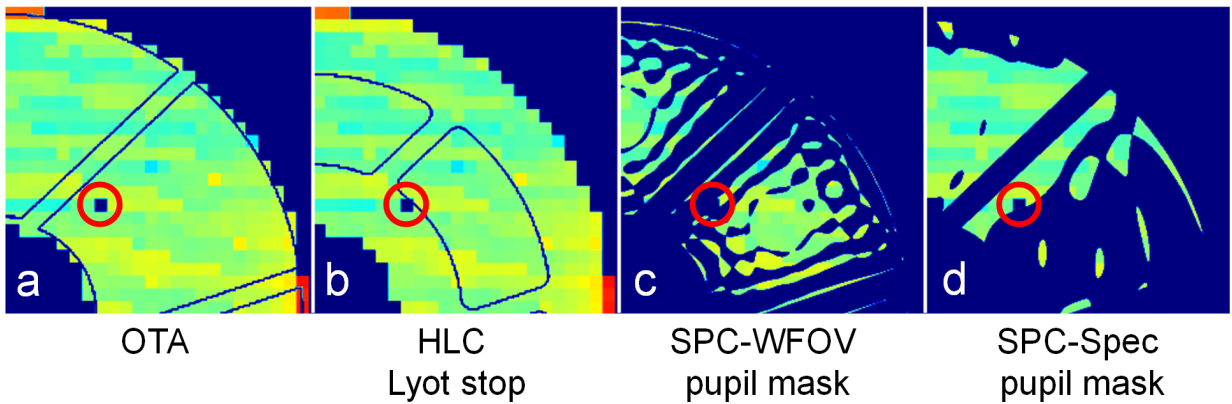

**Figure 58. Maps of maximum actuator stroke over one-quarter of DM1 with obscuration patterns superposed and the dead actuator circled. (a) The OTA obscurations outlined; (b) the HLC Lyot stop openings outlined; (c) the SPC-WFOV pupil mask; (d) the SPC-Spec pupil mask.**

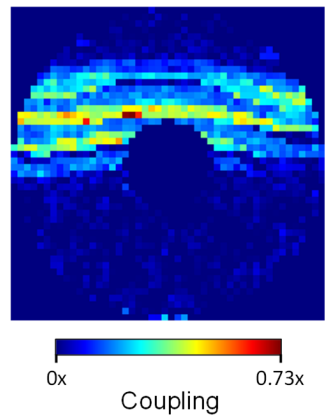

**Figure 59. Map of flight DM2 coupling showing how much each actuator is electrically coupled to the one above it.**

### 7.1.2 Actuator coupling

Almost half of DM2 actuators are electrically coupled in one direction. An actuator's piston will be the sum of its commanded stroke and a fraction of the commanded stroke of the actuator above it, depending on where it is (Figure 59). DM1 does not have any significant coupling. In the model, the effect is reproduced by shifting the map of the commanded strokes down one row, multiplying by the VSG2-measured coupling factor, and adding the result to the unshifted stroke map.

The actuators on DM2 that are used to compensate for the dead one on DM1 are within the zone of high coupling. Experiments show that not including coupling in the control model for HLC degrades the contrast by $\sim 5 \times 10^{-10}$ for the same number of EFC iterations as when including it. For the SPC-WFOV, however, not including it causes a nearly 10× degradation for the same regularization schedule, likely due to the increased sensitivity of that mode to higher-order errors in the control model. Therefore, the CGI flight control model includes coupling in all modes to address this issue.

### 7.1.3 Actuator influence functions

The influence function of each actuator was measured in VSG2 and averaged into a single representative function for each DM that is used by the PROPER DM model (Figure 60). The width of an influence function is dependent on the facesheet thickness – a thicker facesheet, like that on DM2, results in a broader function. As discussed in Section 6.3.2, deviations between the actual and assumed influence functions can lead to limitations in the achievable dark hole contrast, especially for larger field-of-view coronagraphs like the SPC-WFOV.

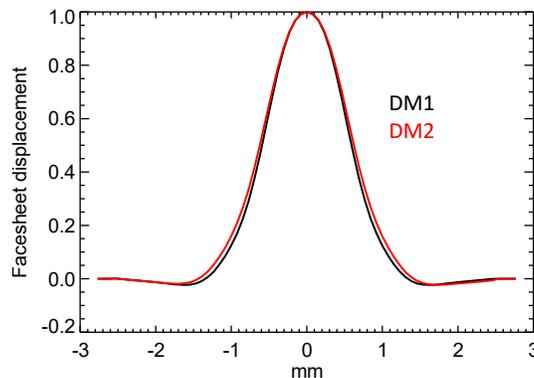

**Figure 60. Cross-sections of the derived actuator influence functions of the flight DMs. DM2, with a thicker facesheet, has a slightly broader function than DM1. The actuator separation is 0.991 mm.**

### 7.1.4 Cylindrical surface deformations in the flight DMs

The low-order patterns seen on the flight DMs (Figure 61) can be decomposed into defocus (Z4) and astigmatism (Z6) terms, and their cause has been traced to the electrical connections. Pins are soldered to conductive pads on the back of the DM blocks, and the array is reinforced with a layer of epoxy. The ribbon cables from the driver electronics are connected to these via sockets embedded in rectangular header blocks, also made of epoxy. Each header connects to 50 pins across by 2 pins high. Structural modeling after discovery of the patterns showed that when the DMs are exposed to a vacuum or dry atmosphere, the epoxy shrinks as it desorbs water over a span of ~40 days. The reinforcement layer shrinks in all directions and introduces defocus. Because the headers are aligned horizontally, they pull the sides of the DM inward, creating a convex cylindrical deformation. Similar DMs that have been used in the testbeds do not have this surface change due to desorption, but their electrical connection method was deemed not suitable for flight given the vibrations expected during launch. The flight connector approach successfully went through random vibration and thermal cycling tests, but for schedule reasons its performance was not fully characterized using a complete DM in vacuum prior to the flight units.

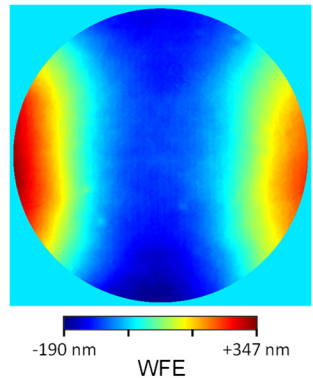

**Figure 61. VSG2-measured wavefront error of the illuminated region of flight DM1 in an unpowered state after drift had converged (>50 days). Surface error is -½ WFE, so the surface deformation is convex.**

The VSG2 measurements show that the pattern is sensitive to the DM bias voltage, and so it has been characterized as bias-dependent Z4 and Z6 values. After dry-out, the corresponding WFE aberration coefficients (-2× the surface deformation) at 0 V are Z4, Z6 = 47, 104 nm RMS for DM1 and 83, 97 nm RMS for DM2 (measured over a 46.3 mm diameter area). Each aberration increases by ~23 nm RMS from 0 to 40 V. That equates to a combined P-V WFE of Z6 = 1210 nm just from the DMs (desiccated with 40 V bias). The revised DM model adds the Z4 and Z6 aberrations interpolated from the measured values for a specified bias. Together with the measured gains and mid-spatial-frequency surface error maps, the model reproduces the VSG measurements well (Figure 62).

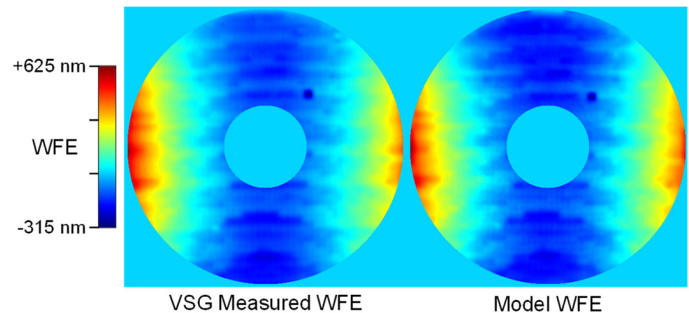

**Figure 62. WFE (-2× surface error) of the illuminated region of flight DM1 with 40V applied to all actuators (left) as measured using the VSG2 and (right) simulated using the revised model. The dark spot in the upper right is due to the dead actuator. The wavefront is dominated by the cylindrical deformation of the DM surface. The horizontal banding is due to adjacent actuators having similar gains.**

### 7.1.5  Other surface errors

Two other VSG2-measured surface errors are included in the DM model (Figure 63). First is the 0V surface error with Z1 – Z6 fitted and removed, which reveals the polishing errors having an RMS SFE of 6.6 nm. Second is the surface "wrinkle" that occurs due to inter-actuator stresses, so it cannot be removed by the DM actuators themselves. It is a dominant residual aberration after the wavefront has been flattened in testbed experiments (see Figure 4a of Ruane et al.[92]). This increases with voltage (1.2 nm RMS SFE at 40V), so it is applied relative to the specified bias. The impact of these is generally inconsequential on the achieved contrast, but they are included for completeness.

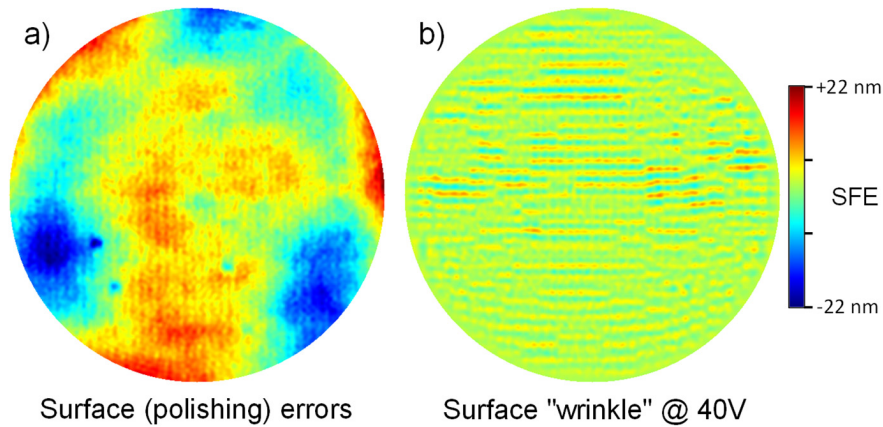

Figure 63. (a) Surface error at 0V with Z1-Z6 removed.; (b) The surface "wrinkle" present at 40V bias.

### 7.1.6  Performance with as-measured flight DMs: can a good dark hole still be dug?

The large surface deformations, gain variations, dead/weak actuators, stroke limitations, finite stroke (DAC) resolution, and neighbor rules present in the flight DMs comprise a significant set of deviations from ideal that may prevent achieving a good dark hole. The Z4 component can be corrected with the FCM, but that still leaves >1 μm P-V WFE of Z6. The project had to determine whether the DMs could correct this much error by themselves, in addition to system errors and the HLC patterns, or if some other compensation must be applied. Again, without a testbed, this had to be evaluated with modeling.

Using the updated flight DM model, simulations were run for each of the three coronagraphs. As there must be sufficient positive and negative stroke relative to the mean level to accommodate stroke, a bias of 40 V – 50 V is required. However, the higher the bias, the greater the dead actuator will stand out, so the lowest possible bias is favored. The simulations began with flattening out Z6 with the DMs and compensating Z4 by moving the FCM. The flat-system-WFE solution and the HLC DM pattern (if in HLC mode) were added to the DMs, after which EFC was used to dig the dark hole.

Under these conditions, the HLC solutions have practically the same contrast ($3 \times 10^{-9}$) at 40 or 50 V, despite having nearly 40 actuators on each DM that hit the stroke limits at 40 V and only 10 at 50 V. The SPC-Spec mode appears insensitive to the bias, but the SPC-WFOV definitely is. At 40 V its mean dark hole contrast is $4 \times 10^{-9}$, but at 50 V it is $2 \times 10^{-8}$. So, as long as the bias is wisely chosen and there are no significant sources of WFE beyond what is already assumed in the models, the DMs can correct themselves and the system errors without requiring other means of compensation, but it leaves essentially no stroke margin.

### 7.1.7 Z6 compensation

Even though the simulations suggest that the DMs can correct their deformations themselves for the assumed total system WFE, there is little stroke headroom to tolerate any additional, unforeseen aberrations. There is clearly a desire to regain margin by correcting the deformations by other means. Z6 could be reduced by (1) physically altering the headers, (2) altering the alignments of CGI OAPs 1-4 to generate an opposite amount of aberration, or (3) changing the prescription of one of the other optics in CGI (specifically, the fold mirror between OAPs 3 and 4). The first solution is risky and could damage the DMs, of which there are only three available candidates and two are needed. The second is relatively safe, can be largely optimized during integration, and can also compensate Z4, though it complicates alignment. The third requires a commitment to the corrective prescription and cannot be modified after integration (except for replacing it with the original flat optic).

The CGI project chose to interactively adjust the OAP alignments during integration in February 2023 while monitoring the system aberrations with an interferometer. Since this occurred with unpowered DMs in air with a dry nitrogen purge, they were not fully desiccated and did not have the full Z4 and Z6 aberrations that would be seen in vacuum with a bias. The ambient temperature was 19.5° C, compared to the flight setting of ~25° C. These differences result in a partial compensation relative to the expected on-orbit state. After alignment, the measured system (FSM through OAP8) Z6 WFE was -6 nm RMS. The estimated on-orbit value from both DMs, after correction for bias (40V), temperature, and desorption, is Z6 = 66 nm RMS (323 nm P-V) WFE. Simulations with this lower aberration have no actuators that exceed the stroke limits (except the dead/weak ones), providing an improved contrast of $1 \times 10^{-9}$ with the HLC. Thus, the DM Z4 and Z6 changes after dryout have been addressed at the optical bench level through OAP realignment, maintaining CGI contrast performance and preserving stroke margin.

### 7.1.8 Actuator "creep"

A CGI DM actuator consists of layers of electrodes sandwiched in a ceramic substrate. When the applied voltage changes, the ceramic expands or contracts, but it does not achieve its final dimensions instantaneously (Figure 64). It continues to morph slowly over tens of days as it relaxes, an effect termed *creep*. Measurements in the VSG2 of the flight DMs suggest that the actuators do not come to a final rest until tens of days after a voltage is applied. After 16 days, the actuators have moved by 10% of their initial offset. The effect is also historically additive – if the voltages are again changed, the induced creep progresses in addition to the prior trend.

There are a number of times when the CGI DM voltages are adjusted by significant amounts: flattening the DMs, flattening the system WFE, adding the initial HLC DM pattern, creating the dark hole solution for each mode, and switching coronagraph modes. This will also happen every time CGI or the DM electronics are power cycled (e.g., due to a fault protection response). Each of these will initiate creep that will cause WFE instability over days, up to tens of nm for the largest voltage changes. The significance of creep on performance is dependent on the overall deviations between  actuators, which is both small compared to Figure 64 and follows a well-known linear vs. *log*(time) characteristic of the ceramic. For a future dedicated coronagraphic flagship mission, where modes may change often, this may be a concern. However, in the case of CGI, given its contrast requirements and observation cadence, there is sufficient time for the creep to converge with proper operational planning.

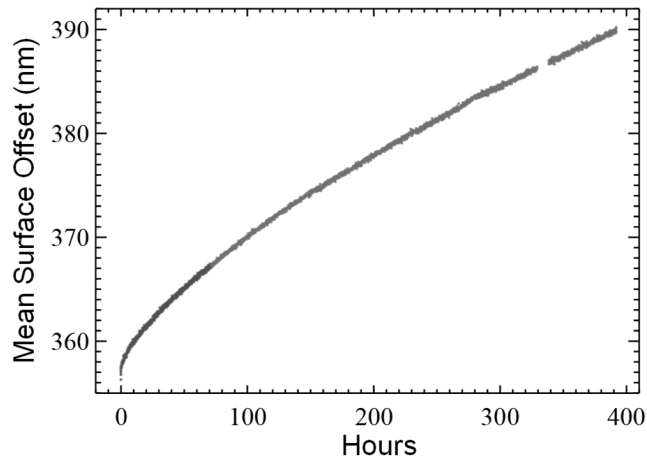

**Figure 64. Plot of "creep" in DM1 measured with the VSG. Starting with 0 V and zero surface offset on a dried-out DM1 (73 days after establishing a vacuum), a uniform voltage bias was applied on all actuators at t = 0 h. The mean surface offset was then measured over time. This represents the most drastic change possible – most changes to the DM will be much smaller, with proportionally lower creep.**

Creep can be partially compensated in two ways. The first is to use the measured creep decay function and adjust the commanded values so that the desired strokes will be reached after convergence. This still requires waiting weeks after a large change in the DM patterns. The second method is to iterate to the desired solution by putting on an amplified pattern followed by a reduced one, repeatedly alternating the cycle while decreasing the amplification. The goal is to drive opposite creep trends that cancel out each other. This technique was not tested on the flight DMs, but it has been on similar devices in the testbeds, achieving a factor of 2 to 3 reduction in creep. Neither of these options is currently baselined for CGI flight operation.

We also assessed the impact of DM creep during the integration and testing of CGI. The allocated time in the thermal vacuum chamber does not allow for 40 days to pass until creep stops. This scenario has been evaluated using modeling[93] incorporating the measured creep function and without using any of the compensation methods described above. It begins with flattening the wavefront of the combined telescope optical simulator and CGI system, which sets creep in motion. One day later the HLC DM pattern is added, introducing another creep trend. After an additional 3 days, EFC is run to generate the dark hole. After EFC, the simulations show that the dark hole mean contrast degrades by ~7×10⁻⁹ after 10 hours; waiting 13 days instead of 3 before EFC reduces the contrast change to ~4×10⁻⁹. Thus, the integration and test schedule allows verification of CGI requirements in the presence of creep.

# 8  Stability and Structural, Thermal, & Optical Performance (STOP) modeling

The modeling described so far has been concerned with optimizing the static contrast between the dark hole speckles and any potential field sources (exoplanets). However, after a deep hole is dug, an exoplanet or circumstellar disk may still be fainter than the remaining background, making identification and accurate measurements difficult or impossible. Post-processing algorithms[52] are used to subtract the speckles from the images based on the assumption that they remain largely stable over time.

Since speckles are instrumental and unrelated to the sky, the image of another, isolated star can be subtracted in the computer (the *reference differential imaging technique*, RDI, also known as *classical subtraction*). This, however, introduces potential speckle mismatches due to differences in stellar spectra and diameters. An alternative is to use images of the same star observed at different roll orientations of the telescope (*angular differential imaging*, or ADI). In this case, the astronomical sources will appear to rotate about the star as seen on the detector while the speckle pattern appears fixed. This avoids any spectral mismatches but is not optimal for an extended object such as a disk, portions of which may overlap at the different orientations. Sources near the IWA may also self-subtract if the roll angles are small. Both RDI and ADI have been used successfully for high-contrast imaging on ground[52] and space telescopes[94,95], including more advanced implementations of them that essentially optimize the reference image via mode selection.

The use of post-processing establishes two primary limiting noises: shot noise and subtraction residuals. There is an upper allowable limit on the speckle intensity such that shot noise does not dominate over the exoplanet signal, even with completely static speckles that would allow perfect subtraction. This means digging as high a contrast dark hole as possible. Changes in speckle morphology due to wavefront instabilities will lead to subtraction residuals that create a background filled with remnants that may be mistaken for a planet or disk.

The major sources of wavefront disturbances relevant to CGI are thermally induced structural changes and pointing errors. These are introduced from both inside of CGI (e.g., mechanism motions) and externally by the telescope and spacecraft (e.g., solar-illumination-dependent structural deformations, reaction wheel vibrations). The observatory has interface requirements that specify alignment and wavefront error stabilities of the beam being delivered to the coronagraph. CGI itself has internal stability requirements extending down to the component level; most of these flow from a speckle stability goal of ~$5 \times 10^{-9}$ contrast in HLC Band 1 for a $V_{mag} \leq 5$ star.

## 8.1  Disturbances

Stationed at L2, *Roman* avoids significant thermal emission from the Earth, leaving the Sun and internal components as the primary heat sources. The slews used for RDI and the rolls for ADI alter the received solar illumination, causing thermal imbalances and structural deformations. *Roman* lacks a large sunshield like that used for the *James Webb Space Telescope*[96]. Instead, the solar panels serve that function, and the barrel is insulated, though not as elaborately as proposed for future dedicated missions. A variety of heaters are located on the telescope and instrument carrier to maintain thermal stability. The primary mirror, for example sits in a thermally controlled tub.

CGI itself is thermally well-isolated from the telescope and the outer environment with multi-layer insulation and heaters controlling the temperature of the bench and optical components, including the DMs. Motions of the PAMs produce the largest thermal disturbances inside CGI.

*Roman* uses six reaction wheels, varying their speeds to orient the observatory and maintain pointing against solar torque. They introduce vibrations that result in rapidly oscillating wavefront variations (jitter), most significantly in the form of pointing (wavefront tip/tilt) errors. The observatory has an allowable drift requirement of ≤10 mas per hour with ≤12 mas RMS per-axis pointing jitter (fine guidance is provided by WFI). Because CGI requires more stringent positioning of the star on the FPM, the LOCAM and FSM are used to correct pointing at a 1 kHz rate. The post-FSM pointing requirement is <0.57 mas RMS jitter for at least 70% of the observing time.

Besides pointing jitter, the vibrations induce low-order WFE jitter that is not measured by LOCAM, which has a correction rate of 0.1 Hz for Z4 – Z11. Modeling has shown that these oscillations change the dark hole contrast by <$10^{-11}$ and are thus not important to CGI performance. So, from hereon only pointing jitter is evaluated.

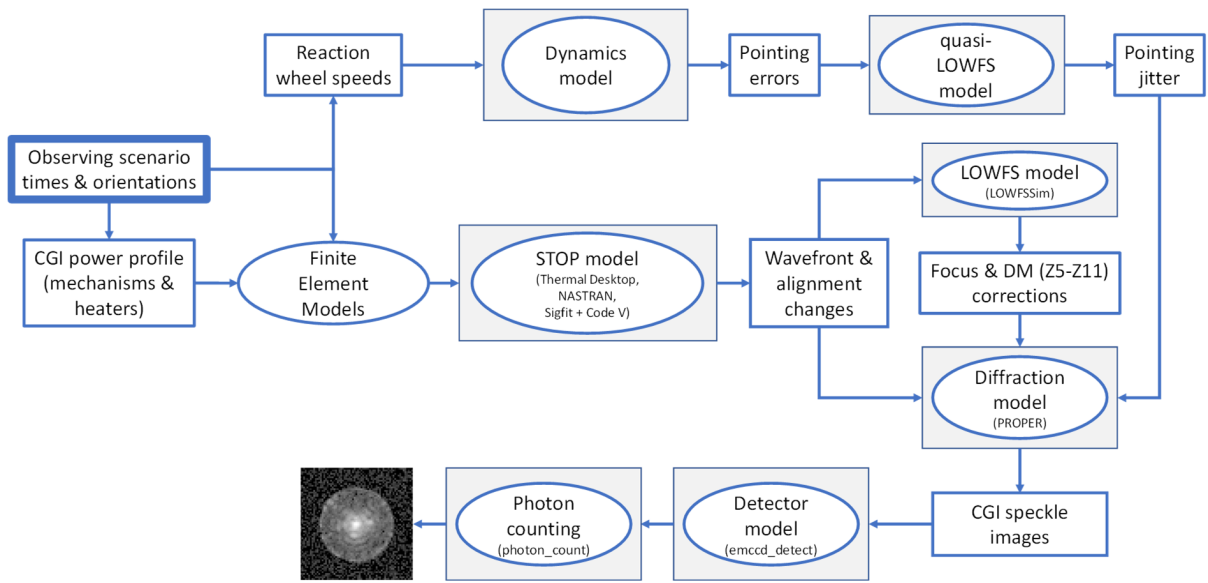

**Figure 65. Flowchart of the steps involved in generating the time series of CGI dark hole speckle images for an observing scenario.**

### 8.2 STOP modeling

Because post-processing is an integral component of the potential CGI scientific programs, performance estimates depend critically on representing the time-dependent changes in the system. One of CGI's most significant technological contributions to future coronagraphic missions is the development and exercise of highly detailed STOP modeling (Structural, Thermal, and Optical Performance) with realistic observing scenarios (Figure 65). The CGI STOP analysis is a combination of models: 1) a finite element model (FEM) is used to define the thermal states of the spacecraft, observatory, and instrument; 2) separate FEMs define the material properties of the structures to compute deformations caused by the thermal and vibrational perturbations; and 3) an optical ray trace model is used to convert the structural deformations into optical displacements and the resulting wavefront changes. These results are inputs to the LOCAM and diffraction models to create a time series of images that can be used to test post-processing algorithms and predict the scientific performance.

The *Roman* project, managed by the NASA Goddard Space Flight Center (GSFC), established modeling requirements with specified commercial modeling software packages. This ensures that the same validated, industry-standard methods are used by both GSFC and JPL and operate on the same inputs. The STOP FEMs for the *Roman* spacecraft, observatory, and WFI are defined by the Integrated Modeling team at GSFC in collaboration with the telescope contractor, L3Harris. The CGI STOP FEMs are generated at JPL and integrated into the full system model by GSFC. JPL is also responsible for defining the CGI observing scenario (OS), including schedules of spacecraft slews and rolls and operations of electronics and mechanisms. Upon receiving the fully integrated model, JPL runs it through its STOP modeling pipeline, IMPipeline[97]. In parallel, GSFC generates the OS-appropriate reaction wheel speeds, from which JPL computes post-FSM pointing jitter.

### 8.2.1 Observing scenarios

The OS is the primary input to the STOP models. It defines where and when in the sky the spacecraft is pointing, which in turn describe the solar incidence angles, slews and rolls (wheel speeds), and mechanism motions (e.g., filter changes). A CGI OS is designed to evaluate the system's stability for both RDI and ADI observations, so it includes both reference star imaging and rolls on the science target. It is intended to be generally representative, not prescriptive, of the sequences and exposures that would be used for a single-band exoplanet imaging program, essentially a "day in the life" of the coronagraph. Actual observing programs may differ in total length, exposure settings, and staging (a spectrographic program would be many times longer to account for the lower per-pixel flux due to the dispersed light on the detector).

The typical reference star will be bright (V = -0.1 to 3), and the science target, chosen from a list of known exoplanet hosts, will be fainter (V ≥ 5). In all the CGI scenarios evaluated, the science target has been 47 UMa, a G1V star (V = 5.03) around which multiple extrasolar planets have been detected via radial velocity measurements. Earlier scenarios (OS8 and before) used η UMa (B3V, V = 1.86) as the reference, while later ones used ζ Pup (O4I, V = 2.25); the change will be explained later.

The fixed solar panel on *Roman* limits the available field of view at any given time to a 72°-wide annulus on the sky perpendicular to the spacecraft-Sun line (the rotation around this line is unconstrained). There is a finite window of time during the year that a given pair of reference and target stars can be observed. This defines the solar incidence angles over the span of the OS.

The OS begins with the telescope pointing at some location within the High Latitude Survey, a region of the sky where WFI will spend about one-third of the *Roman* mission's time searching for signs of cosmic acceleration. A total of 120 hours is spent here, allowing the spacecraft (and thermal model) to reach thermal equilibrium. The telescope is then slewed to the reference star (a slew and star acquisition take about 30 minutes), and 50 hours of HOWFS is conducted to restore the dark hole to a nominal state. It is assumed that a dark hole solution was derived in an earlier calibration program, and only a small number of iterations are required at the beginning of an OS to recover from any system changes that may have occurred since the last coronagraphic observations. This stage is not included in the diffraction modeling – it is asserted that the observations begin with a good dark hole solution.

Next begins a repeated cycle of imaging of the reference and target stars (Figure 66). First, 45 min of reference star images are taken, and then the telescope is slewed to the target star. The target is imaged at back-and-forth angles (+13°, -13°, +13°, and -13°) for 105 min at each roll; limiting the amount of time spent at one orientation reduces the amount of drift caused by rolling the telescope. It takes 15 min to roll and reacquire the star. Finally, the reference star is observed again for 45 min. Adjacent reference and target observations are taken at the same roll orientations to minimize disturbances (the roll angle otherwise does not matter for the reference star). This cycle can be repeated as many times as needed to reach the desired total exposure time, with two cycles and one downlink/uplink span fitting within a 24 hour period. The last set of reference observations from the prior cycle are used as the first set in the next cycle, when possible.

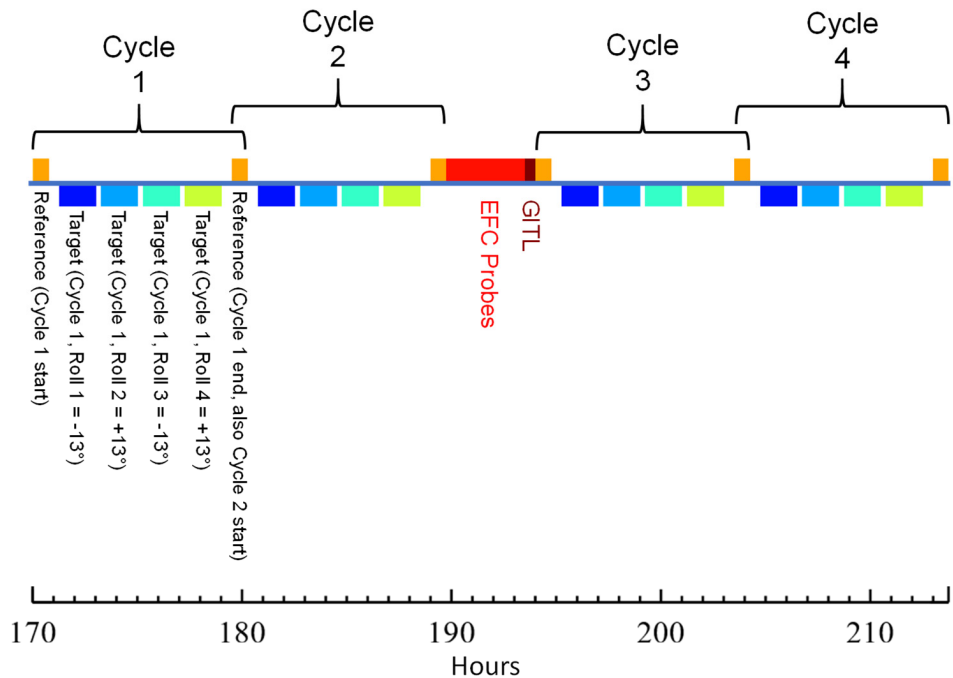

**Figure 66. Observing Scenario 11 timeline. Roll angles are offsets relative to the default (solar normal) angle.**

The observations in one cycle are self-contained and independent of the others (excluding shared spans of reference star observations). This allows for less strenuous stability requirements than would exist for a single long chain of observations. In the simplest case of post-processing, the mean of the reference star images within a cycle forms an estimate of the target star's dark hole over the ~8 hours of imaging. The bookending of the cycle with reference star images establishes a wavefront stability requirement over those 10 hours. If the entire OS were instead composed of a single, long, serial sequence of imaging (e.g., reference, then the target at +13°, then the target at -13°), then the stability requirement would have to extend over the entire span, potentially multiple days, and would be much more difficult to guarantee.

The speckle morphology must remain extremely stable within a cycle but can vary by a greater amount among cycles, if the mean speckle brightness remains below a limit defined by shot noise (though the more stable it is across cycles, the better). A slow, uncorrected drift in wavefront error may lead to exceeding this, so periodic WFC may be required. In earlier scenarios (pre-OS11), time was allocated between each cycle to execute one iteration of probing and EFC on the reference star to restore the dark hole, if needed. This assumed that WFC calculations would occur autonomously onboard CGI. However, the CGI project descoped this capability in 2020, resulting in a savings in electronics, mass, and power. Instead, probe images will be transmitted to the ground, where all WFC calculations will occur, and then DM settings will be uploaded ("ground-in-the-loop", GITL[98]). During presumed science operations, there will be only one 30-minute period every 24 hours for download/WFC/upload, allowing for just a single iteration of EFC for wavefront maintenance. After 2 cycles in OS11, there is an allocation for 3.5 hours of probing on the reference star followed by 0.5 hour for GITL WFC.

Over the past years a variety of scenarios have been developed, the latest being OS11. Some were abandoned in favor of a new OS before a STOP model could be run. Only a few realistic scenarios (OS6[99], OS9, OS11) were run through the full end-to-end simulation, from STOP modeling to speckle field time series.

### 8.2.2 Thermal modeling

The first STOP stage is thermal modeling. An AutoCAD® drawing defines the FEM, including material properties and heater control loops. The schedules of solar incidence angles and internal CGI heat sources (PAM movements, computer operations) are integrated into the FEM, which is then run through Thermal Desktop® (TD). The temperature of each node in OS11 is reported at 15 min intervals over the OS, though the thermal solver actually computes at 1 min timesteps (longer timesteps were initially tried, but they led to unstable results). This high cadence, large number of nodes (~100,000), and large number of rays (35,000 per node), result in long execution times, taking 13 days to compute the OS11 thermal model.

The OS11 thermal modeling results show that the primary mirror, sitting at the bottom of the telescope tube in its thermally controlled tub, remains very stable (Figure 67a). In comparison, the secondary mirror supports are quite sensitive to changes in solar illumination of the observatory (Figure 67b), and the corresponding changes in the OTA alignment end up driving most of the low-order aberration variations (the secondary mirror itself is thermally stable to better than 1 mK). This is not surprising given that the thin struts are over 2 m long. There are multiple heaters on the major components of the observatory, including the secondary support struts, support bipods, and IC. Within CGI the CFAM motor is the largest source of variable heat production (Figure 67c). It is used over OS11 to switch between filters for target acquisition, broadband imaging, and HOWFS probe imaging. The temperature changes in nearby components, such as the FSAM (Figure 67d) and the top of the optics bench (Figure 67e), are dominated by the heat produced while switching filters. The temperatures of components, such as the OAPs (Figure 67f), that are more distant from the CFAM or close to the CGI enclosure are correlated more with the solar illumination changes. CGI has heaters and temperature sensors on the optical bench, DMs, and camera; there are just sensors on the FSM, FCM, and alignment mechanisms.

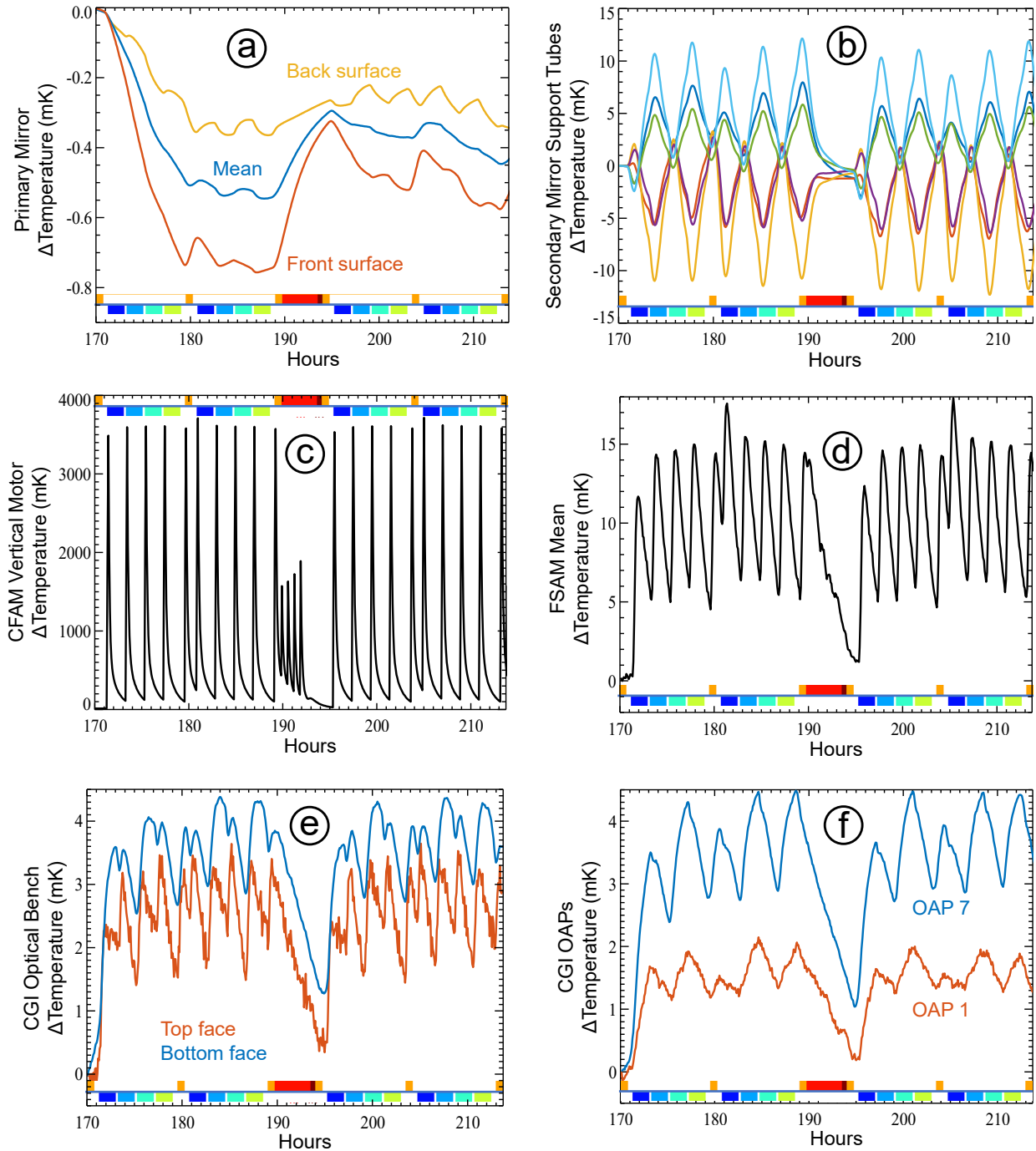

**Figure 67.** OS11 temperature changes relative to $t = 170$ h predicted by thermal modeling for (a) the front and back primary mirror surfaces, and the mean primary temperature; (b) the six secondary mirror support tubes (each a different color); (c) the CFAM vertical displacement motor; (d) the FSAM mean; (e) the means for the top and bottom faces of the CGI optical bench; (f) OAPs 1 and 7 means.

### 8.2.3 Structural modeling with thermal disturbances

A separate, multi-million-node structural FEM defines nearly every component in the telescope, instrument, and spacecraft. Material properties such as density, elastic modulus, and coefficient of thermal expansion are included. While the same geometry is used to construct the thermal and structural models, their nodal count and exact locations differ between the two. As such, a linear interpolation maps the outputs of the thermal model to the nodes of the structural FEM. This step takes approximately 3 days to run for OS11 due to the size of models, and the necessary checks that are needed to verify the mapping accuracy.

### 8.2.4 Optical ray tracing

The output of the structural model, as implemented in Nastran, is a list of node positions versus time. SigFit® is used to convert the nodal displacements into surface deformations (for the OTA and TCA optics) and rigid-body motions of the optics represented in a CodeV optical prescription of the system. These are then ray-traced to get wavefront changes and optical component displacements. This stage takes 9 hours to run for OS11.

This is the first stage where a modeling uncertainty factor (MUF) is applied. As the name suggests, this is used to add margin to a result when there are potential errors in the assumed characteristics or behaviors (e.g., uncertainties in material properties). The *Roman* project has adopted a 2× MUF for the structural model, which is applied by SigFit, so that the displacements are twice as large as computed (there is no thermal MUF). This has a cascading effect on the modeling stages that follow. The larger displacements mean increased wavefront error changes and optical surface displacements from the ray trace model, and those in turn result in greater speckle changes from the diffraction model.

The ray trace produces tables of WFE changes (described as Zernike coefficients) and critical surface offsets (interfaces, DMs and masks). Displacements of upstream optics will lead to offsets of the beam on the FPM, so at each timestep the pointing in this circumstance is corrected by tilting the entire observatory in the prescription. Other low-order errors are not compensated here, leaving those for the LOCAM and diffraction models.

The aberration changes in OS11 (Figure 68) are dominated by the relative displacements between the primary and secondary mirrors; their separation varies by ~2 nm over the timespan, which drives the focus variations. The excited directional aberrations (astigmatism, coma) are aligned with the sun-facing side of the observatory and are strongly correlated with the rolls. Focus (Z4) and coma (Z7, Z8) vary by ~60 pm and astigmatism (Z5, Z6) by ~40 pm. Aberrations Z9 and above vary by only a few picometers or less over the nearly two days of OS11. While optical surface deformations are computed, only wavefront error changes up to Z45 are used to generate the speckle fields. Over OS11, for instance, the primary mirror surface varies by no more than 4 pm RMS for spatial frequencies >Z45.

To match the available alignment options in the PROPER model, beam displacements are separated into offsets relative to the chief ray on individual optics due to IC-CGI shear or bench deformations (represented as shifts of their associated surface error maps or masks and the DM centers). These differences matter, especially regarding the DMs (Figure 69). The contrast sensitivity to a differential beam offset between the two DMs (caused by a bench deformation, for instance) is ~10× greater than if the beam was offset by the same amount on both DMs (i.e., due to an IC-CGI shear). Shears on the pupil masks (Lyot stops and SPC masks; Figure 70) also introduce contrast changes. The SPC pupil masks are oversized and fully block the telescope obscurations, even for the maximum OS11 offsets, so their sensitivity to shears in these circumstances is due to alignment changes relative to the DM solutions. The SPC Lyot stop shear sensitivity is miniscule because the dark zone of the pupil is larger than the stop (Figure 24), but it is high for the HLC Lyot stop because the pupil at that plane is filled with light (Figure 2). The DM and mask shears appear similar because they are dominated by IC-CGI interface shears.

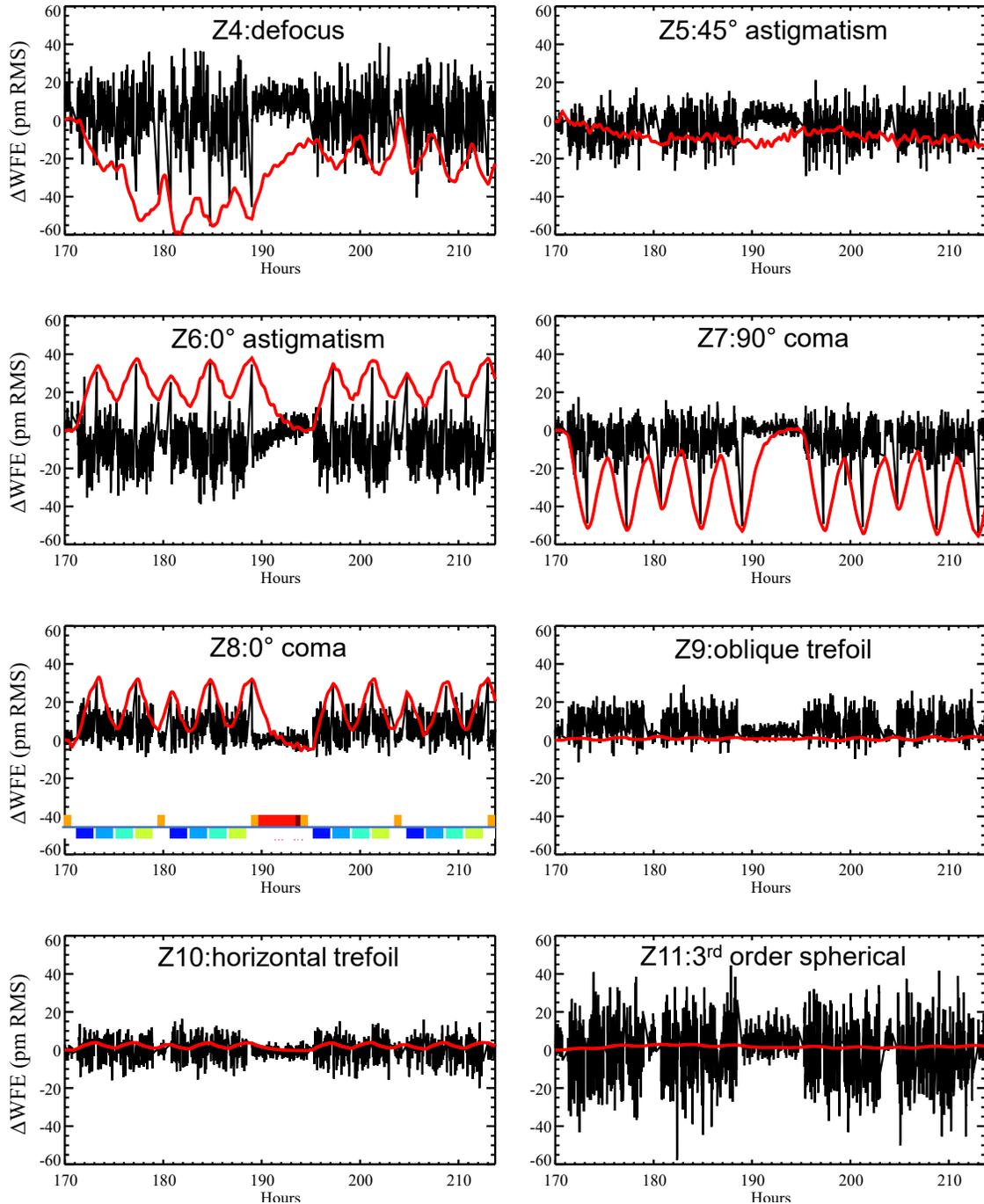

**Figure 68. STOP-model-predicted OS11 low-order wavefront changes relative to** $t$ **= 170 h at the FPM plane (red) and the corresponding LOWFS sensed-and-corrected values (black). The values are picometers RMS of aberration.**

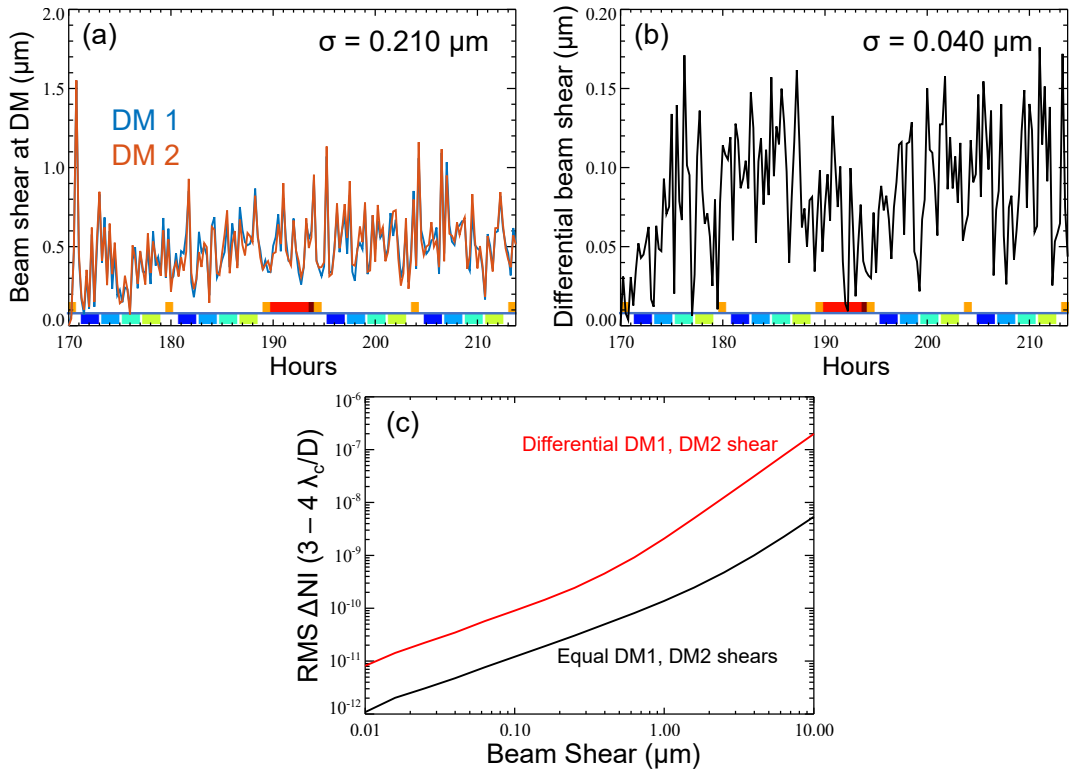

**Figure 69. (a) Predicted total displacements (due to beam offsets and surface shifts) of the chief ray relative to the DM surfaces over OS11. The variations are dominated by beam shear from upstream optics and the IC-CGI interface; (b) displacement of DM2 relative to DM1; (c) open-loop contrast sensitivities to common (black) and differential (red) beam shears on DM1 and DM2.**

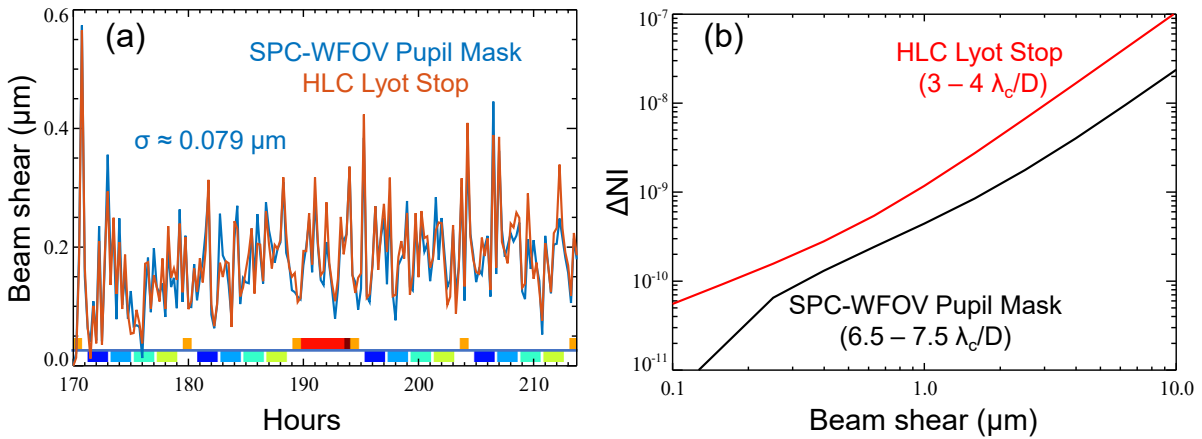

**Figure 70. (a) Predicted total displacements (due to beam offsets and mask shifts) of the chief ray relative to the SPC-WFOV pupil mask (blue) and HLC Lyot stop (orange); (b) open-loop contrast sensitivities to offsets of the beam on the SPC-WFOV pupil mask (Band 4) and the HLC Lyot stop (Band 1) in an aberrated system with EFC-derived DM solutions.**

### 8.2.5 Integrated modeling pipeline

The use of specialized packages for each step in the STOP process is common, but it typically means that one analyst runs one stage, collects the results, and sends them to another analyst to run the next stage, etc. This can lead to large delays. At JPL this sequence has been automated into an integrated modeling pipeline, IMPipeline[97]. It is based on Luigi, an open-source Python framework for complex batch processing. It provides facilities for web interfaces, process visualization, workflow management, and dependency resolution. In cases where a modeling tool only accepts inputs through a graphical user interface (e.g., Thermal Desktop), the Python pywin32 package is used to automate the interactions. The STOP chain can be run using one configuration file. The final output is a list of Zernike aberrations (up to Z45) and positions of critical components relative to the chief ray (including shear of the CGI relative to the IC, masks, DMs, etc.) versus time. These are used as inputs to the LOCAM and diffraction models.

### 8.2.6 Incomplete and CGI-only models

Over most of the development cycle, the thermal and structural FEMs used for computing OS time series results lacked a complete representation of CGI. Up through OS8 (2019), a full model of the instrument (including heater control loops) was not ready, as materials and designs were not finalized. In those cases, CGI was represented only as a rigid box attached to the IC, so only wavefront changes and alignments up to the CGI entrance were captured. In OS9 (2020), mostly complete CGI FEMs existed but were not yet integrated into the full system model. The full model was run to get the aberrations and misalignments up to the IC-CGI interface, and the CGI model was run separately to simulate internal disturbances (e.g., PAM motions), with the results combined afterwards. By OS11 (2021), the full model included the high-fidelity CGI representation. During the design process, the CGI-only thermal and structural models were repeatedly exercised using a significantly reduced node count version of the observatory (diffraction modeling was not included).

### 8.2.7 Structural modeling with dynamic disturbances

The structural disturbances caused by vibrations from the reaction wheels are computed using a non-automated chain of procedures separate from IMPipeline. The first step is creating a schedule of wheel speeds that enables the slews, rolls, and tracking needed in an OS (Figure 71a). This is created by an analyst at GSFC, who manually tunes it to keep each of the six wheels below 5 revolutions/sec during CGI observations to avoid introducing large vibrations (the WFI can tolerate larger pointing errors). These serve as the input to a dynamic structural model that includes detailed representations of the wheels.

The outputs of the dynamics model are power spectral density (PSD) profiles that represent the source image displacement magnitudes for a span of vibrational frequencies, one PSD per time step. The PSDs are multiplied by frequency-dependent modeling uncertainty factors (MUFs) of $3\times - 10\times$. These apply margins to accommodate for uncertainties in the behaviors of the wheels and the responses of the structural elements. The pointing corrections made by the LOCAM-derived commands to the FSM are then included as a frequency-dependent filter applied to the PSDs, accounting for the stellar brightness. The PSDs are then integrated over frequency to produce tables of time-integrated, Gaussian-distributed pointing errors with X and Y RMS widths as seen at the FPM, sampled at 0.25 sec intervals (Figure 71b). The total RMS jitter assumed at each timestep is the root-sum-square of these values over 1 minute and a 0.3 mas RMS additional contribution from assumed measurement errors.

In addition to pointing offset jitter, the dynamics model also produces estimates of wavefront jitter, specifically rapid Z4 – Z11 oscillations due to vibration-induced optical misalignments. The ensemble of these, however, amount to less than a few picometers at most, so they are not included in the time series simulations.

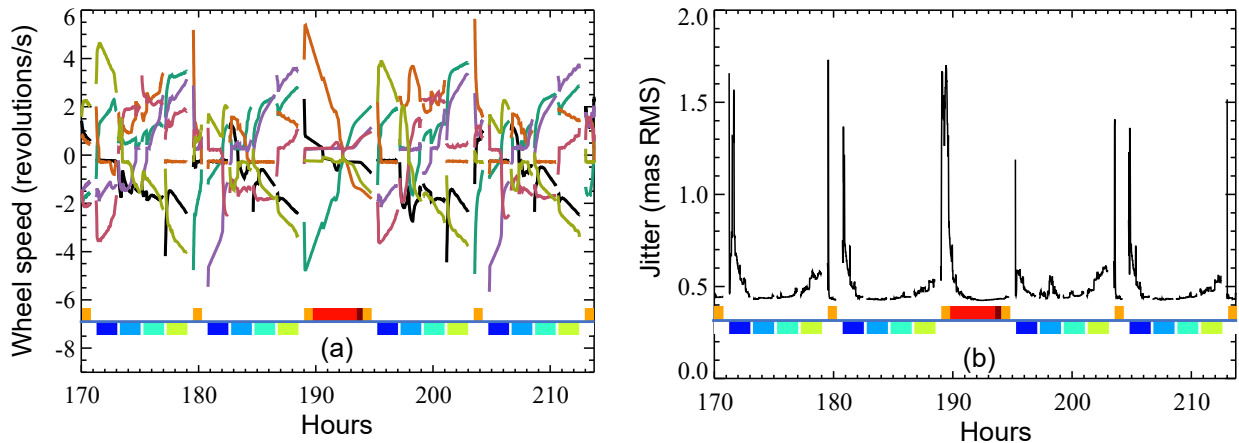

**Figure 71. (a) Rotational speeds of the six reaction wheels over the OS11 timespan (different color per wheel). (b) Predicted FSM-corrected pointing jitter for OS11 (root-sum-square of the X and Y jitters). In both plots there are gaps during slews and rolls.**

### 8.3 *The surprising case of OS8 and tailoring observations to minimize WFE changes*

As a slight detour, it is illuminating to review an earlier observing scenario and how the STOP results from it led to a change in the selection of the reference star for subsequent scenarios. Like OS11, OS8 from 2019 included multiple cycles, each beginning with imaging a reference star (1 h), followed by four back-and-forth rolls of ±12.5° on the target star (2 h each roll), and ending with more images of the reference (1 h). At that time, autonomous onboard WFC was assumed, so a 4 h refresh of the dark hole using EFC on the reference star was done after every cycle; with calculations now occurring on the ground, there is less frequent refreshing. The science target was 47 UMa, as it is for OS11, but the reference star was η UMa.

The STOP results for OS8 were startling (Figure 72). In previous scenarios, 3$^{rd}$ order spherical aberration (Z11) varied by only a few pm, and higher order sphericals (Z22, Z37) were in the numerical noise. However, in OS8, they all varied by tens of pm (Figure 72b). This was especially problematic because while the LOCAM can sense and enable corrections of Z4 – Z11, it is unable to do so for Z22 and Z37 without software changes. As Figure 5 shows, the HLC sensitivities to spherical aberrations are high, and these uncorrected wavefront changes would result in the contrast varying by nearly 10$^{-8}$ when <10$^{-9}$ is desired. Given that prior scenarios used the same stars, it was initially puzzling why the variations increased.

The large wavefront changes were correlated with the primary mirror temperature (Figure 72c). Despite being in a temperature-controlled tub, it was varying by nearly 20 mK; in comparison, the primary varies by <1 mK in OS11 (Figure 67a). The cause was an unexpected (to the CGI team) change in the telescope's aperture cover (Figure 73). In prior designs it was a solid door that also functioned as a sunshield for the entrance aperture, but it was changed to a large, deployable structure (Deployable Aperture Cover, DAC) to provide greater stray light rejection. The new configuration, as represented in the model, was not as thermally isolated as the prior designs. When the telescope was slewed between the reference and target stars, the change in solar pitch caused a thermal variation on the inner surface of the DAC that was radiatively coupled to the inner surface of the telescope barrel opposite to it. This, in turn, was radiatively coupled to the front surface of the primary mirror, which responded by curling, as measured by the spherical aberrations.

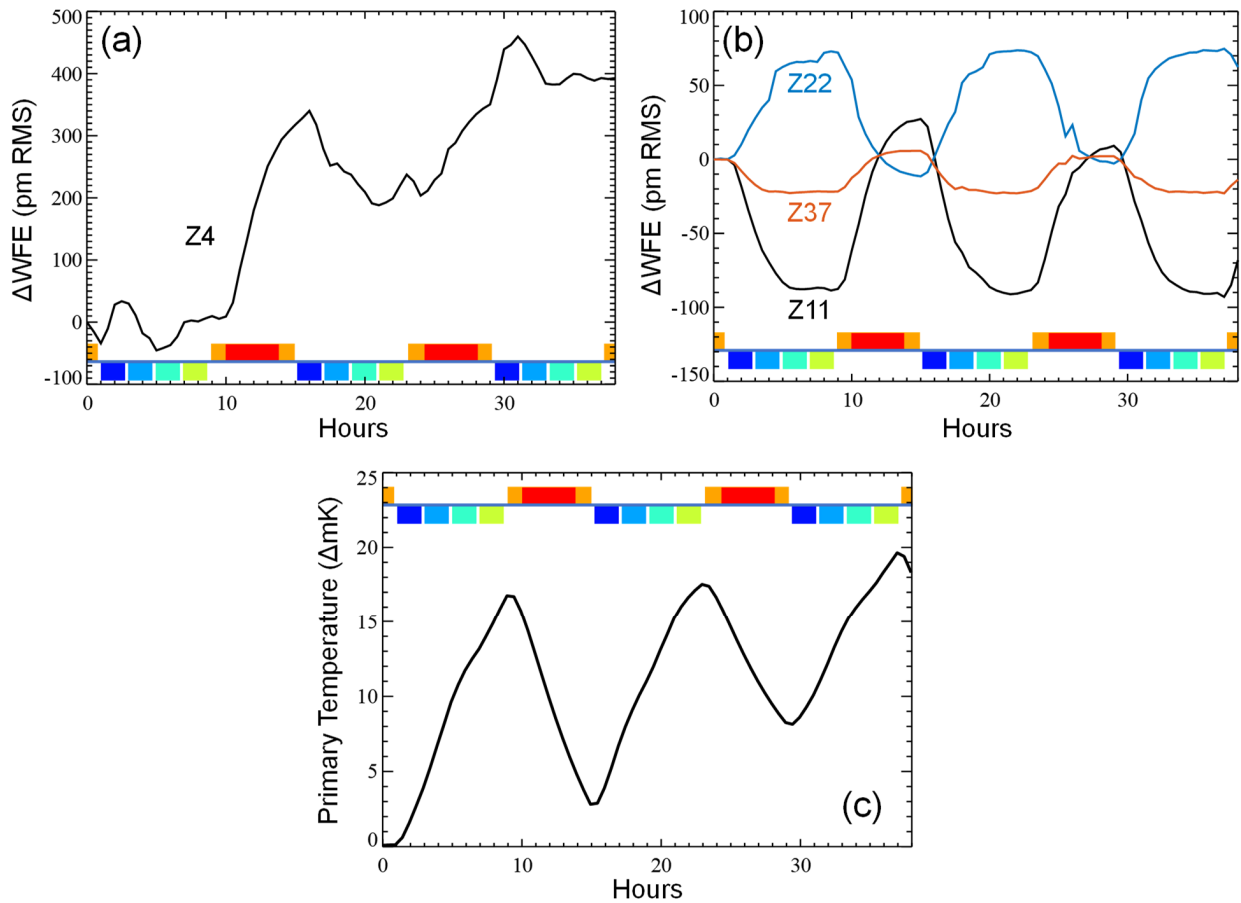

**Figure 72. STOP model results for OS8. (a) Focus (Z4) change relative to the first reference star image time. (b) 3rd (Z11), 5th (Z22), and 7th (Z37) order spherical aberration changes. (c) Temperature change in the primary mirror. The observation legends correspond to the same colors as used for OS11, except the reference star (orange span) is η UMa rather than ζ Pup.**

While the thermal hot-spotting on the DAC would later be reduced by revisions to the design and improvements in the model, the CGI team was able to avoid the large temperature changes in the next scenario, OS9, by choosing a different reference star. η UMa was originally chosen as the reference because it is nearby to 47 UMa in the sky (24° separation). However, the solar pitch difference between the two stars is ~15°, and this is large enough to drive the thermal effects seen in the STOP model. In a list of potential science targets and compatible reference stars, the median solar pitch is ~3.5°, so 15° is a rather extreme outlier. In prior scenarios this was an advantage as it demonstrated the system's high tolerance to pitch, but in the updated design it became a liability. ζ Pup was subsequently chosen as the reference star for later scenarios. Even though it is 166° from 47 UMa, the solar pitch difference is just 3.5°. This reduced the variations in the spherical aberrations by a magnitude or more in OS9 (Figure 74). This demonstrates the utility of STOP modeling for optimizing the operations of CGI and future missions.

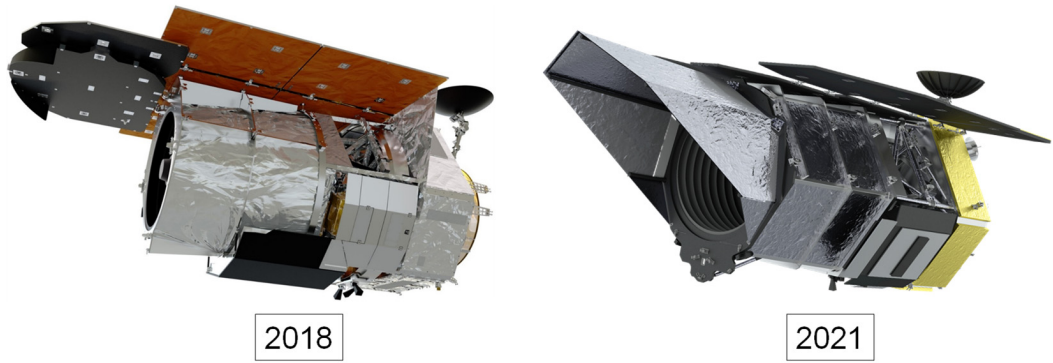

**Figure 73. Configurations of the Roman observatory in 2018 and 2021. Note that the aperture cover/sunshield changed from a solid door to a deployable, kite-like structure.**

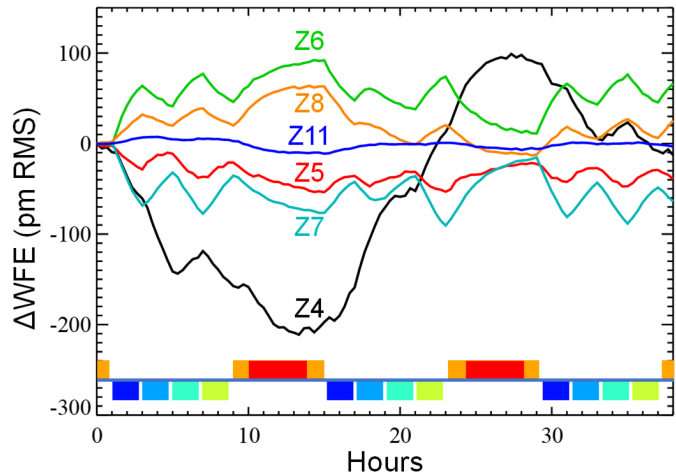

**Figure 74. STOP model results for OS9 for focus (Z4), astigmatism (Z5, Z6), coma (Z7, Z8), and 3rd order spherical (Z11) relative to the first reference star image time. Trefoil and 5th-7th order spherical aberrations vary by <15 pm and are not plotted.**

### 8.4 LOWFS simulation

The last step before image generation is simulating the measurements of the low-order aberration changes using the LOWFS. These are used in the diffraction model to set the FCM and DM1; the sensing and correction of pointing errors (wavefront tip and tilt) are not included and are instead added later as post-FSM-correction jitter. The Z4 – Z45 aberration curves from the STOP model are resampled to 10 second timesteps and fed to LOWFSSim. The code generates a broadband image at the LOCAM and realizes 10,000 EMCCD frames for that image. Each frame is converted to Zernike change estimates using a model wavefront control matrix, and the 10,000 estimates are averaged. This assumes the true wavefront error is quasi-static over 10 second intervals and avoids generating 10,000x as many broad-band LOCAM images. The changes in OS11 are sub-picometer at this scale. The sensed coefficients are fed to the baseline Z4 and Z5-Z11 control algorithms which have 0.0016Hz of temporal bandwidth and the change in the FCM and DM1 simulated. When used to simulate dark hole images, a 100 second running average of the post-correction Zernikes is used.

The differences between the OS11 measured and actual aberrations are shown in Figure 68. At first glance, the LOCAM-derived corrected values are much noisier than the original ones. This is largely a consequence of the control parameters, such as bandwidth, being tuned for the wavefront variations allowed in the stability error budget rather than the much lower predicted OS11 values. The noise is of high temporal frequency, but the values over a 10 hour cycle average to zero. It will be shown that even with this unoptimized tuning, the low-order corrections provide more stable speckle fields near the IWA. With larger variations, the advantage would be even greater.

## 8.5 Image generation

After the inputs have been assembled (aberrations, optical shifts, LOCAM-derived corrections, jitter), the diffraction model is exercised. The initial dark hole is dug, and then the system is perturbed over time.

### 8.5.1 Defining the initial optical state

The goal of the OS simulations is to provide an estimate of how the dark hole speckles will vary over time to evaluate the effectiveness of post-processing algorithms. As was demonstrated in Section 3.1.2, the sensitivity to WFE changes is dependent on the ambient dark hole intensity, so a relevant time series simulation requires an EFC-derived solution that is realistic and not overly optimistic. The diffraction model is defined with an assortment of assumed misalignments and fabrication errors, as listed in Table 6. The known errors reflect measured FPM and previous testbed errors, while the unknown are estimates of measurement errors based on testbed experience. The system WFE map used in the control model and for flattening the wavefront is the computed monochromatic phase error at the FPM exit pupil at the central bandpass wavelength. This is a stand-in for the phase-retrieved map, which was not available at the time of the time series generation. Phase retrieval measurement errors are represented by multiplying the phase maps by a factor of 1.1. The design obscuration pattern is used without any amplitude errors in the control model.

The OS11 time series images were generated prior to the significant measured compensation of the DM surface deformations during instrument integration by adjusting the OAP alignments (Section 7.1.7). Instead, it was assumed that the OAP alignments would only correct about 25% of the astigmatism, so the two DMs were set to generate 185 nm RMS of Z6 WFE, compared to the post-integration expectation of 66 nm RMS.

For the HLC, two versions of the image time series are generated, with "default" and "conservative" cases. The PROPER models used for both are identical except the conservative one has 1.5× increased polarization-dependent aberrations, representing uncertainty in the ray-trace-predicted telescope polarization WFE, and ~2× increased contrast sensitivities to those aberrations (obtained by multiplying them by an additional $\sqrt{2}$). Separate EFC solutions are derived for each case and are used as the initial state of the system at the first timestep. The contrast of the final EFC iteration in the conservative case is not always significantly worse than for the optimistic one, so the DM solution from an intermediate iteration is used that provides ~2× worse mean contrast.

**Table 6. System static misalignments and fabrication errors included in the optical model for OS 11.**

| Error | Type | Amount |
|---|---|---|
| DM1 X,Y offsets | known | +0 mm, +0 mm |
| DM1 additional X,Y offsets | unknown | +0.1 mm, +0.1 mm |
| DM1 clocking offset | known | +0.11° |
| DM1 clocking offset uncertainty | unknown | +0.1° |
| DM2 X,Y offsets | known | -0.05 mm, -0.05 mm |
| DM2 additional X,Y offsets | unknown | -0.05 mm, -0.05 mm |
| DM2 clocking offset | known | -0.11° |
| DM2 clocking offset uncertainty | unknown | -0.1° |
| DM spatial scale uncertainty | unknown | 0.3% |
| FPM clocking offset | known | +1° |
| FPM clocking offset uncertainty | unknown | +0.14° |
| FPM dielectric bias uncertainty | unknown | -10 nm |
| FPM dielectric height uncertainty | unknown | 5% |
| FPM dielectric misalignment | unknown | 0.5 μm |
| FPM nickel bias uncertainty | unknown | 5% |
| FPM offset uncertainty | unknown | 0.82 μm |
| FPM spatial scale uncertainty | unknown | 0.4% |
| FPM offset along optical axis | unknown | 70 μm |
| CGI shear relative to IC | known | 0.14% of beam diameter |
| Phase retrieval error | unknown | 10% |

**Table 7. Time-variable features included in the optical model for OS 11.**

| Errors |
|---|
| FSM-corrected pointing jitter |
| Z4 – Z37 aberration changes |
| Secondary mirror X,Y,Z displacements |
| TCA optics X,Y displacements |
| IC-CGI interface X,Y displacements (at FSM) |
| Z4 correction with FCM |
| Z5 – Z11 correction with DM1 |
| DM1 & DM2 X,Y offsets |
| DM1 & DM2 thermally-induced surface change |
| Shaped pupil mask X,Y displacements |
| Lyot stop X,Y displacements |

*8.5.2  Generating the time series images*

The STOP-predicted alignment shifts, aberrations, and the LOCAM-derived Z4 – Z11 corrections (Table 7) are introduced into the system at 1 minute intervals. At each timestep the wavefront is propagated through the full system. This stage is very computationally intensive – there are 1830 timesteps in OS11. The HLC simulations use 7 wavelengths across the bandpass, and with 4 polarization components there are a total of 51240 E-fields generated. The SPC-WFOV uses 11 wavelengths (its larger field requires finer wavelength sampling to produce smooth speckles at the field edge). The SPC-Spec simulations, which are intended as inputs to a separate spectrographic simulator, require much finer spectral sampling and cover a broader bandpass, so 31 wavelengths are used to produce 226,920 E-fields. In all cases the fields are sampled at $0.1\ \lambda_c/D$ resolution (the SPC-Spec E-field file is 137 GB in size, with single-precision complex values).

Using the techniques described in Section 5.7, at each timestep tens of pointing-offset images are generated to represent jitter and finite stellar diameter ($D = 0.4$ mas for the reference star, 0.9 mas for the target). For the conservative case, pointing offsets are multiplied by 1.6, which approximates a ~2× increase in contrast sensitivity to tip/tilt errors. The final HLC and SPC-WFOV intensity images are combined over polarization and wavelength (with appropriate spectral weighting) and then binned to the detector pixel resolution. These images, along with those from OS6 and OS9, are available to the community[100].

Including the STOP and diffraction modeling, it takes weeks to generate the suite of OS11 images (the bulk of which is running the thermal model), so it is an expensive endeavor that can only be done once every year or two. Note: The OS11 sequence includes a span in the middle to refresh the dark hole solution using one iteration of EFC, in case the system drifts were large enough to require it. In the time series generated here, the solution was left alone because of the high stability of the system.

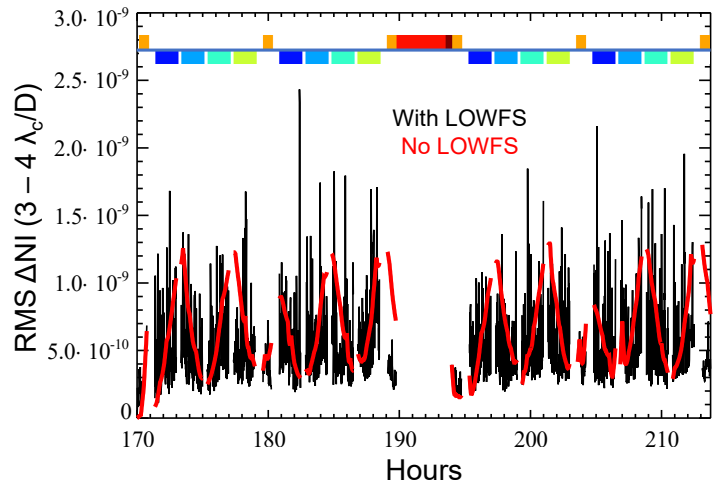

**Figure 75. OS11 HLC Band 1 RMS contrast change (default case) near the IWA relative to the image at $t = 170$ h with 1 minute cadence. The red line is without control of Z4 – Z11 variations with LOWFS and black is with. Jitter and detector noise are not included.**

### 8.5.3  OS11 HLC speckle variations

In the absence of detector noise, the speckle variations in OS11 are dominated either by LOCAM measurement noise or pointing jitter, depending on the time. Figure 75 plots at each timestep the RMS of the difference relative to the first timestep of pixels near the IWA for the HLC OS11 default series, with and without compensation of $Z4 - Z11$ using LOCAM (jitter is omitted). The non-LOWFS plot shows smooth, cyclic variations that are mainly due to the uncorrected coma, to which the HLC is particularly sensitive (Figure 5). The corrected values have high-temporal-frequency noise in comparison, largely an artifact of the LOCAM settings having been tuned for the greater aberration variations allowed in the stability error budget.

At first glance it may appear that the LOCAM is not providing any benefit besides tip/tilt correction. A video of the variations, showing the effects of correction, is provided in Figure 76 and demonstrates its utility. The critical metric is the mean stability within each roll/cycle. Figure 77 plots the mean of the target star images within each roll and within each cycle subtracted by the mean of the reference star images bounding each cycle (i.e., applying RDI post-processing). Over three of the four cycles the results with LOWFS are equal to or better than the non-LOWFS ones. In the presence of larger aberration variations, the advantages of using LOCAM would be more apparent.

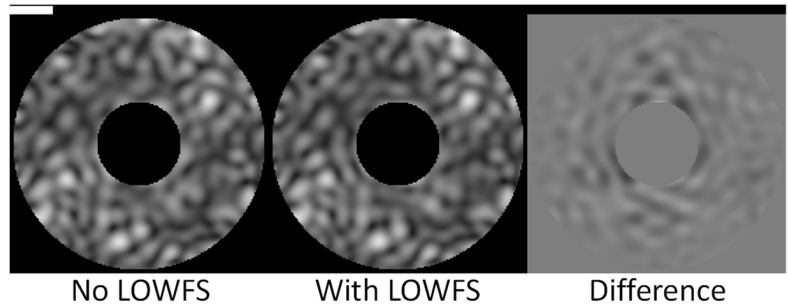

No LOWFS      With LOWFS      Difference

**Figure 76. A frame from the video, available with the online publication, that shows the HLC Band 1 speckle variations over OS11, without and with LOCAM corrections of $Z4 - Z11$ (no jitter or detector noise is included). The square-root of the intensity is shown in the two images on the left, and the difference image is shown with a linear stretch, independently scaled. The bar at top indicates the relative frame number within the OS11 distribution. (MP4, 6.1 MB).**

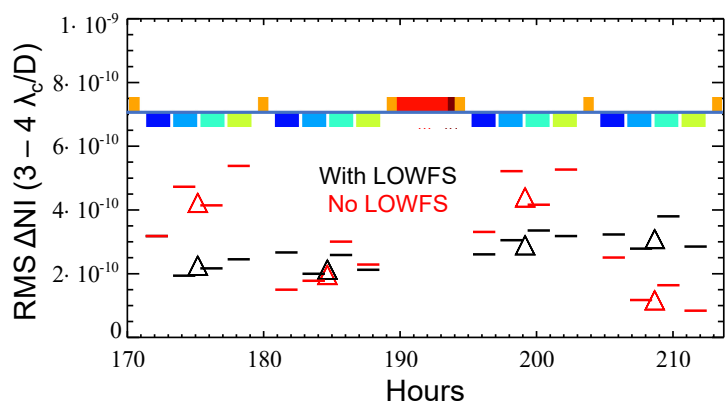

**Figure 77. OS11 HLC Band 1 RMS contrast change (default case) near the IWA. Each line segment represents the mean of the target images during that roll orientation subtracted by the mean of the bounding reference star images. The triangles represent the mean of all target images within a cycle subtracted by the mean of the bounding reference star images. The red symbols are without compensation of $Z4 - Z11$ variations with LOWFS and black are with. Jitter and detector noise are not included.**

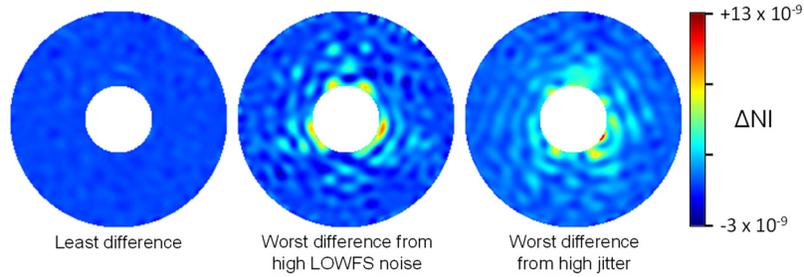

**Figure 78. Maps of OS11 HLC Band 1 contrast changes relative to $t$ = 170 h: (left) the timestep with the least RMS variation; (middle) the largest variation due to high LOWFS noise; (right) the largest variation due to high jitter. The annulus shown is $r = 3 - 9.7\ \lambda_c/D$. Detector noise is not included.**

When jitter is included, the short-timescale speckle variations continue to be dominated by LOWFS noise over the majority of OS11, except during those short intervals when the reaction wheel speeds are high. During hours 171 and 189, for example, the wheel speeds are near their maximum and the resulting jitter is relatively large (Figure 71). Figure 78 shows the best and worst-case differences between the first and subsequent timesteps. Note that the difference due to LOWFS noise has more isolated speckles that may be mistaken for exoplanets, while that due to jitter has more arc-shaped residuals. This may be due to the correction of low-order aberrations with a high-order DM.

Figure 79 replicates the RDI analysis used for Figure 77, except it now includes both LOWFS and jitter variations and shows results for the default and conservative cases. As one would expect by doubling the contrast sensitivity to jitter, the conservative case RMS contrast of ~5.5 × 10⁻¹⁰ is about twice that of the default.

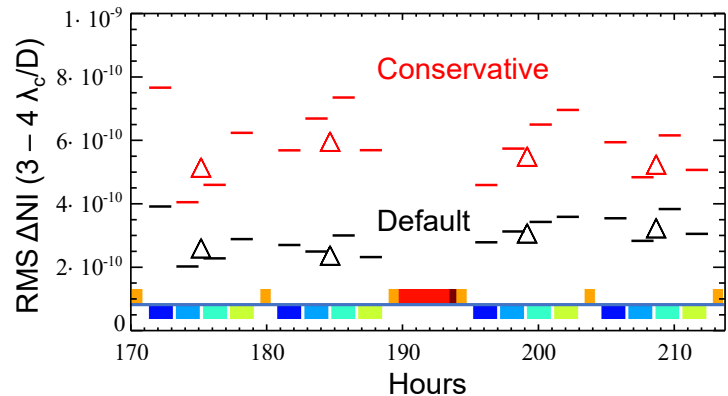

**Figure 79. OS11 HLC Band 1 RMS contrast change near the IWA for the default and conservative cases, including LOWFS and pointing jitter but no detector noise. See Figure 77 for a description of the symbols.**

### 8.5.4 OS11 SPC speckle variations

The results for the SPC modes show similar behaviors as the HLC, with jitter dominating the time variations in both modes. More interesting is the chromatic variations of the speckles over wavelength in SPC-Spec (Figure 80), as they may confuse extraction of exoplanet spectral features. In ground-based spectrographs and those on *HST* and *JWST*, the speckles are dominated by phase variations, so they behave in a generally predictable way over a limited bandpass – the speckle field will appear to grow with increasing wavelength. In CGI, however, the post-EFC dark hole is a complex mixture of phase and amplitude contributions, and the hole appears to "boil" with changing wavelength.

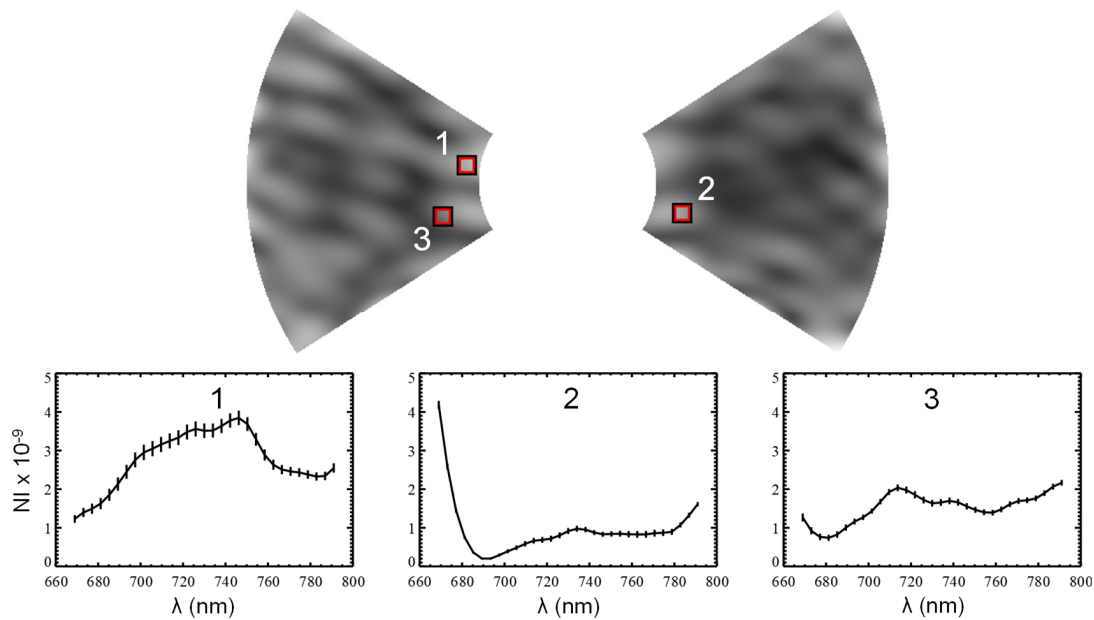

**Figure 80. The mean brightness variations over OS11 versus wavelength are plotted within three different regions of the SPC-Spec Band 3 field (default case, flat input spectrum). The error bars indicate the standard deviation of the variation over OS11.**

### 8.6 Post-processing of simulations

Once the noiseless images have been generated, they can be interpolated spatially to match the detector sampling and over time to match the framerate. The results can then be sent to the detector model and then through a photon-counting algorithm (recall that photon counting is not done on the detector). These final images (Figure 81) can finally be processed to remove the stellar signal.

Classical RDI (Figure 82) is the simplest of the post-processing techniques. It is baselined in the analytical CGI performance error budget as its effect on the final instrumental background can be reasonably predicted from derived coronagraphic sensitivities and specified stability parameters. In a perfectly static system, RDI would be the optimal method, and the result would be limited purely by detector noise rather than target/reference speckle mismatches. More advanced algorithms[101,102] have been developed to deal with unstable speckles, most using principal component analysis (PCA) to derive from the time series itself the speckle modes corresponding to the most significant wavefront instability contributors. They combine weighted modes to create an optimal reference field to subtract at each timestep. $\kappa_{pp}$ is included in the error budget as the post-processing gain, with a value >1 indicating an improvement over RDI.

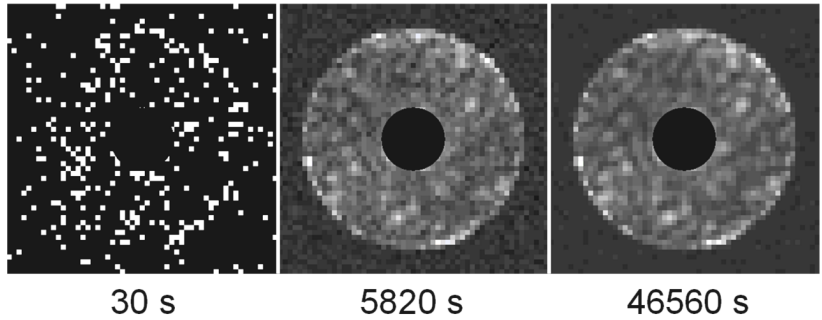

**Figure 81. Photon-counted OS11 dark hole images of the target star, 47 UMa, for the HLC Band 1 (default case, -13° roll). A frame rate of 30 s and gain of 5000 was assumed. The single frame on the left shows one-photon counts. The others show the sum of photon-counted frames for one of the roll intervals within a cycle (5820 s) and over all cycles (46560 s). The central $r = 2.8\ \lambda_c/D$ has been masked. Two planets have been inserted (see Figure 82 for planet locations).**

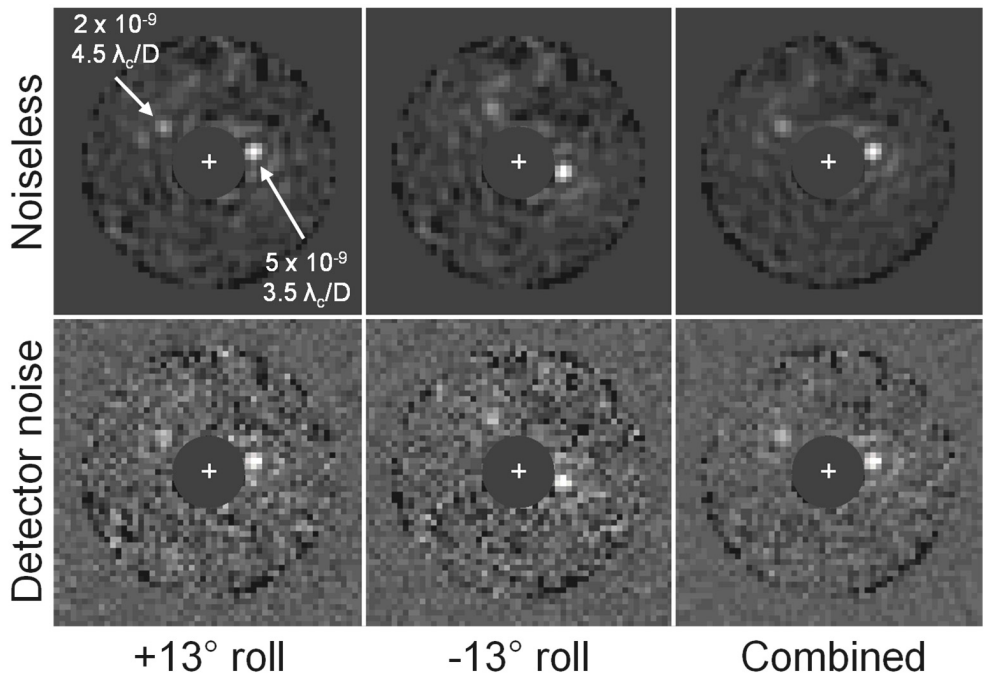

**Figure 82. RDI results for OS11 HLC Band 1 images (default case), with and without detector noise. Each roll orientation was processed separately, and the results were registered via interpolation and combined. Two planets at the indicated flux ratios and separations from the star are included.**

While the improvement from RDI can be predicted, $\kappa_{pp}$ is not as easily computed, so it must be derived from simulations. PCA post-processing[102] has been applied to the earlier HLC OS9 time series, and the results show that in the absence of detector noise the $\kappa_{pp}$ is $1.7 - 2.0$ while with noise it is $1.2 - 1.5$. Because these advanced methods derive the modes from the time variation in the fields, the low signal and high stability confuses any speckle variations with detector noise. On ground-based telescopes and *HST*, which are significantly less stable than the predictions for *Roman* and have worse contrasts, the advanced techniques can provide greater improvement factors.

The prior time series results have been used for other post-processing studies as well. The extractions of an exoplanet spectra from both simulated IFS[103] and slit[104] data have been demonstrated. The slit data results showed that classical RDI produced better results than PCA RDI, which provided a $\kappa_{pp}$ of 0.8 (ADI post-processing was not investigated). These earlier simulations have also been used as part of the CGI Community Exoplanet Imaging Challenge[105], which engaged various teams to try their post-processing algorithms on simulations in which exoplanets were embedded. They competed to produce the most accurate estimates of planet photometry and astrometry.

The imaging of circumstellar dust disks has been investigated[106], with and without using the polarizers. Rapid methods for simulating the effect of the field-dependent PSF has been developed[107,108], and disk models have been made available to the community[109] for additional studies.

In addition to science simulations, the OS9 results have provided the basis for an exploration[110] of alternative methods of dark hole maintenance that do not require repeated visits to the reference star and instead rely on modulations of the DM patterns.

## 9   Lessons learned and conclusions

As the first advanced space coronagraph, CGI has had the time and resources to develop and exercise fully detailed diffraction and STOP models. This experience provides the foundation for planning the next generation of dedicated high contrast missions like *Habitable Worlds Observatory*. We discuss here some of the knowledge gained, including some non-technical details, such as resource allocations.

### 9.1  Diffraction modeling lessons

Numerical simulations were first used for CGI in the downselect process, during which a representative diffraction model with realistic optical aberrations was created to evaluate the performance of the various proposed coronagraphs. With a common system model and wavefront control algorithm, the designs could be fairly competed against each other. After downselect, the Integrated Modeling team was formally established by the project as part of the Systems Engineering group. For the first few years the team was focused primarily on diffraction and wavefront control modeling, validating the models against testbed experiments, and validating the CGI error budgets against the models. The studies concentrated on mask evaluation and wavefront control optimization, where various instrumental parameters and techniques were explored and requirements were established. At this stage, up to 4 full-time diffraction modelers were needed to handle these tasks. Finding analysts with a combination of diffraction and wavefront control modeling experience took some effort, as these tasks are not nearly as common or standardized as structural, thermal, or optical ray-trace modeling, so some on-the-job training was required. Most analysts were fluent in Matlab, followed by Python and then IDL; PROPER being available in all these languages facilitated use of a common model (it seems likely that future missions will rely more on the freely-available Python). Over the past decade the pool of those with relevant experience has grown substantially, with coronagraphy and the associated wavefront control techniques becoming more common topics of graduate-level studies at universities. Besides diffraction modelers, there was frequent need for experienced ray tracing analysts, so dedicated personnel would be needed for a flagship-level mission.

Once the CGI project reached the implementation phase, by which time the major trade studies had concluded, the diffraction modeling group downsized to 2 analysts at ~70% time each, and it will remain so through instrument integration and testing. The diffraction modelers have some long-term tasks, such as generation of the observing scenario time series or calculating sensitivities for the performance error budgets. However, most of their work involves rapid analyses to assess risks or problems, such as potential or measured component imperfections. It is thus important to have experienced analysts available even when no specific long-term modeling tasks are forecasted.

It is important for the modeling team to maintain a common, version-controlled model. This was complicated by using the three different languages supported by PROPER. While an official model would be regularly updated and provided in each language, each analyst would inevitably modify the code to implement whatever mask change, misalignment, or DM defect that they may be tasked with studying at that moment. The complications of rapidly incorporating these changes in the other languages became burdensome, with delays in implementation making cross-validation within the team difficult. Whenever a new version of the official model was issued (due to a layout change, for instance), each analyst would then have to decide to integrate their latest modifications into the new code or purge them. The alternative would be to strictly enforce that all modifications be implemented in the official model in all languages at the same time, something that would not be practical in many cases where rapid responses are needed. Future missions should consider a compromise plan, where analysts may modify their copy of the model temporarily but then those changes must either be formally incorporated into the official model or purged from any models used later. It is important, that each model, modified or not, used to generate reported results be archived to reproduce those later if needed. If, as expected, Python becomes the de-facto computational environment, then the difficulties of mixed-language models will be irrelevant.

Over the years on the CGI project, numerous modeling codes, both public and proprietary, have been used for different analyses. When the codes are mixed and matched, it has proven important to verify agreement between them in terms of inputs, intermediate results such as conversion of DM commands into OPD maps, thin film calculations, and so on. Agreement between models will be even more important for *HWO*, with its more demanding contrast and stability requirements.

While diffraction modeling is specialized and often precludes quick results, it has proven to be an instrumental tool in predicting and understanding the behavior of CGI. There have been times when diffraction-based results have at first appeared contradictory to simpler and faster analytical or statistical models, and reconciliation revealed missing contributions to the analytical models requiring additional terms. Any analytical model, such as the performance error budget, needs verification against a detailed numerical one. The diffraction models also help set expectations for and troubleshoot testbed and later instrument test results, and it is worth ensuring that modelers gain testbed experience.

## 9.2 STOP modeling lessons

As the *Roman* and CGI projects progressed into their preliminary design phases, STOP modeling was introduced, including development of the CGI Integrated Modeling Pipeline. While the diffraction and STOP modeling efforts were initially coordinated under one task, it was decided that having separate groups, each led by a subject expert, would be more efficient. The structural and thermal modelers (one person each) were shared between the modeling and Mechanical/Thermal/Structural (MTS) team, and an analyst versed in Sigfit and CodeV was employed part-time, along with the pipeline developer. The STOP team lead was responsible for analyzing the results, including providing the outputs of the OS STOP runs to the diffraction modeling team. The dynamic (jitter) modeling was separately undertaken by the CGI Pointing Acquisition and Control Element team.

While the diffraction modelers usually worked independently of the overall *Roman* project modeling effort led by GSFC, the STOP group was much more involved, given that they had to deal with mechanical interfaces and the thermal variations from the observatory and follow the same Math Model Guidelines. As discussed in Section 8.2, the integrated modeling team at GSFC was responsible for creating the observatory structural and thermal models, including integrating the telescope model from the vendor and the CGI model from the JPL team. CGI, as a technology demonstrator, cannot impose requirements on the observatory, so the JPL team was not fully immersed into the *Roman* project's engineering processes, leading to occasional issues (e.g., Section 8.3). The *Roman* project, however, readily accommodated CGI requests when they did not impact schedule or budget. For a flagship coronagraphic mission, where the telescope requirements will be set by the coronagraphic performance, a strong collaboration between the observatory and coronagraph design and modeling teams will be critical.

As the observatory and CGI progress through their various design, fabrication, and testing stages, the fidelity of the observatory and CGI STOP models increased significantly. In the early stages, they included many structural elements with undefined properties (including zero thermal expansion coefficients), and thermal control loops and pointing error corrections took time to optimize, sometimes resulting in significantly different outcomes. Studies[111,112] of future potential coronagraphic missions have undertaken limited actual or notional STOP modeling, both thermal and dynamic, to predict wavefront stability. Based on the evolution seen in both the observatory and CGI STOP models over time, the results of these simpler models should include large margins, with some MUFs even greater than those applied by CGI.

The primary limitation for generating more observing scenarios is the time required for STOP modeling. Whereas the diffraction model can be sped up using specifically optimized code running on GPUs, the need to run the STOP model on validated, industry-standard (though not necessarily speed-optimized) software, such as Thermal Desktop, will continue to draw out the OS simulation time to many days.

## 9.3  Conclusions

CGI represents the culmination of nearly three decades of effort to put into space a coronagraph capable of imaging and characterizing $\sim 10^{-9}$ contrast giant exoplanets. It is a steppingstone to future missions such as *HWO* that will search for signs of life on $\sim 10^{-10}$ contrast terrestrial planets. It has advanced many critical technologies: obscured-system coronagraph design, DM characterization and control, mask fabrication, wavefront sensing and control, and space-qualified photon-counting detectors. While officially a technology demonstrator, upon the completion of its associated tasks, it is hoped that it will become a scientific instrument capable of unequaled exoplanet and circumstellar disk science.

Given the limitations of testbed experiments and pre-launch tests (including the inability to evaluate performance with the telescope on the ground), the CGI project has expended unprecedented resources over the past decade to conduct high fidelity, coordinated numerical diffraction and STOP modeling. These have allowed investigations of wavefront control optimization, tolerances to fabrication errors, and sensitivities to thermally-and-dynamically-induced misalignments and optical aberrations. They have been used to produce sets of simulated time sequences of coronagraphic observations available for the community to evaluate post-processing algorithms, determine science performance, and plan observations. The propagation software and models (e.g., PROPER, LOWFSSim, CGISim, etc.) have been made publicly available and can be used as examples for the future modeling that will be needed for *HWO*, which will be even more difficult given its likely segmented system and more stringent contrast requirements.

## Appendix A: List of acronyms

| | |
|---|---|
| ADI | Angular differential imaging |
| AR | Anti-reflection (coating) |
| CCD | Charge coupled device |
| CFAM | Color filter alignment mechanism |
| CGI | Coronagraph instrument |
| DAC | Digital-to-analog converter or Deployable aperture cover |
| DM | Deformable mirror |
| DPAM | Dispersion/polarization alignment mechanism |
| EFC | Electric field conjugation |
| EMCCD | Electron-multiplied charge coupled device |
| EXCAM | Exoplanetary camera |
| FALCO | Fast linearized coronagraph optimizer |
| FCM | Focus control mechanism |
| FEA | Finite element analysis |
| FEM | Finite element model |
| FFT | Fast Fourier transform |
| FPAM | Focal plane alignment mechanism |
| FPM | Focal plane mask |
| FRN | Flux ratio noise |
| FSAM | Field stop alignment mechanism |
| FSM | Fast steering mirror |
| GITL | Ground-in-the-loop (wavefront control) |
| GSFC | Goddard Space Flight Center |
| HLC | Hybrid Lyot coronagraph |
| HOWFSC | High-order wavefront sensing & control |
| HST | Hubble Space Telescope |
| HWO | Habitable Worlds Observatory |
| IC | Instrument carrier |
| IDL | Interactive data language |
| IFS | Integral field spectrograph |
| IWA | Inner working angle |
| JPL | Jet Propulsion Laboratory |
| JWST | James Webb Space Telescope |
| LOCAM | Low-order wavefront senor camera |
| LOWFS | Low-order wavefront sensor |
| LSAM | Lyot stop alignment mechanism |
| MFT | Matrix Fourier transform |
| MUF | Modeling uncertain factor |
| NI | Normalized intensity |
| OAP | Off-axis parabola |
| OS | Observing scenario |
| OTA | Optical telescope assembly |
| OWA | Outer working angle |
| PCA | Principal component analysis |
| PDI | Polarization differential imaging |
| PIAACMC | Phase-induce amplitude apodization complex mask coronagraph |
| POMA | Pick-off mirror assembly |
| PR | Phase retrieval |
| PSD | Power spectral density |
| PSF | Point spread function |
| RDI | Reference differential imaging |
| RMS | Root-mean-square |
| SFE | Surface error |
| SNR | Signal-to-noise ratio |
| SPC | Shaped pupil coronagraph |
| SPC-Spec | Shaped pupil coronagraph for spectroscopy |
| SPC-WFOV | Shaped pupil coronagraph for wide-field imaging |
| STOP | Structural, thermal, & optical (modeling) |
| SVD | Singular value decomposition |
| TCA | Tertiary collimator assembly |
| TD | Thermal Desktop® |
| TTR | Technology demonstration threshold requirement |
| VSG | Vacuum surface gauge |
| WFC | Wavefront control |
| WFSC | Wavefront sensing and control |
| WFE | Wavefront error |
| WFI | Wide-field instrument |
| WFIRST | Wide-field infrared survey telescope |

## Appendix B: Wavefront control modes

One can gain some intuition regarding the impact of model errors on and behavior of wavefront control, including field weighting and regularization by looking at the modes (basis functions) that form the least-squares solution. These can be obtained using the singular value decomposition (SVD) of the Jacobian, $G$:

$$SVD(G) = U\Sigma V^{\mathrm{T}} \qquad (25)$$

The DM pattern and field changes for each mode are in $V$ and $U$, respectively, while $\Sigma$, the singular values, can be considered weights on the modes (regularization modifies $\Sigma$).

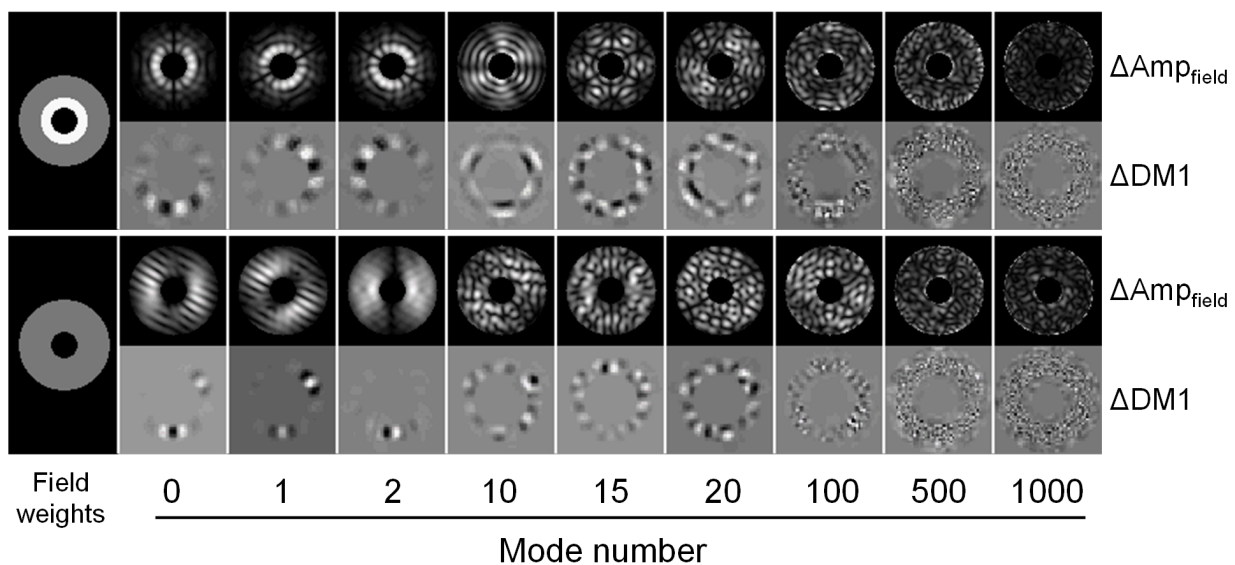

**Figure 83. SVD-derived basis functions for the HLC Band 1, up to mode 1000. In the bottom panel the field weighting is uniform, while the top panel shows the functions when field weights are 2× higher out to $r = 5$ $\lambda_c/D$. The top row of each panel shows the amplitude change at final focus for each mode, while the row below it shows the DM1 pattern change. Each image is individually scaled in intensity.**

Figure 83 shows the modes for the HLC in Band 1, with and without localized field weighting. SVD sorts them from strongest (mode 0, the largest amplitude changes relative to DM stroke) to weakest. The lower modes are sometimes called "easy" and the higher ones "hard", referring to how much stroke is required to alter the field. Some basic morphological characteristics are evident. The first modes have DM and field patterns having lower-spatial-frequency features, with small DM amplitudes producing field changes with relatively large amplitudes and extents. As the mode number increases, both the DM and field changes increase in spatial frequency, eventually becoming dominated by pixel-to-pixel variations; actuators in regions of the pupil blocked by the Lyot stop are activated, providing chromatic control by modifying the field prior to the FPM. At high modes, the sensitivities to model parameters like detector orientation become critical given the small field structures. Hence, there is a need for regularization to suppress these higher modes so that they do not pollute the overall solution.

Increasing the weights on field locations near the IWA results in notably different modes. Even with weighting, the DM pattern for the lowest mode is multiple spatial frequency cycles. The lowest mode shown in the weighted case of Figure 83 corresponds approximately to Z43. Lower spatial frequency aberrations, such as astigmatism (Z5,6) or coma (Z7,8), are significantly suppressed by the FPM and so do not have a corresponding mode. If one of these, say Z6, is the only aberration present, EFC will not find the globally optimal solution by putting -Z6 on the DM, but instead will introduce a less-optimal mix of the available modes. This emphasizes the need for flattening the wavefront using phase retrieval measurements prior to running EFC to correct low order aberrations.

The modes for SPC-Spec and SPC-WFOV are shown in Figure 84 and Figure 85, respectively.

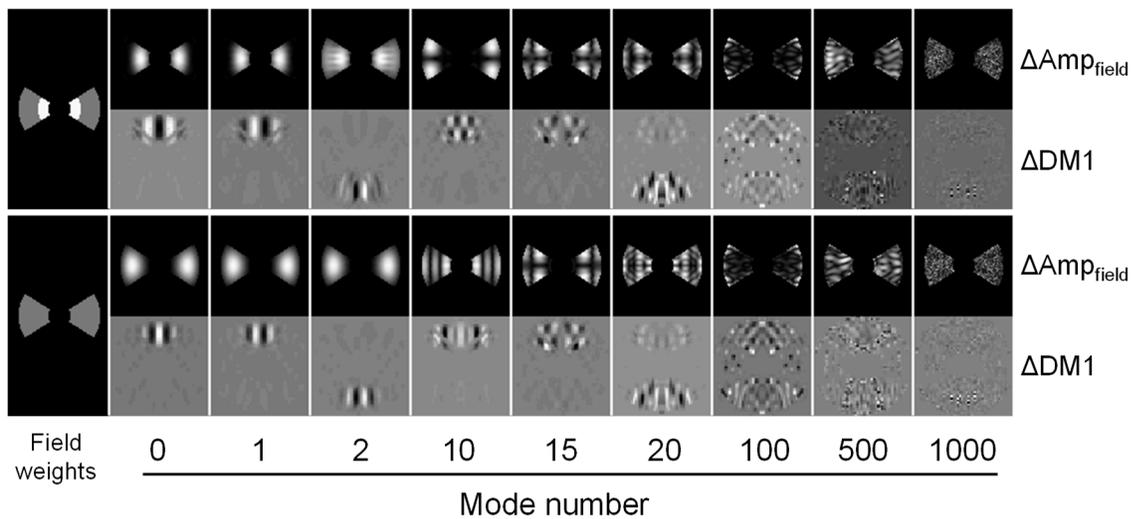

**Figure 84. SVD-derived basis functions for the SPC-Spec Band 3, up to mode 1000.**

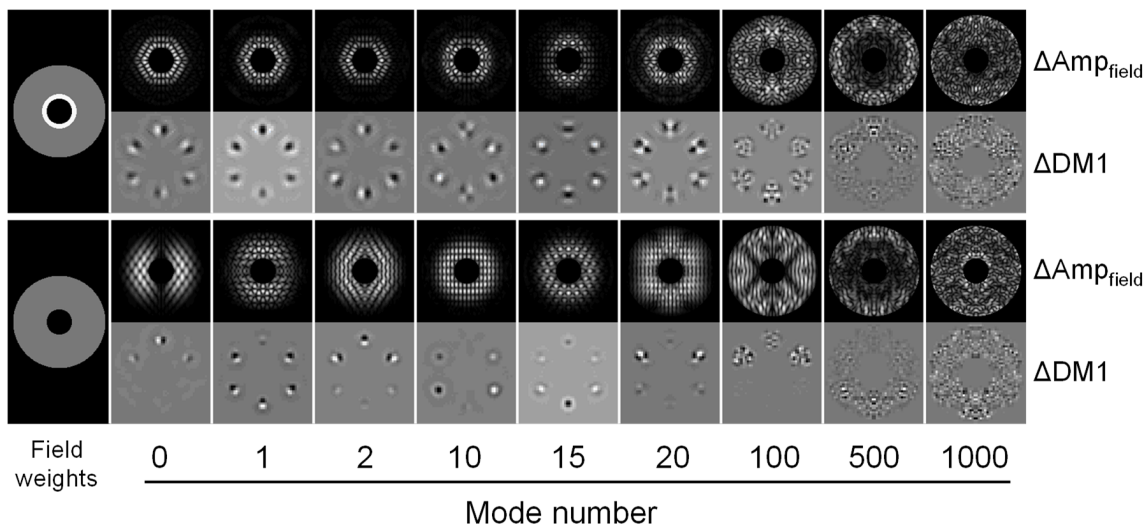

**Figure 85. SVD-derived basis functions for the SPC-WFOV Band 4, up to mode 1000.**

**Code, Data, and Materials Availability**

Unless otherwise specified herein for specific packages (e.g., PROPER, CGISim, LOWFSSim) and results (OS11 time series simulations), the input data and codes used to produce the presented results are not publicly available due to proprietary and export control constraints.

**Acknowledgments**

The authors thank the NASA/Goddard *Roman Space Telescope* Integrated Modeling Team, led by Alice Liu, for their thorough work on the observatory STOP and jitter models. We thank the following current and former members of the CGI team (at JPL unless otherwise noted) for their efforts on the various modeling, science planning, and instrument design, calibration, and testing: David Arndt, Vanessa Bailey, Robert Effinger, Gary Gutt, Josh Kempenaar, Kevin Ludwick (UAH), Luis Marchen, Sam Miller (UAH), Brian Monecelli, Patrick Morrissey, Charles Noecker, Erkin Sidick, and Hong Tang. We also thank the two anonymous referees who provided valuable comments and identified some errors. This work was carried out at the JPL, California Institute of Technology, under a contract with the National Aeronautics and Space Administration (Grant No. 80NM0018D0004).

**References**

[1] R. Soummer, et al. "Orbital motion of HR 8799 b, c, d using Hubble Space Telescope data from 1998: Constraints on inclination, eccentricity, and stability", *Astrophys. J.* **741**(1) 55 https://doi.org/10.1088/0004-637X/741/1/55

[2] J. Wang, et al., "Atmospheric Monitoring and Precise Spectroscopy of the HR 8799 Planets with SCExAO/CHARIS", *Astronon. J.* **164** 143 https://doi.org/10.3847/1538-3881/ac8984 (2022).

[3] B. Mennesson, et al., "The Roman Space Telescope coronagraph technology demonstration: current status and relevance to future missions," Proc. SPIE, 12180 121801W https://doi.org/10.1117/12.2629176 (2022).

[4] B. Gaudi, et al., "The Habitable Exoplanet Observatory (HabEx)," Proc. SPIE, 11115 111150M https://doi.org/10.1117/12.2530036 (2019).

[5] A. Roberge, M. Bolcar, K. France, "Telling the story of life in the cosmos: the LUVOIR telescope concepts," Proc. SPIE, 11115 111150O https://doi.org/10.1117/12.2530475 (2019).

[6] National Academies of Sciences, Engineering, and Medicine, "Pathways to Discovery in Astronomy and Astrophysics for the 2020s," The National Academies Press. https://doi.org/10.17226/26141 (2021).

[7] D. Spergel et al., "Wide-Field InfraRed Survey Telescope-Astrophysics Focused Telescope Assets WFIRST-AFTA Final Report," https://arxiv.org/abs/1305.5422, https://doi.org/10.48550/arXiv.1305.5422 (2013).

[8] I. Poberezhskiy et al., "Roman coronagraph instrument: engineering overview and status," Proc. SPIE, 12180 121801X https://doi.org/10.1117/12.2630540 (2022).

[9] J. Krist, "End-to-end numerical modeling of AFTA coronagraphs," Proc. SPIE, 9143 91430V https://doi.org/10.1117/12.2056759 (2014).

[10] J. Krist, B. Nemati, B. Mennesson, "Numerical modeling of the proposed WFIRST-AFTA coronagraphs and their predicted performances," *J. Astron. Telesc. Instrum. Syst.*, **2**(1) 011003 https://doi.org/10.1117/1.JATIS.2.1.011003 (2015).

[11] W. Traub, et al., "Science yield estimation for AFTA coronagraphs," Proc. SPIE, 9143 91430N https://doi.org/10.1117/12.2054834 (2014).

[12] J. Trauger, D. Moody, J. Krist, B. Gordon, "Hybrid Lyot Coronagraph for WFIRST-AFTA: coronagraph design and performance metrics," *J. Astron. Telesc. Instrum. Syst.* **2**(1), 011013, https://dx.doi.org/10.1117/1.JATIS.2.1.011013 (2016).

[13] A. Carlotti, N. J. Kasdin, R. Vanderbei, "Shaped pupil coronagraphy with WFIRST-AFTA," Proc. SPIE, 8864 886410 https://doi.org/10.1117/12.2024096 (2013).

[14] B. Kern, et al., "Phase-induced amplitude apodization complex mask coronagraph mask fabrication, characterization, and modeling for WFIRST-AFTA," *J. Astron. Telesc. Instrum. Syst.,* **2**(1) 011014 https://doi.org/10.1117/1.JATIS.2.1.011014 (2016).

[15] B.-J. Seo, et al., "Hybrid Lyot coronagraph for WFIRST: high-contrast broadband testbed demonstration," Proc. SPIE, 10400 104000F https://doi.org/10.1117/12.2274687 (2017).

[16] B.-J. Seo, et al., "Hybrid lyot coronagraph for WFIRST: high contrast testbed demonstration in flight-like low flux environment," Proc. SPIE, 10698 106982P https://doi.org/10.1117/12.2314358 (2018).

[17] D. Marx, et al., "Shaped pupil coronagraph: disk science mask experimental verification and testing," Proc. SPIE, 10698 106981E https://doi.org/10.1117/12.2312602 (2018).

[18] F. Shi, et al., "WFIRST low order wavefront sensing and control dynamic testbed performance under the flight like photon flux," Proc. SPIE, 10698 106982O https://doi.org/10.1117/12.2312746 (2018).

[19] E. Cady, et al., "Shaped pupil coronagraphy for WFIRST: high-contrast broadband testbed demonstration," Proc. SPIE, 10400 104000E https://doi.org/10.1117/12.2272834 (2017).

[20] H. Zhou, et al., B. Nemati, J. Krist, E. Cady, C. Prada, B. Kern, I. Poberezhskiy, "Closing the contrast gap between testbed and model prediction with WFIRST-CGI shaped pupil coronagraph," Proc. SPIE, 9904 990419 https://doi.org/10.1117/12.2232211 (2016).

[21] H. Zhou, J. Krist, B.-J. Seo, B. Kern, E. Cady, I. Poberezhskiy, "Roman CGI testbed HOWFSC modeling and validation," Proc. SPIE, 11443 114431W https://doi.org/10.1117/12.2561087 (2020).

[22] B. Mennesson, et al., "WFIRST-AFTA coronagraph performance: feedback from post-processing studies to overall design," Proc. SPIE, 9605 96050D https://doi.org/10.1117/12.2189403 (2015).

[23] J. Krist, B. Nemati, B. Mennesson, "Numerical modeling of the proposed WFIRST-AFTA coronagraphs and their predicted performances," *J. Astron. Telesc. Instrum. Syst.* **2**(1) 011003 https://doi.org/10.1117/1.JATIS.2.1.011003 (2016).

[24] F. Shi, et al., "Low-order wavefront sensing and control for WFIRST-AFTA coronagraph," *J. Astron. Telesc. Instrum. Syst.* **2**(1) 011021 https://doi.org/10.1117/1.JATIS.2.1.011021 (2016).

[25] A. J. Riggs, et a., "Flight mask designs of the Roman Space Telescope coronagraph instrument," Proc. SPIE, 11823 118231Y https://doi.org/10.1117/12.2598599 (2021).

[26] T. Whitman, et al., "Roman optical telescope assembly (OTA) build and integration progress," Proc. SPIE, 12180 121801N  https://doi.org/10.1117/12.2630105 (2022).

[27] J. Cavaco & A. Wirth, "Deformable mirror designs for extreme AO (XAO)", Proc. SPIE, 9148 914823 https://doi.org/10.1117/12.2057374 (2014).

[28] T. Groff, et al., "Roman Space Telescope CGI: prism and polarizer characterization modes," Proc. SPIE, 11443 114433D https://doi.org/10.1117/12.2562925 (2021).

[29] P. Morrissey, et al. "Flight photon counting electron multiplying charge coupled device development for the Roman Space Telescope coronagraph instrument", *J. Astron. Telesc. Instrum., Syst.* **9**(1) 016003 https://doi.org/10.1117/1.JATIS.9.1.016003 (2023).

[30] E. Bloemhof, J. Wallace, "Simple broadband implementation of a phase contrast wavefront sensor for adaptive optics," *Optics Express* **12**(25) 6240 (2004).

[31] E. Bendek, et al., "Enabling binary stars high-contrast imaging on the Roman Space Telescope coronagraph instrument," Proc. SPIE, 11823 118231I https://doi.org/10.1117/12.2594992 (2021).

[32] G. Ruane, et al., "Wavefront sensing and control in space-based coronagraph instruments using Zernike's phase-contrast method," *J. Astron. Telesc. Instrum. Syst.* **6**(4) 045005. https://doi.org/10.1117/1.JATIS.6.4.045005 (2020).

[33] A. J. Riggs, et al., "Fast linearized coronagraph optimizer (FALCO) I: a software toolbox for rapid coronagraphic design and wavefront correction," Proc. SPIE, 10698 106982V. https://doi.org/10.1117/12.2313812 (2018).

[34] https://github.com/ajeldorado/falco-matlab and https://github.com/ajeldorado/falco-python

[35] G. Ruane, et al., "Review of high-contrast imaging systems for current and future ground- and space-based telescopes I: coronagraph design methods and optical performance metrics," Proc. SPIE, 10698 106982S https://doi.org/10.1117/12.2312948 (2018).

[36] J. Llop-Sayson, et al., "Coronagraph design with the electric field conjugation algorithm," *J. Astron. Telesc. Instrum. Syst.* **8**(1) 015003 https://doi.org/10.1117/1.JATIS.8.1.015003 (2022).

[37] K. Balasubramanian, et al., "Critical characteristics of coronagraph masks influencing high contrast performance," Proc. SPIE, 1117 111171H https://doi.org/10.1117/12.2530825 (2019).

[38] A.J. Riggs, G. Ruane, B. Kern, "Directly constraining low-order aberration sensitivities in the WFIRST coronagraph design," Proc. SPIE, 1117 111170F https://doi.org/10.1117/12.2529588 (2019).

[39] E. Bloemhof, "Suppression of Speckle Noise by Speckle Pinning in Adaptive Optics", *Astrophys. J.* **582**(1) L59 https://doi.org/10.1086/346100 (2003).

[40] E. Douglas, et al., "Simulating the effects of exozodiacal dust in WFIRST CGI observations," Proc. SPIE 1117 111170K https://doi.org/10.1117/12.2529488 (2019).

[41] A. Carlotti, N. J. Kasdin, R. Vanderbei, "Shaped pupil coronagraphy with WFIRST-AFTA," Proc. SPIE, 8864 886410 https://doi.org/10.1117/12.2024096 (2013).

[42] K. Balasubramanian, et al., "WFIRST-AFTA coronagraph shaped pupil masks: design, fabrication, and characterization," *J. Astron. Telesc. Instrum. Syst.* **2**(1) 011005 https://doi.org/10.1117/1.JATIS.2.1.011005 (2016).

[43] A. J. Riggs, et al., "Shaped pupil coronagraph design improvements for the WFIRST coronagraph instrument," Proc. SPIE, 10400 104000O https://doi.org/10.1117/12.2274437 (2017).

[44] J. Gersh-Range, A.J. Riggs, N. Kasdin, "Flight designs and pupil error mitigation for the bowtie shaped pupil coronagraph on the Nancy Grace Roman Space Telescope," *J. Astron. Telesc. Instrum. Syst.* **8**(2) 025003 https://doi.org/10.1117/1.JATIS.8.2.025003 (2022).

[45] R. Soummer, et al. "High-contrast imager for complex aperture telescopes (HiCAT): 5. First results with segmented-aperture coronagraph and wavefront control", Proc. SPIE, 10698 106981O https://doi.org/10.1117/12.2314110 (2018).

[46] S. Shaklan, J. Green, "Reflectivity and optical surface height requirements in a broadband coronagraph.1.Contrast floor due to controllable spatial frequencies," *Applied Optics* **45**(21) 5143-5153 https://doi.org/10.1364/AO.45.005143 (2006).

[47] J. Krist, C. Burrows, "Phase-retrieval analysis of pre- and post-repair Hubble Space Telescope images," Applied Optics **34**(22) 4951-4964 https://doi.org/10.1364/AO.34.004951 (1995).

[48] A. Caillat, et al., "Super polished mirrors for the Roman Space Telescope CoronaGraphic Instrument: design, manufacturing, and optical performances," Proc. SPIE, 12180 121801Y https://doi.org/10.1117/12.2627240 (2022).

[49] J. Goodman, "Introduction to Fourier Optics, Second Edition." McGraw Hill, New York. (1988).

[50] J. Breckinridge, W. Lam, R. Chipman, "Polarization Aberrations in Astronomical Telescopes: The Point Spread Function", *Pub. Astron. of the Pacific* **127**(951), 445 https://doi.org/10.1086/681280 (2015).

[51] C. Thalmann, et al., "SPHERE ZIMPOL: overview and performance simulation", Proc. SPIE 7014 70143F https://doi.org/10.1117/12.789158 (2008).

[52] K. Follette, "An Introduction to High Contrast Differential Imaging of Exoplanets and Disks", *Proc. Astron. Soc. of the Pacific*, https://arxiv.org/abs/2308.01354 (accepted; 2023).

[53] D. Marx, et al., "Prediction and evaluation of the image of the WFIRST coronagraph pupil at the shaped-pupil mask," Proc. SPIE, 11443 1144339 https://doi.org/10.1117/12.2562629 (2020).

[54] G. Lawrence, "Optical Modeling," in *Applied Optics and Optical Engineering*, Vol. XI, Eds. R. Shannon, J. Wyant, Eds., pp. 125-200, Academic Press, (1992).

[55] J. Krist, "PROPER: an optical propagation library for IDL," Proc. SPIE, 6675 66750P https://doi.org/10.1117/12.731179 (2007).

[56] https://proper-library.sourceforge.net/

[57] R. Belikov, N. J. Kasdin, R. Vanderbei, "Diffraction-based Sensitivity Analysis of Apodized Pupil-mapping Systems," *Astrophys. J.* **652**(1) 833 https://dx.doi.org/10.1086/507941 (2006).

[58] J. Krist, et al., "Assessing the Performance of Internal Coronagraphs Through End-to-End Modeling," TDEM Study Final Report, Jet Propulsion Laboratory, https://exoplanets.nasa.gov/exep/technology/TDEM-awards/ (2013).

[59] R. Soummer, et al., "Fast computation of Lyot-style coronagraph propagation," *Optics Express*, **15**(24) 15935 https://doi.org/10.1364/OE.15.015935 (2007).

[60] https://sourceforge.net/projects/cgisim/

[61] https://github.com/wfirst-cgi

[62] B. Nemati, "Photon counting and precision photometry for the Roman Space Telescope Coronagraph," Proc. SPIE, 11443 114435F https://doi.org/10.1117/12.2575983 (2020).

[63] E. Douglas, et al., "A review of simulation and performance modeling tools for the Roman coronagraph instrument," Proc. SPIE, 11443 1144338 https://doi.org/10.1117/12.2561960 (2020).

[64] K. Milani, E. Douglas, J. Ashcraft, "Updated simulation tools for Roman coronagraph PSFs," Proc. SPIE, 11819 118190E https://doi.org/10.1117/12.2594807 (2021).

[65] https://poppy-optics.readthedocs.io/en/latest/

[66] R. Gerchberg, W. Saxton, "A Practical Algorithm for the Determination of Phase from Image and Diffraction Plane Pictures," *Optik*, **35**(2) 237-246 (1972).

[67] D. Misell, "A method for the solution of the phase problem in electron microscopy," *J. of Physics D:Applied Physics* **6**(1) https://dx.doi.org/10.1088/0022-3727/6/1/102 (1973).

[68] J. Krist & C. Burrows, "Phase retrieval analysis of pre-and-post repair Hubble Space Telescope Images", *Applied Optics* 34, 4951 https://doi.org/10.1364/AO.34.004951 (1995).

[69] A. Give'on, B. Kern, S. Shaklan, "Pair-wise, deformable mirror, image plane-based diversity electric field estimation for high contrast coronagraphy," Proc. SPIE, 8151 815110 https://doi.org/10.1117/12.895117 (2011).

[70] T. Groff, et al., "Methods and limitations of focal plane sensing, estimation, and control in high-contrast imaging," *J. Astron. Telesc. Instrum. Syst.* **2**(1) 011009 https://doi.org/10.1117/1.JATIS.2.1.011009 (2016).

[71] E. Cady, S. Shaklan, "Measurements of incoherent light and background structure at exo-Earth detection levels in the High Contrast Imaging Testbed," Proc. SPIE 9143 914338 https://doi.org/10.1117/12.2055271 (2014).

[72] B.-J. Seo, et al. "Hybrid lyot coronagraph for WFIRST: high contrast testbed demonstration in flight-like low flux environment," Proc. SPIE, 10698 106982P https://doi.org/10.1117/12.2314358 (2018).

[73] H. Zhou, J. Krist, B. Nemati, "Diffraction modeling of finite subband EFC probing on dark hole contrast with WFIRST-CGI shaped pupil coronagraph," Proc. SPIE, 9911 99111S https://doi.org/10.1117/12.2232129 (2016).

[74] A. Give'on, et al., "Broadband wavefront correction algorithm for high-contrast imaging systems," Proc. SPIE, 6691 66910A https://doi.org/10.1117/12.733122 (2007).

[75] J. Llop-Sayson, et al., "Coronagraph design with the electric field conjugation algorithm," *J. Astron. Telesc. Instrum. Syst.* **8**(1) 015003 https://doi.org/10.1117/1.JATIS.8.1.015003 (2022).

[76] E. Sidick, et al., "Optimizing the regularization in broadband wavefront control algorithm for WFIRST coronagraph," Proc. SPIE, 10400 1040022 https://doi.org/10.1117/12.2274440 (2017).

[77] D. Marx, et al., "Electric field conjugation in the presence of model uncertainty," Proc. SPIE, 10400 104000P https://doi.org/10.1117/12.2274541 (2017).

[78] H. Zhou, et al., "High accuracy coronagraph flight WFC model for WFIRST-CGI raw contrast sensitivity analysis," Proc. SPIE, 10698 106982M https://doi.org/10.1117/12.2313719 (2018).

[79] D. Marx, et al., "Shaped pupil coronagraph: disk science mask experimental verification and testing," Proc. SPIE, 10698 106981E https://doi.org/10.1117/12.2312602 (2018).

[80] S. Will, et al., "Wavefront control with algorithmic differentiation on the HiCAT testbed," Proc. SPIE, 11823 118230V https://doi.org/10.1117/12.2594283 (2021).

[81] S. Will, T. Groff, J. Fienup, "Jacobian-free coronagraphic wavefront control using nonlinear optimization," *J. Astron. Telesc. Instrum. Syst.* **7**(1) 019002 https://doi.org/10.1117/1.JATIS.7.1.019002 (2021).

[82] Dube, B., et al., "Exascale integrated modeling of low-order wavefront sensing and control for the Roman Coronagraph instrument," *J. Optical Society of America A* **39**(12) C133 https://doi.org/10.1364/JOSAA.472364 (2022).

[83] https://github.com/nasa-jpl/lowfssim

[84] https://github.com/brandondube/prysm

[85] B. Nemati, et al., "Method for deriving optical telescope performance specifications for Earth-detecting coronagraphs," *J. Astron. Telesc. Instrum. Syst.* **6**(3) 039002 https://doi.org/10.1117/1.JATIS.6.3.039002 (2020).

[86] B. Nemati, et al. "The Analytical Performance Model and Error Budget for the Roman Coronagraph" (submitted, 2023).

[87] https://roman.gsfc.nasa.gov/science/milestone_archive.html

[88] I. Poberezhskiy, et al., "Technology development towards WFIRST-AFTA coronagraph", Proc. SPIE 9143 91430P https://doi.org/10.1117/12.2060320 (2014).

[89] E. Sidick, et al., "Sensitivity of WFIRST coronagraph broadband contrast performance to DM actuator errors," Proc. SPIE, 10400 1040006 https://doi.org/10.1117/12.2274421 (2017).

[90] H. Zhou, et al., "Wavefront control performance modeling with WFIRST shaped pupil coronagraph testbed", Proc. SPIE 10400 1040005 https://doi.org/10.1117/12.2274391 (2017).

[91] E. Bendek, C. Prada, "Advancing deformable mirror controllers for exoplanet imaging," Proc. SPIE, 12180 121802E https://doi.org/10.1117/12.2630707 (2022).

[92] G. Ruane, et al., "Broadband vector vortex coronagraph testing at NASA's high contrast imaging testbed facility," Proc. SPIE, 12180 1218024 https://doi.org/10.1117/12.2628972 (2022).

[93] H. Zhou, et al. "Roman Coronagraph HOWFSC Modeling: Case Study and Latest Ground and In-Orbit Raw Contrast Prediction," Proc. SPIE, Manuscript in preparation (2023).

[94] J. Krist, et al., "Hubble Space Telescope Advanced Camera for Surveys Coronagraphic Imaging of the AU Microscopii Debris Disk", *Astron. J.* **129**(2) 1008 https://doi.org/10.1086/426755 (2005).

[95] J. Krist, et al., "Hubble Space Telescope Observations of the HD 202628 Debris Disk", *Astron. J.* **144**(2) 45 https://doi.org/10.1088/0004-6256/144/2/45 (2012).

[96] J. Gardner, et al. "The James Webb Space Telescope Mission", Pup. Astron. Soc. of the Pacific **135** 068001 https://doi.org/10.1088/1538-3873/acd1b5 (2023).

[97] N. Saini, et al., "IMPipeline: an integrated STOP modeling pipeline for the WFIRST coronagraph," Proc. SPIE, 10400 1040008 https://doi.org/10.1117/12.2274260 (2017).

[98] I. Poberezhskiy, et al., "Roman space telescope coronagraph: engineering design and operating concept," Proc. SPIE, 11443 114431V https://doi.org/10.1117/12.2563480 (2020).

[99] J. Krist, et al., "WFIRST coronagraph flight performance modeling," Proc. SPIE, 10698 106982K https://doi.org/10.1117/12.2310043 (2018).

[100] https://roman.ipac.caltech.edu/sims/Coronagraph_public_images.html

[101] R. Soummer, L. Pueyo, and J. Larkin, "Detection and Characterization of Exoplanets and Disks Using Projections on Karhunen-Loève Eigenimages", *Astrophys. J.* **755**(2) L28 http://dx.doi.org/10.1088/2041-8205/755/2/L28 (2012).

[102] M. Ygouf, et al., "Roman Coronagraph Instrument Post Processing Report – OS9 HLC Distribution," https://roman.ipac.caltech.edu/sims/Coronagraph_public_images.html (2021).

[103] M. Rizzo, et al., "WFIRST CGI integral field spectrograph performance and post-processing in the OS6 observing scenario," Proc. SPIE, 10698 106986U https://doi.org/10.1117/12.2312400 (2018).

[104] M. Ygouf, N. Zimmerman, V. Bailey., "Roman Coronagraph Instrument Post Processing Report – OS9 SPC Distribution," https://roman.ipac.caltech.edu/sims/Coronagraph_public_images.html (2022).

[105] J. Girard, et al., "The Roman exoplanet imaging data challenge: a major community engagement effort," Proc. SPIE, 11443 1144337 https://doi.org/10.1117/12.2561736 (2020).

[106] R. Anche, et al., "Simulations of polarimetric observations of debris disks through the Roman Coronagraph Instrument," Proc. SPIE, 12180 1218056 https://doi.org/10.1117/12.2629497 (2022).

[107] E. Douglas, et al., "Simulating the effects of exozodiacal dust in WFIRST CGI observations," Proc. SPIE, 11117 111170K https://doi.org/10.1117/12.2529488 (2019).

[108] K. Milani, E. Douglas, "Faster imaging simulation through complex systems: a coronagraphic example," Proc. SPIE, 11484 1148405 https://doi.org/10.1117/12.2568204 (2020).

[109] https://roman.ipac.caltech.edu/sims/Circumstellar_Disk_Sims.html

[110] L. Pogorelyuk, et al., "Dark hole maintenance with modal pairwise probing in numerical simulations of Roman coronagraph instrument," *J. Astron. Telesc. Instrum. Syst.* **8**(1) 019002 https://doi.org/10.1117/1.JATIS.8.1.019002 (2022).

[111] J. Krist, et al., "Numerical modeling of the Habex coronagraph," Proc. SPIE, 11117 1111705 https://doi.org/10.1117/12.2530462 (2019).

[112] R. Juanola-Parramon, et al., "Modeling and performance analysis of the LUVOIR coronagraph instrument," *J. Astron. Telesc. Instrum. Syst.* **8**(3) 034001 https://doi.org/10.1117/1.JATIS.8.3.034001 (2022).